\documentclass[bibyear]{aa}  
\usepackage{graphicx}
\usepackage{txfonts}
\usepackage{lscape}

\newenvironment{itemlist}{
\begin{list}{}{\setlength{\leftmargin}{10mm}\setlength{\parsep}{0mm}
\setlength{\itemsep}{3mm}} }{\end{list}}
\usepackage{color}

\begin{document}

   \title{Hayabusa-2 Mission Target Asteroid 162173 Ryugu (1999~JU$_3$): Searching for
          the Object's Spin-Axis Orientation\thanks{This work includes space data from
          (i) {\it Herschel}, an ESA space observatory with science instruments provided
              by European-led Principal Investigator consortia and with important
              participation from NASA;
          (ii) Spitzer Space Telescope, which is operated by the Jet Propulsion Laboratory,
               California Institute of Technology under a contract with NASA;
          (iii) AKARI, a JAXA project with the participation of ESA.}
}
   \author{
          T.\ G.\ M\"{u}ller \inst{1},
          J.\ \v{D}urech \inst{2},
          M.\ Ishiguro \inst{3},
          M.\ Mueller \inst{4},
          T.\ Kr\"uhler \inst{1},
          H.\ Yang \inst{3},
          M.-J.\ Kim \inst{5},
          L.\ O'Rourke \inst{6},
          F.\ Usui \inst{7},
          C.\ Kiss \inst{8},
          B.\ Altieri \inst{6}, 
          B.\ Carry \inst{9},
          Y.-J.\ Choi \inst{5},
          M.\ Delbo \inst{10},
          J.\ P.\ Emery \inst{11},
          J.\ Greiner \inst{1},
          S.\ Hasegawa \inst{12},
          J.\ L.\ Hora \inst{13},
          F.\ Knust \inst{1},
          D.\ Kuroda \inst{14},
          D.\ Osip \inst{15},
          A.\ Rau \inst{1},
          A.\ Rivkin \inst{16},
          P.\ Schady \inst{1},
          J.\ Thomas-Osip \inst{15},
          D.\ Trilling \inst{17},
          S.\ Urakawa \inst{18},
          E.\ Vilenius \inst{19},
          P.\ Weissman \inst{20},
          P.\ Zeidler \inst{21}
          }
   \institute{   {Max-Planck-Institut f\"{u}r extraterrestrische Physik,           
                 Giessenbachstra{\ss}e, Postfach 1312, 85741 Garching, Germany;
                 tmueller@mpe.mpg.de
                 }
          \and   {Astronomical Institute, Faculty of Mathematics and Physics,      
                 Charles University,
                 V Hole\v{s}ovi\v{c}k\'{a}ch 2, 180 00, Praha 8, Czech Republic;
                 }
          \and   {Department of Physics and Astronomy, Seoul National University,  
                 Gwanak, Seoul 151-742, Korea
                 }
          \and   {Kapteyn Astronomical Institute, Rijksuniversiteit Groningen, 
                  Postbus 800, 9700 AV Groningen, The Netherlands
                 }
          \and   {Korea Astronomy and Space Science Institute, 776 Daedeokdae-ro,  
                 Yuseong-gu, 305-348 Daejeon, Korea
                 }
          \and   {European Space Astronomy Centre (ESAC), European Space Agency,   
                 28691 Villanueva de la Ca\~nada,
                 Madrid, Spain
                 }
          \and   {Center for Planetary Science, Graduate School of Science,        
                 Kobe University, 7-1-48, Minatojima-Minamimachi, Chuo-Ku,
                 Kobe 650-0047, Japan
                 }
          \and   {Konkoly Observatory, Research Center for Astronomy and           
                 Earth Sciences, Hungarian Academy of Sciences;
                 Konkoly Thege 15-17, H-1121 Budapest, Hungary
                 }
          \and   {IMCCE, Observatoire de Paris, UPMC Paris-06,                     
                 Universit\'e Lille1, UMR8028 CNRS, 77 Av.\
                 Denfert Rochereau, 75014 Paris, France
                 }
          \and   {Laboratoire Lagrange, UNS-CNRS, Observatoire de la               
                 C\^ote d'Azur, Boulevard de l'Observatoire-CS 34229,
                 06304 Nice Cedex 4, France
                 }
          \and   {Earth and Planetary Science Department \& Planetary              
                 Geosciences Institute, University of Tennessee, Knoxville,
                 TN 37996, USA
                 }
          \and   {Institute of Space and Astronautical Science,                    
                 Japan Aerospace Exploration Agency,
                 3-1-1 Yoshinodai, Sagamihara, Kanagawa 229-8510, Japan
                 }
          \and   {Harvard-Smithsonian Center for Astrophysics, 60 Garden Street,   
                 MS 65, Cambridge, MA 02138-1516, USA
                 }
          \and   {Okayama Astrophysical Observatory, National Astronomical         
                 Observatory of Japan, Honjo 3037-5, Kamogata, Asakuchi,
                 Okayama 719-0232, Japan
                 }
          \and   {Carnegie Observatories, Las Campanas Observatory, Casilla 60,    
                 La Serena, Chile
                 }
          \and   {Johns Hopkins University Applied Physics Laboratory,             
                 11101 Johns Hopkins Rd., Laurel, MD 20723, USA
                 }
          \and   {Northern Arizona University, Department of Physics and           
                 Astronomy, Bldg. 19, Rm. 209, Flagstaff, AZ 86011,
                 United States
                 }
          \and   {Bisei Spaceguard Center, Japan Spaceguard Association,           
                 1716-3 Okura, Bisei-cho, Ibara, Okayama 714-1411, Japan
                 }
          \and   {Max-Planck-Institut f\"ur Sonnensystemforschung,                 
                 Justus-von-Liebig-Weg 3,
                 37077 G\"ottingenMPS, Germany
                 }
          \and   {Planetary Science Institute, 1700 East Fort Lowell,              
                 Suite 106, Tucson, AZ 85719, USA
                 }
          \and   {Astronomisches Rechen-Institut, Zentrum f\"ur                    
                 Astronomie der Universit\"at Heidelberg,
                 M\"onchhofstr. 12-14, 69120 Heidelberg, Germany
                 }
          }
   \date{Received ; accepted }
\abstract{The JAXA Hayabusa-2 mission was approved in 2010 and launched on December 3, 2014.
The spacecraft will arrive at the near-Earth asteroid 162173 Ryugu (1999~JU$_{3}$) in 2018 where
it will perform a survey, land and obtain
surface material, then depart in December 2019 and return to Earth in December 2020.
We observed Ryugu with the Herschel Space Observatory
in April 2012 at far-infrared thermal wavelengths, supported by several ground-based
observations to obtain optical lightcurves. We reanalysed previously published Subaru-COMICS
and AKARI-IRC observations and merged them with a Spitzer-IRS data set. In
addition, we used a large set of Spitzer-IRAC observations obtained in the period January
to May, 2013. The data set includes two complete rotational lightcurves and a series of ten
"point-and-shoot" observations, all at 3.6 and 4.5\,$\mu$m.
The almost spherical shape of the target together with the insufficient lightcurve quality
forced us to combine radiometric and lightcurve inversion techniques in different ways
to find the object's spin-axis orientation, its shape and to improve the quality
of the key physical and thermal parameters.
Handling thermal data in inversion techniques remains challenging: thermal inertia,
roughness or local structures influence the temperature distribution on the surface.
The constraints for size, spin or thermal properties therefore heavily depend   on
the wavelengths of the observations.
We find that the solution which best matches our data sets leads to this C class asteroid
having a retrograde rotation with a spin-axis orientation of ($\lambda$ = 310$^{\circ}$ - 340$^{\circ}$;
$\beta$ = -40$^{\circ}$ $\pm$ $\sim$15$^{\circ}$) in ecliptic coordinates, an effective diameter
(of an equal-volume sphere)
of 850 to 880\,m, a geometric albedo of 0.044 to 0.050 and
a thermal inertia in the range 150 to 300\,J\,m$^{-2}$\,s$^{-0.5}$\,K$^{-1}$.
Based on estimated thermal conductivities of the top-layer surface in the range
0.1 to 0.6\,W\,K$^{-1}$\,m$^{-1}$, we calculated that the grain sizes are approximately equal to between $1$ and $10$ mm.
The finely constrained values for this asteroid serve as a
`design reference model', which is currently used for various planning, operational
and modelling purposes by the Hayabusa2 team.}
\authorrunning{M\"uller et al.}
\titlerunning{162173 Ryugu: Search for the spin-axis orientation}
\maketitle

\section{Introduction}

Remote observations and in-situ measurements of asteroids are considered
highly complementary in nature: remote sensing shows the global picture, but transforming
measured fluxes in physical quantities frequently depends upon model
assumptions to describe surface properties. In-situ techniques measure
physical quantities, such as\ size, shape, rotational
properties, geometric albedo or surface details, in a more direct way. However, in-situ techniques
are often limited in spatial/rotational/aspect coverage (flybys) and
wavelength coverage (mainly visual and near-IR wavelengths).
Mission targets are therefore important objects for a
comparison of properties derived from disk-integrated measurements
taken before arrival at the asteroid with those produced as output of
the in-situ measurements. The associated benefits are obvious: 
(i) the model techniques and output accuracies for remote, disk-integrated
observations can be validated (e.g., M\"uller et al.\ \cite{mueller14} for
the Hayabusa mission target 25143~Itokawa or O'Rourke et al.\ \cite{orourke12}
for the Rosetta flyby target 21~Lutetia); (ii) the model techniques can then be applied
to many similar objects which are not included in interplanetary mission
studies, but easily accessible by remote observations.
The pre-mission observations are also important for determining the object's
thermal and physical conditions in support for the construction
of the spacecraft and its instruments, and to prepare flyby, orbiting and
landing scenarios.

The JAXA Hayabusa-2 mission, approved in 2010, was successfully launched
on Dec.\ 3, 2014. It is expected to arrive at the asteroid
\object{162173 Ryugu} in 2018, survey the asteroid for a year
and a half, then land and obtain surface material, and finally depart
in December 2019, returning to Earth in December 2020.

  \begin{table*}[h!tb]
    \begin{center}
    \caption{Summary of previously published thermal and physical properties of 162173 Ryugu
             \label{tbl:previous_results}}
    \begin{tabular}{lcclcl}
      \hline
      \hline
      \noalign{\smallskip}
        D$_{eff}$ [km] & p$_V$ & shape & spin properties (fixed) & $\Gamma$ [Jm$^{-2}$s$^{-0.5}$K$^{-1}$] & Reference \\
      \noalign{\smallskip}
      \hline
      \noalign{\smallskip}
       0.92 $\pm$ 0.12 & 0.063$^{+0.020}_{-0.015}$ & a/b=1.21, b/c=1.0 & prograde, obliquity 0$^{\circ}$, P$_{sid}$=7.62722\,h & $>$500 & Hasegawa et al.\ (\cite{hasegawa08}) \\
      \noalign{\smallskip}
       0.90 $\pm$ 0.14 & 0.07 $\pm$ 0.01           & spherical shape & (1) equatorial view, retrograde & $>$150 & Campins et al.\ (\cite{campins09}) \\
       \multicolumn{1}{c}{$\prime \prime$}         & \multicolumn{1}{c}{$\prime \prime$}     & \multicolumn{1}{c}{$\prime \prime$} & (2) $\lambda_{ecl}$=331$^{\circ}$, $\beta_{ecl}$=+20$^{\circ}$; P$_{sid}$=7.62720\,h & 700 $\pm$ 200 & \multicolumn{1}{c}{$\prime \prime$} \\
      \noalign{\smallskip}
       0.87 $\pm$ 0.03 & 0.070 $\pm$ 0.006         & spherical shape & $\lambda_{ecl}$= 73$^{\circ}$, $\beta_{ecl}$=-62$^{\circ}$, P$_{sid}$=7.63\,h & 200-600 &
       M\"uller et al.\ (\cite{mueller11a}) \\
      \noalign{\smallskip}
       1.13 $\pm$ 0.03 & 0.042 $\pm$ 0.003 & polyhedron & $\lambda_{ecl}$= 73$^{\circ}$, $\beta_{ecl}$=-62$^{\circ}$, P$_{sid}$ not given & 300 $\pm$ 50 &
       Yu et al.\ (\cite{yu2014}) \\
     \noalign{\smallskip}
     \hline
    \end{tabular}
    \end{center}
  \end{table*}

For various Hayabusa-2 planning, operational and modelling activities, it is
crucial to know at least the basic characteristics of the mission target asteroid.
Previous publications (Table~\ref{tbl:previous_results}) 
presented shape solutions close to a sphere and a rotation
period of approximately 7.63\,h, but a range of possible solutions for Ryugu's
spin properties which were then tested against visual lightcurves and various
sets of thermal data, using different thermal models and assumptions
for Ryugu's surface properties:

\begin{itemlist}
\item[$\bullet$] Hasegawa et al.\ (\cite{hasegawa08}) assumed an equator-on observing
                 geometry (prograde rotation) for their radiometric analysis and fitted
                 a small set of thermal measurements (AKARI, Subaru).
\item[$\bullet$] Abe et al.\ (\cite{abe08}) found ($\lambda$, $\beta$)$_{ecl}$ =
                 (331.0$^{\circ}$, +20$^{\circ}$) and (327.3$^{\circ}$, +34.7$^{\circ}$),
                 indicating a prograde rotation. The solutions were based on applying
                 two different methods (epoch and amplitude methods) to the available
                 set of visual lightcurves.
\item[$\bullet$] Campins et al.\ (\cite{campins09}) were using the Abe et al.\ (\cite{abe08})
                 spin-axis solution, but also tested an extreme case of an equatorial retrograde geometry
                 ($\lambda$, $\beta$)$_{ecl}$ = (80$^{\circ}$, -80$^{\circ}$) against a
                 single-epoch Spitzer-IRS spectrum.
\item[$\bullet$] M\"uller et al.\ (\cite{mueller11a}) derived three possible solutions
                 for the spin-axis orientation based on a subset of the currently
                 existing data, but assuming a very high (and probably unrealistic)
                 surface roughness: ($\lambda$, $\beta$)$_{ecl}$ = (73$^{\circ}$, -62$^{\circ}$),
                 (69.6$^{\circ}$, -56.7$^{\circ}$), and (77.1$^{\circ}$, -30.9$^{\circ}$).
\item[$\bullet$] Yu et al.\ (\cite{yu2014}) reconstructed a shape model (from low-quality MPC
                 photometric points) under the assumption
                 of a rotation axis orientation with (73$^{\circ}$, -62$^{\circ}$)$_{ecl}$
                 and re-interpreted previously published thermal measurements.
\end{itemlist}

The radiometric studies have been performed using ground and space-based observations
(Table~\ref{tbl:previous_results} and references therein).
Disk-integrated thermal observations from ground (Subaru) and space (AKARI, Spitzer)
were combined with studies on reflected
light (light curves, phase curves and colours). Most studies agree on the object's
effective diameter of $\approx$900\,m, a geometric V-band albedo of 6-8\%,
an almost spherical shape (related to its low lightcurve amplitude) with a siderial
rotation period of approximately 7.63\,h
and a thermal inertia in the range 150 - 1000\,J\,m$^{-2}$\,s$^{-0.5}$\,K$^{-1}$.
A low-resolution near-IR spectrum (Pinilla-Alonso et al.\ \cite{pinilla13}) confirmed
the primitive nature of the C-type object Ryugu.
Two independent studies on the rotational characterisation of the Hayabusa2
target asteroid (Lazzaro et al.\ \cite{lazzaro13}; Moskovitz et al.\ \cite{moskovitz13})
found featureless spectra with very little variation, indicating a nearly homogeneous
surface.
However, one key element necessary for detailed mission planning and a final
radiometric analysis was still not settled: the object's spin-axis orientation.

The shape and spin properties of an asteroid are typically derived 
from inversion techniques (Kaasalainen \& Torppa \cite{kaasalainen01};
Kaasalainen et al.\ \cite{kaasalainen01a}) on the basis of multi-aspect
light curve observations. This procedure was previously applied to
\object{162173 Ryugu} and the results were presented by
M\"uller et al.\ (\cite{mueller11a}). We repeated the analysis this time
using the large, recently obtained  set of visual lightcurves.
The full data set of lightcurves includes measurements taken between July 2007 and July 2012,
covering a wide range of phase and aspect angles. But the very shallow light curve
amplitudes and the insufficient quality of many observations did not allow us
to derive a unique solution for the object's spin-axis orientation. Wide
ranges of pro- and retrograde orientations combined with different shape
models are compatible (in the least-square sense) with the combined
data set of all available lightcurves.

This forced us to combine lightcurve inversion techniques with radiometric
methods in a new way to find the object's spin-axis orientation, its shape
and to improve the quality of the key physical and thermal parameters
of 162173 Ryugu.

In Section~\ref{sec:obs} we present new thermal observations obtained
by Herschel-PACS, re-analysed and re-calibrated AKARI-IRC and
ground-based Subaru observations and Spitzer-IRAC observations at
3.6 and 4.5\,$\mu$m. Ground-based, multi-band visual observations are described
in Section~\ref{sec:grond}.
In Section~\ref{sec:search} we describe our new approach to solve
for the object's properties. First, we present the search for the object's
spin-axis orientation using only the thermal measurements in
combination with a spherical shape model (Section~\ref{sec:spherical_shape}).
In a more sophisticated second step (Section~\ref{sec:convex_all}) we use all
thermal and visual-wavelength photometric data together and allow for
more complex object shapes in our search
for the spin axis.
In Section~\ref{sec:tpm} we use the best shape and spin-axis information
to derive additional physical and thermal properties,
and then discuss the results.
We conclude in Section~\ref{sec:conclusion} by presenting the
derived object properties and discuss our experience in
combining lightcurve inversion and radiometric techniques,
which is applicable to other targets and will help in
defining better observing strategies.

\section{Thermal observations of 162173 Ryugu}
\label{sec:obs}

\subsection{Herschel PACS observations}
\label{sec:pacs}

The European Space Agency's (ESA) Herschel Space Observatory (Pilbratt et al.\
\cite{pilbratt10}) performed observations from the 2$^{nd}$ Lagrangian point
(L2) at 1.5\,$\times$\,10$^{6}$\,km from Earth during the operational phase
from 2009 to 2013. It has three science instruments on board covering the far-infrared
part of the spectrum not accessible from the ground.
The Photodetector Array Camera and Spectrometer (PACS; Poglitsch et al.\ \cite{poglitsch10})
was used to observe 162173 Ryugu as part of the "Measurements of 11 Asteroids \& Comets"
program (MACH-11, O'Rourke et al.\ \cite{orourke14}).
PACS observed the asteroid in early April of 2012 for approximately 1.3\,h, split into two
separate measurements and taken in solar-system-object tracking mode.
The target at this time moved at a Herschel-centric apparent speed of
34$^{\prime \prime}$/h, corresponding to 19.3$^{\prime \prime}$ movement
between the mid-times of both observations.
The observations were performed in the 70/160\,$\mu$m filter combination to get
the best possible S/N in both bands. We selected seven repetitions in each
of the two scan-directions for a better characterisation of the background
and therefore a more accurate object flux.

The PACS measurements were reduced and calibrated in a standard way
as part of the Herschel data pipeline processing.
Further processing was then performed as follows. We produced single
repetition images from both scan direction measurements:
scanA1...scanA7, scanB1...scanB7, not correcting for the apparent
motion of the target (it is slow enough that the movement is not
visible in a single 282s repetition). We then subtracted
from each scanA\_n image the respective, single repetition scanB\_n
image: diff\_1 = scanA1-scanB1,...diff\_7=scanA7-scanB7, producing
differential ("diff") images.

At this point, we co-added the diff images in such a way that each
diff image was shifted by the corresponding apparent motion,
relative to the first diff image. We produced the double-differential
image and then performed the photometry and determined the
noise using the implanted source method (Kiss et al.\ \cite{kiss14}
and references therein) on the final image.
It was not feasible to extract the data from the red (160\,$\mu$m)
image due to the strongly enhanced cirrus background at that wavelength. 
The final 160\,$\mu$m differential image had an estimated confusion
noise level of approximately 7\,mJy, more than a factor of two higher than the
expected source flux.

The final derived flux was aperture and colour corrected to obtain
monochromatic flux densities at the PACS reference wavelengths.
The colour correction value for 162173~Ryugu of 1.005
in the blue band (70\,$\mu$m) is based on a thermophysical model spectral
energy distribution (SED),
corresponding to an approximately 250\,K black-body curve (Poglitsch et al.\
\cite{poglitsch10}).

The flux calibration was verified by a set of five high-quality fiducial
stars ($\beta$And, $\alpha$Cet, $\alpha$Tau, $\alpha$Boo and $\gamma$Dra),
which have been observed multiple times in the same PACS observing mode
as our observations (Balog et al.\ \cite{balog14}) and which led
to an absolute flux accuracy of 5\% for standard PACS photometer observations.
Table~\ref{tbl:herschel_obs} provides the Herschel observation data set,
accompanying information and results.

  \begin{table*}[h!tb]
    \begin{center}
    \caption{Herschel PACS observations of 162173~Ryugu as part of
             the GT1\_lorourke programme executed on operational day OD\,1057
             under the observation identifier OBSID 1342243716 \& 1342243717.
             The data were taken with the mini-scan map mode at a scan speed of
             20$^{\prime \prime}$/s and scan angles, with respect to the instrument,
             of 70$^{\circ}$ and 110$^{\circ}$, in the blue (70\,$\mu$m)
             and red (160\,$\mu$m) bands simultaneously. Note, that phase angles
             $\alpha$ are positive before, and negative after opposition.
             $\lambda_c$ is the central reference wavelength and FD is the
             monochromatic and colour-corrected flux density at $\lambda_c$.
             \label{tbl:herschel_obs}}
    \begin{tabular}{ccrrrr}
      \hline
      \hline
      \noalign{\smallskip}
      OD & OBSID & Start time & Duration [s] & Bands & Scan-angle \\
      \noalign{\smallskip}
      \hline
      \noalign{\smallskip}
      1057 & 1342243716 & 2012-04-05T00:48:20 & 1978 & 70/160 & 70$^{\circ}$ \\
      1057 & 1342243717 & 2012-04-05T01:22:21 & 1978 & 70/160 & 110$^{\circ}$ \\
      \noalign{\smallskip}
      \hline
    \end{tabular}
    \begin{tabular}{crrrrrr}
      \hline
      \hline
      \noalign{\smallskip}
      JD mid-time & r [AU] & $\Delta$ [AU] & $\alpha$ [$^{\circ}$] & $\lambda_c$ [$\mu$m] & FD [mJy] & $\sigma$ [mJy] \\ 
      \noalign{\smallskip}
      \hline
      \noalign{\smallskip}
      2456022.55719 & 1.2368 & 0.4539 & +50.4 &  70.0 & 9.47 & 1.80 \\
      \multicolumn{1}{c}{$\prime \prime$} & \multicolumn{1}{c}{$\prime \prime$} & \multicolumn{1}{c}{$\prime \prime$} & \multicolumn{1}{c}{$\prime \prime$} & 160.0 & \multicolumn{1}{c}{$<$7.0} & \multicolumn{1}{c}{---} \\
     \noalign{\smallskip}
     \hline
    \end{tabular}
    \end{center}
  \end{table*}

\subsection{Re-analysis of AKARI-IRC observations}
\label{sec:akari}

  \begin{table*}[h!]
    \begin{center}
    \caption{AKARI-IRC observations of 162173~Ryugu: the colour-corrected flux densities FD at reference wavelengths $\lambda_c$,
             together with the observational error ($\sigma$) and the absolute flux error ($\sigma_{abs}$).
             \label{tbl:akari_obs}}  
    \begin{tabular}{crrrrrrr}
      \hline
      \hline
      \noalign{\smallskip}
      JD mid-time & r [AU] & $\Delta$ [AU] & $\alpha$ [$^{\circ}$] & Band $\lambda_c$ [$\mu$m] & FD [mJy] & $\sigma$ [mJy] & $\sigma_{abs}$ [mJy]\\ 
      \noalign{\smallskip}
      \hline
      \noalign{\smallskip}
       2454236.53572 & 1.414394 & 0.992030 & +45.63 &  L15 15.0 & 7.61 & 0.20 & 0.43 \\  
       2454236.53718 & 1.414394 & 0.992019 & +45.63 &  L24 24.0 & 7.37 & 0.25 & 0.45 \\  
     \noalign{\smallskip}
     \hline
    \end{tabular}
    \end{center}
  \end{table*}

The AKARI observations were included in work
by Hasegawa et al.\ (\cite{hasegawa08}) and also used by M\"uller et al.\
(\cite{mueller11a}) and amount to a single-epoch data set from the IRC instrument
with measurements at 15 and 24\,$\mu$m. These measurements were reanalysed
with the 2015 release of the imaging data reduction toolkit\footnote{AKARI IRC imaging
toolkit version 20150331} (Egusa et al.\ \cite{egusa16}).
The flux calibration is described by Tanabe et al.\ (\cite{tanabe08}).
The new L15 flux is approximately 7\% lower than the previous value in Hasegawa et al.\
(\cite{hasegawa08}), while the L24 flux is almost identical (Table~\ref{tbl:akari_obs}).

\subsection{Re-analysis of Subaru-COMICS observations}
\label{sec:subaru}

The Ryugu observations were described in
detail by Hasegawa et al.\ (\cite{hasegawa08}) and also used
by M\"uller et al.\ (\cite{mueller11a}). Here, we re-analysed
all data with a more representative hand\-ling of the variable
atmospheric conditions during the five hours (10:30 - 15:30 UT)
of observations on the 28th August, 2007. Using the CFHT
skyprobe\footnote{\tt http://www.cfht.hawaii.edu/cgi-bin/elixir/\-skyprobe.pl?plot\&mcal\_20070828.png},
we found that the sky was generally stable to an accuracy of $<$ $\sim$0.05\,mag, but
sporadically attenuated by $>$0.1\,mag, probably caused by the passage of 
thin clouds. Although the skyprobe operates at optical wavelengths, it certainly
affected the N-band photometry as well.

For the data reduction we followed the latest version (from Nov.\ 2012) of the
COMICS cookbook\footnote{\tt http://www.naoj.org/Observing/DataReduction/\-Cookbooks/\-COMICS\_CookBook2p2E.pdf}.
Here we put special emphasis on the construction of time-varying sky flats, a very
critical element for the final accuracy of the derived fluxes. With this new element
we could recover the flux of a standard star, placed at different detector positions
and observed multiple times during our campaign, on a 3\% level.
The monitoring of the calibration star 66~Peg (HD\,220363) allowed us
to establish the instrumental magnitudes at the times of the Ryugu
observations. Finally, we conducted aperture photometry for the standard
star and our target with different aperture sizes. Colour
corrections are typically only on a 1-3\% level (Hasegawa et al.\ \cite{hasegawa08};
M\"uller et al.\ \cite{mueller04b}), but depend on the object's spectral energy distribution
and the atmospheric conditions and were not applied. Based on
the quality of the stellar model for 66~Peg (Cohen et al.\ \cite{cohen99}), and uncertainties due
to colour, aperture and atmosphere issues, we added a 5\% error to account for the absolute
flux calibration error
in the various N-band filters. The updated flux densities, errors and observational
circumstances are listed in Table~\ref{tbl:subaru_obs}.

  \begin{table*}[h!tb]
    \begin{center}
    \caption{Subaru-COMICS colour-corrected flux densities FD at reference wavelength $\lambda_c$ of 162173~Ryugu after
             our new reduction and calibration scheme (including 5\% error for the
             absolute flux calibration).
             \label{tbl:subaru_obs}}
    \begin{tabular}{crrrrrr}
      \hline
      \hline
      \noalign{\smallskip}
      JD mid-time & r [AU] & $\Delta$ [AU] & $\alpha$ [$^{\circ}$] & $\lambda_c$ [$\mu$m] & FD [mJy] & $\sigma$ [mJy] \\
      \noalign{\smallskip}
      \hline
      \noalign{\smallskip}
  2454341.00663 & 1.28725 & 0.30667 & +22.29 &  8.8  &     59 &   8 \\  
  2454341.04788 & 1.28714 & 0.30654 & +22.28 &  8.8  &     51 &   6 \\  
  2454341.09713 & 1.28702 & 0.30639 & +22.27 &  8.8  &     55 &   8 \\  
  2454341.10150 & 1.28701 & 0.30637 & +22.26 &  8.8  &     54 &  17 \\  
  2454340.98996 & 1.28729 & 0.30673 & +22.29 &  9.7  &     76 &  26 \\  
  2454340.98371 & 1.28730 & 0.30675 & +22.30 & 10.5  &     85 &  12 \\  
  2454340.96392 & 1.28735 & 0.30681 & +22.30 & 11.7  &    109 &  12 \\  
  2454340.96842 & 1.28734 & 0.30680 & +22.30 & 11.7  &    111 &  12 \\  
  2454340.99529 & 1.28728 & 0.30671 & +22.29 & 11.7  &    116 &  11 \\  
  2454341.01079 & 1.28724 & 0.30666 & +22.29 & 11.7  &    110 &  11 \\  
  2454341.02829 & 1.28719 & 0.30660 & +22.28 & 11.7  &    116 &  14 \\  
  2454341.03246 & 1.28718 & 0.30659 & +22.28 & 11.7  &     98 &  13 \\  
  2454341.03612 & 1.28717 & 0.30658 & +22.28 & 11.7  &    106 &  12 \\  
  2454341.05558 & 1.28713 & 0.30652 & +22.28 & 11.7  &     90 &  15 \\  
  2454341.05917 & 1.28712 & 0.30651 & +22.28 & 11.7  &     93 &  11 \\  
  2454341.07333 & 1.28708 & 0.30646 & +22.27 & 11.7  &    104 &  14 \\  
  2454341.08925 & 1.28704 & 0.30641 & +22.27 & 11.7  &    104 &  17 \\  
  2454341.04563 & 1.28715 & 0.30655 & +22.28 & 12.4  &    119 &  15 \\  
     \noalign{\smallskip}
     \hline
    \end{tabular}
    \end{center}
  \end{table*}

\subsection{Warm Spitzer observations in 2013}

\object{Ryugu} was the target of an extensive photometric
observation program (Mueller et al.\ \cite{muellerm12}; \cite{muellerm13})
in early 2013 using the Infrared Array Camera (IRAC, Fazio et al.\ \cite{fazio04}) onboard
the Spitzer Space Telescope (Werner et al.\ \cite{werner04}).
The observations include ten "point-and-shoot" measurements consisting of short
standard IRAC measurements that were spaced by several days up to a few
weeks between January 20 and May 29, 2013, and two complete lightcurves,
each using IRAC's channels 1 and 2 at nominal wavelengths of 3.550\,$\mu$m and
4.493\,$\mu$m, respectively\footnote{The mnemonic Spitzer IRAC channel designations
are 3.6\,$\mu$m and 4.5\,$\mu$m.}.
The point-and-shoot observations were taken between 20th January  and 29th May, 2013,
covering a phase angle range between -54$^{\circ}$ and -102$^{\circ}$.
The lightcurve observations were taken on 10/11th February and 2nd May, 2013
at phase angles of -84$^{\circ}$ and -85$^{\circ}$, respectively.
Each lightcurve observation lasted approximately 8 hours, which is a little longer than a
full rotation period of Ryugu.
Observational details are given in Table\ \ref{tbl:spitzer_obs}.

IRAC's channels 1 and 2 observe the sky simultaneously, with non-overlapping
field of views (FOV) that are a number of arc\,minutes offset from one another.
As in previous IRAC observations of near-Earth objects (see, e.g. Trilling
et al.\ \cite{trilling10}), we manually set up a dither pattern in which 
channels 1 and 2 alternate being on target (off-target
frames are discarded in the data analysis). While this incurs additional
overheads due to telescope slews, it enables quasi-simultaneous sampling
between the two channels.
The target was dithered on different parts of the FOV to minimize the impact
of any pixel-to-pixel gain differences. The standard photometric measurements
(point-and-shoot) took approximately 10\,min each. For technical reasons, each of the two lightcurve
observations (lc1a/b \& lc2a/b) had to be split in two separate observations
Astronomical Observation Requests (``AORs'' in Spitzer terminology) that
were scheduled back-to-back, with a gap of $\sim$7.5\,min between them.
In the first lightcurve epoch, a 12\,s frame time was used. Between the two AORs, there were 465 on-target
frames per channel.
The observation setup in the second lightcurve epoch was identical, except that
two consecutive 6\,s frames were used to avoid saturation.
The Moving Object mode was used tracking at asteroid rates.
Target movement during individual frames was at the sub-arcsecond level and
hence much smaller than IRAC's pixel scale of 1.8$^{\prime \prime}$.
Over the duration of a lightcurve, however, the target moved by tens of
arc\,minutes, that is, several IRAC FOV widths.

\begin{table*}
\begin{center}
\caption{
\label{tbl:spitzer_obs}
   Spitzer/IRAC "Map PC" 3.6\,$\mu$m and 4.5\,$\mu$m observations of 162173~Ryugu as part of
   the Spitzer Proposal ID \#90145.
   Above: observation start and end times (UTC) of all
   AORs.  The AOR keys \& labels are related to observations in 
   the \emph{Spitzer Heritage Archive}.
   Below: fluxes in the two photometric channels, averaged per epoch. 
   Ephemeris information ($r$, $\Delta$, $\alpha$)
   is given for observation midtime, with positive phase angles $\alpha$ before and
   negative after opposition. The latter is calculated
   as the average of the midtimes of the analysed data frames in both channels
   not corrected for light travel time. The colour-corrected monochromatic flux
   densities FD at reference wavelengths $\lambda_c$ (3.55\,$\mu$m \& 4.49\,$\mu$m) are given. The flux uncertainty includes
   the $\sim$5\% calibration uncertainty. $<$lc1$>$ and $<$lc2$>$ are lightcurve-averaged flux densities.}
    \begin{tabular}{clrrr}
      \hline
      \hline
      \noalign{\smallskip}
      AORKEY & AOR label & Start time & End time & Duration [hms] \\
      \noalign{\smallskip}
      \hline
      \noalign{\smallskip}
      48355840 & IRACPC\_1999\_JU3\_lc1a & 2013-02-10 20:07:16 & 2013-02-11 01:03:33 & 04h56m17s \\
      48355584 & IRACPC\_1999\_JU3\_lc1b & 2013-02-11 01:10:47 & 2013-02-11 04:03:00 & 02h52m13s \\
      48355328 & IRACPC\_1999\_JU3\_lc2a & 2013-05-02 11:47:00 & 2013-05-02 16:55:06 & 05h08m06s \\
      48355072 & IRACPC\_1999\_JU3\_lc2b & 2013-05-02 17:03:38 & 2013-05-02 19:52:28 & 02h48m50s \\
      \noalign{\smallskip}
      \hline
      \noalign{\smallskip}
      47925760 & IRACPC\_1999\_JU3-p1c   & 2013-01-20 02:01:04 & 2013-01-20 02:10:37 & 00h09m33s \\
      47926016 & IRACPC\_1999\_JU3-p1d   & 2013-01-27 23:00:59 & 2013-01-27 23:10:34 & 00h09m35s \\
      47926272 & IRACPC\_1999\_JU3-p1e   & 2013-01-31 01:09:29 & 2013-01-31 01:19:04 & 00h09m35s \\
      47926528 & IRACPC\_1999\_JU3-p1f   & 2013-02-09 02:24:19 & 2013-02-09 02:33:58 & 00h09m39s \\
      47926784 & IRACPC\_1999\_JU3-p2a   & 2013-04-28 19:24:40 & 2013-04-28 19:34:43 & 00h10m03s \\
      47927040 & IRACPC\_1999\_JU3-p2b   & 2013-05-05 02:37:37 & 2013-05-05 02:47:40 & 00h10m03s \\
      47927296 & IRACPC\_1999\_JU3-p2c   & 2013-05-09 23:41:59 & 2013-05-09 23:51:58 & 00h09m59s \\
      47927552 & IRACPC\_1999\_JU3-p2d   & 2013-05-15 13:42:43 & 2013-05-15 13:52:45 & 00h10m02s \\
      47927808 & IRACPC\_1999\_JU3-p2e   & 2013-05-23 21:16:01 & 2013-05-23 21:26:18 & 00h10m17s \\
      47928064 & IRACPC\_1999\_JU3-p2f   & 2013-05-29 09:45:27 & 2013-05-29 09:55:53 & 00h10m26s \\
   \end{tabular}
    \begin{tabular}{ccrrrrrrr}
      \hline
      \hline
      \noalign{\smallskip}
      & Julian Date & r    & $\Delta$ & $\alpha$     & \multicolumn{4}{c}{monochromatic FD and abs.\ error [mJy]} \\
Label & mid-time    & [AU] &  [AU]    & [$^{\circ}$] & FD$_{3.55}$ & err$_{3.55}$ & FD$_{4.49}$ & err$_{4.49}$ \\
      \noalign{\smallskip}
      \hline
      \noalign{\smallskip}
$<$lc1$>$ & 2456334.50201 & 1.01214 & 0.22957 & -83.6 & 1.30 & 0.07 &  7.21 & 0.39 \\
$<$lc2$>$ & 2456415.15792 & 1.00661 & 0.11129 & -85.0 & 4.65 & 0.25 & 26.00 & 1.40 \\
      \noalign{\smallskip}
      \hline
      \noalign{\smallskip}
p1c & 2456312.58738 & 1.06780 & 0.24935 & -71.6 &  1.21 & 0.07 &  7.01 & 0.39 \\  
p1d & 2456320.46234 & 1.04627 & 0.24442 & -76.0 &  1.12 & 0.07 &  6.42 & 0.35 \\  
p1e & 2456323.55157 & 1.03824 & 0.24181 & -77.7 &  1.19 & 0.07 &  6.58 & 0.36 \\  
p1f & 2456332.60356 & 1.01637 & 0.23202 & -82.6 &  1.33 & 0.08 &  7.20 & 0.40 \\  
p2a & 2456411.31228 & 0.99889 & 0.11258 & -88.9 &  3.94 & 0.22 & 22.06 & 1.20 \\  
p2b & 2456417.61294 & 1.01186 & 0.11093 & -82.3 &  5.38 & 0.30 & 29.00 & 1.57 \\  
p2c & 2456422.49095 & 1.02296 & 0.11138 & -76.6 &  5.01 & 0.28 & 28.16 & 1.52 \\  
p2d & 2456428.07480 & 1.03666 & 0.11392 & -69.9 &  6.42 & 0.35 & 35.48 & 1.92 \\  
p2e & 2456436.38969 & 1.05869 & 0.12205 & -60.2 &  6.95 & 0.38 & 35.25 & 1.91 \\  
p2f & 2456441.91017 & 1.07419 & 0.13051 & -54.5 &  6.10 & 0.34 & 30.23 & 1.63 \\  
     \noalign{\smallskip}
     \hline
    \end{tabular}
    \end{center}
  \end{table*}

The data reduction followed the method used in the ExploreNEOs program
(Trilling et al.\ \cite{trilling10}). Briefly, a mosaic of the field is constructed
from the data set itself and then subtracted from the individual Basic
Calibrated Data (BCD) frames to mitigate contamination from
background sources. Due to the short frame times, trailing of field stars
during an individual BCD can be neglected. We were therefore able to
generate a high signal-to-noise mosaic of the field. Aperture photometry was
then performed on the background-subtracted frames.
A small number of BCDs were rejected because the target was affected by
bad IRAC pixels, cosmic-ray hits, or bright field stars. The derived IRAC fluxes
and noise values from this Spitzer PID 90145 are given in the appendix
(\ref{app:spitzer_ps}, \ref{app:spitzer_lc}) in
Tables~\ref{tbl:spitzer_ps}, \ref{tbl:spitzer_lc1_ch1}, \ref{tbl:spitzer_lc1_ch2},
\ref{tbl:spitzer_lc2_ch1}, and \ref{tbl:spitzer_lc2_ch2}, including the
two lightcurve measurement sequences in full-time resolution.

The measured IRAC fluxes account for the sum of sunlight reflected by the
object and thermally emitted flux, integrated over the IRAC passbands. Due to the
low albedo of our target combined with its low heliocentric distance, the
reflected flux contribution is relatively small.
We estimated that the reflected-light contributions are approximately 2-3\% at 3.55\,$\mu$m
(approximately 10\% at 3.1\,$\mu$ where the bandpass opens) and well below 1\% at 4.49\,$\mu$m.
We did not subtract these contributions, but consider it in the radiometric
analysis in our thermophysical model setup which includes thermal emission and
reflected light simultaneously.
To obtain monochromatic flux densities at the channel 1\&2 reference wavelengths,
we have to colour-correct the thermal fluxes. Here, we used model calculations
of the object's SED (reflected light and thermal emission)
and combined it with the publically available IRAC spectral response
tables\footnote{{\tt http://irsa.ipac.caltech.edu/data/SPITZER/docs/\-irac/calibrationfiles/spectralresponse/};
see also discussion in Hora et al.\ \cite{hora08}}.
The resulting colour-correction factors weakly depend on the selected object properties.
We used a 6\% albedo in combination with a thermal inertia of 200\,J\,m$^{-2}$\,s$^{-0.5}$\,K$^{-1}$
and obtained colour-correction factors\footnote{divide in-band thermal fluxes by
correction factors to obtain monochromatic fluxes} of 1.09 (channel 1)
and 1.04 (channel 2), with an estimated error of approximately 2\% to cover object-SED and
passband uncertainties, as well as differences in the observing geometry.
In addition to the 2\% error for colour correction, we also added a 5\%
error to account for limitations in the absolute flux calibration of the
IRAC channels, diffuse straylight, moving target issues and possible other
calibration changes during the warm part of the Spitzer
mission\footnote{\tt http://irsa.ipac.caltech.edu/data/SPITZER/docs/irac/}, such as
intrapixel sensitivity variations or warm image features.
The final colour-corrected flux densities and absolute flux errors are given
in Table~\ref{tbl:spitzer_obs}. For the two lightcurves, we averaged the observed
fluxes (see Tables~\ref{tbl:spitzer_lc1_ch1}, \ref{tbl:spitzer_lc1_ch2},
\ref{tbl:spitzer_lc2_ch1}, and \ref{tbl:spitzer_lc2_ch2}) over the rotation
period of 7.63\,h, colour-corrected the fluxes and added the calibration
errors as for the point-and-shoot observations.

\subsection{Additional thermal measurements}
\label{sec:thermal}

In addition, we also used Spitzer-IRS measurements.
The single-epoch Spitzer-IRS spectrum was presented by Campins et al.\
(\cite{campins09}) and we used it in its calibrated full version and
also in a rebinned version with 20 points over the entire wavelength
range from 5.2 to 37.7\,$\mu$m (see description in M\"uller et
al.\ \cite{mueller11a}). The absolute IRS flux accuracy is given
with 5-7\% which is added quadratically to the uncertainties in the
reduced spectrum. Later on, in Section~\ref{sec:ti} we also
use the IRS spectrum without the absolute calibration error for
comparing the measured SED slopes with TPM slopes.

\section{Ground-based visual observations of 162173 Ryugu with GROND}
\label{sec:grond}

  \begin{table*}[h!tb]
    \begin{center}
    \caption{GROND data set of observations 162173 Ryugu. Note, that phase angles
             $\alpha$ are positive before and negative after opposition.
             \label{tbl:grond_obslog}}

    \begin{tabular}{lllrrlll}
      \hline
      \hline
      \noalign{\smallskip}
   Julian Date & Y/M/D & H:M:S & obsrunid & seqnum & r[AU] & D[AU] & $\alpha$[$^{\circ}$] \\ 
      \noalign{\smallskip}
      \hline
      \noalign{\smallskip}
2456075.58586 & 2012/5/28 &  02:03:38.7  &  1 & 1 & 1.35995 & 0.34822 & +4.92 \\ 
2456075.59086 & 2012/5/28 &  02:10:50.2  &  1 & 2 & 1.35996 & 0.34822 & +4.91 \\ 
 \\ 
2456087.54855 & 2012/6/ 9 &  01:09:54.8  &  2 & 1 & 1.37879 & 0.36931 & -8.60 \\ 
2456087.55395 & 2012/6/ 9 &  01:17:41.0  &  2 & 2 & 1.37880 & 0.36932 & -8.61 \\ 
2456087.55891 & 2012/6/ 9 &  01:24:50.0  &  2 & 3 & 1.37880 & 0.36934 & -8.62 \\ 
2456087.56394 & 2012/6/ 9 &  01:32:04.8  &  2 & 4 & 1.37881 & 0.36935 & -8.62 \\ 
2456087.56892 & 2012/6/ 9 &  01:39:14.7  &  2 & 5 & 1.37882 & 0.36936 & -8.63 \\ 
2456087.57394 & 2012/6/ 9 &  01:46:28.6  &  2 & 6 & 1.37882 & 0.36937 & -8.63 \\ 
2456087.57900 & 2012/6/ 9 &  01:53:45.3  &  2 & 7 & 1.37883 & 0.36939 & -8.64 \\ 
2456087.58585 & 2012/6/ 9 &  02:03:37.5  &  3 & 1 & 1.37884 & 0.36941 & -8.64 \\ 
2456087.59087 & 2012/6/ 9 &  02:10:50.8  &  3 & 2 & 1.37885 & 0.36942 & -8.65 \\ 
2456087.59590 & 2012/6/ 9 &  02:18:05.6  &  3 & 3 & 1.37885 & 0.36943 & -8.66 \\ 
2456087.60084 & 2012/6/ 9 &  02:25:12.7  &  3 & 4 & 1.37886 & 0.36945 & -8.66 \\ 
2456087.60582 & 2012/6/ 9 &  02:32:23.2  &  3 & 5 & 1.37887 & 0.36946 & -8.67 \\ 
2456087.61082 & 2012/6/ 9 &  02:39:34.5  &  3 & 6 & 1.37888 & 0.36947 & -8.67 \\ 
2456087.61585 & 2012/6/ 9 &  02:46:49.3  &  3 & 7 & 1.37888 & 0.36949 & -8.68 \\ 
2456087.62087 & 2012/6/ 9 &  02:54:03.1  &  3 & 8 & 1.37889 & 0.36950 & -8.68 \\ 
 \\ 
2456087.67131 & 2012/6/ 9 &  04:06:40.8  &  4 & 1 & 1.37896 & 0.36964 & -8.74 \\ 
2456087.67915 & 2012/6/ 9 &  04:17:58.6  &  4 & 2 & 1.37897 & 0.36966 & -8.75 \\ 
2456087.68411 & 2012/6/ 9 &  04:25:06.7  &  4 & 3 & 1.37898 & 0.36967 & -8.75 \\ 
2456087.68906 & 2012/6/ 9 &  04:32:14.7  &  4 & 4 & 1.37899 & 0.36969 & -8.76 \\ 
2456087.69403 & 2012/6/ 9 &  04:39:24.0  &  4 & 5 & 1.37899 & 0.36970 & -8.76 \\ 
2456087.69899 & 2012/6/ 9 &  04:46:33.1  &  4 & 6 & 1.37900 & 0.36972 & -8.77 \\ 
2456087.70740 & 2012/6/ 9 &  04:58:39.7  &  5 & 1 & 1.37901 & 0.36974 & -8.78 \\ 
2456087.71614 & 2012/6/ 9 &  05:11:14.3  &  6 & 1 & 1.37903 & 0.36977 & -8.79 \\ 
2456087.72124 & 2012/6/ 9 &  05:18:34.9  &  6 & 2 & 1.37903 & 0.36978 & -8.79 \\ 
2456087.72618 & 2012/6/ 9 &  05:25:42.0  &  6 & 3 & 1.37904 & 0.36979 & -8.80 \\ 
2456087.73113 & 2012/6/ 9 &  05:32:50.0  &  6 & 4 & 1.37905 & 0.36981 & -8.80 \\ 
2456087.73609 & 2012/6/ 9 &  05:39:58.4  &  6 & 5 & 1.37905 & 0.36982 & -8.81 \\ 
2456087.74102 & 2012/6/ 9 &  05:47:04.5  &  6 & 6 & 1.37906 & 0.36984 & -8.82 \\ 
2456087.75022 & 2012/6/ 9 &  06:00:18.8  &  7 & 1 & 1.37907 & 0.36986 & -8.82 \\ 
2456087.75519 & 2012/6/ 9 &  06:07:28.8  &  7 & 2 & 1.37908 & 0.36988 & -8.83 \\ 
2456087.76014 & 2012/6/ 9 &  06:14:36.0  &  7 & 3 & 1.37909 & 0.36989 & -8.84 \\ 
2456087.76512 & 2012/6/ 9 &  06:21:46.0  &  7 & 4 & 1.37909 & 0.36991 & -8.84 \\ 
2456087.77008 & 2012/6/ 9 &  06:28:54.5  &  7 & 5 & 1.37910 & 0.36992 & -8.85 \\ 
2456087.77505 & 2012/6/ 9 &  06:36:04.4  &  7 & 6 & 1.37911 & 0.36994 & -8.85 \\ 
2456087.78008 & 2012/6/ 9 &  06:43:18.7  &  7 & 7 & 1.37912 & 0.36995 & -8.86 \\ 
2456087.78509 & 2012/6/ 9 &  06:50:31.9  &  7 & 8 & 1.37912 & 0.36997 & -8.86 \\ 
 \\ 
2456088.68616 & 2012/6/10 &  04:28:04.2  &  8 & 1 & 1.38039 & 0.37253 & -9.81 \\ 
2456088.69113 & 2012/6/10 &  04:35:13.9  &  8 & 2 & 1.38039 & 0.37255 & -9.82 \\ 
2456088.69613 & 2012/6/10 &  04:42:25.6  &  8 & 3 & 1.38040 & 0.37256 & -9.82 \\ 
2456088.70113 & 2012/6/10 &  04:49:37.7  &  8 & 4 & 1.38041 & 0.37258 & -9.83 \\ 
2456088.70717 & 2012/6/10 &  04:58:19.8  &  9 & 1 & 1.38041 & 0.37260 & -9.84 \\ 
2456088.71345 & 2012/6/10 &  05:07:22.1  &  9 & 2 & 1.38042 & 0.37262 & -9.84 \\ 
2456088.71843 & 2012/6/10 &  05:14:32.4  &  9 & 3 & 1.38043 & 0.37263 & -9.85 \\ 
2456088.72343 & 2012/6/10 &  05:21:44.4  &  9 & 4 & 1.38044 & 0.37265 & -9.85 \\ 
2456088.72831 & 2012/6/10 &  05:28:45.8  &  9 & 5 & 1.38044 & 0.37266 & -9.86 \\ 
2456088.73326 & 2012/6/10 &  05:35:53.8  &  9 & 6 & 1.38045 & 0.37268 & -9.87 \\ 
2456088.73817 & 2012/6/10 &  05:42:58.0  &  9 & 7 & 1.38046 & 0.37269 & -9.87 \\ 
2456088.74315 & 2012/6/10 &  05:50:08.1  &  9 & 8 & 1.38046 & 0.37271 & -9.88 \\ 
2456088.74946 & 2012/6/10 &  05:59:13.5  & 10 & 1 & 1.38047 & 0.37273 & -9.88 \\ 
2456088.75444 & 2012/6/10 &  06:06:23.5  & 10 & 2 & 1.38048 & 0.37274 & -9.89 \\ 
2456088.75932 & 2012/6/10 &  06:13:25.6  & 10 & 3 & 1.38049 & 0.37276 & -9.89 \\ 
2456088.76423 & 2012/6/10 &  06:20:29.8  & 10 & 4 & 1.38049 & 0.37277 & -9.90 \\ 
2456088.76919 & 2012/6/10 &  06:27:37.9  & 10 & 5 & 1.38050 & 0.37279 & -9.90 \\ 
2456088.77419 & 2012/6/10 &  06:34:49.9  & 10 & 6 & 1.38051 & 0.37280 & -9.91 \\ 
2456088.77913 & 2012/6/10 &  06:41:56.4  & 10 & 7 & 1.38051 & 0.37282 & -9.91 \\ 
2456088.78409 & 2012/6/10 &  06:49:05.5  & 10 & 8 & 1.38052 & 0.37283 & -9.92 \\ 
2456088.79049 & 2012/6/10 &  06:58:18.0  & 11 & 1 & 1.38053 & 0.37285 & -9.93 \\ 
2456088.79539 & 2012/6/10 &  07:05:21.7  & 11 & 2 & 1.38054 & 0.37287 & -9.93 \\ 
2456088.80039 & 2012/6/10 &  07:12:33.7  & 11 & 3 & 1.38054 & 0.37288 & -9.94 \\ 
2456088.80530 & 2012/6/10 &  07:19:38.2  & 11 & 4 & 1.38055 & 0.37290 & -9.94 \\ 
     \noalign{\smallskip}
     \hline
    \end{tabular}
    \end{center}
  \end{table*}

The Gamma-Ray burst Optical-Near-ir Detector (GROND, Greiner et al.\
\cite{greiner08}) is mounted on the MPI 2.2m telescope at the ESO La
Silla observatory (Chile). GROND observes in four optical-
($g$$^\prime$, $r$$^\prime$, $i$$^\prime$, $z$$^\prime$) and three near-IR ($J$, $H$,
$K$) filters, simultaneously. The observations of 162173 Ryugu
were performed in pointing mode with four dithers carried out
every minute to place the source again in the centre of the $5.4\times5.4$ arcmin$^2$
FOV of the optical Charge-Coupled Devices (CCDs).

A first short set of observations was taken on  27/28th May, 2012, for 2
$\times$ 4\,min, followed by $\approx5$\,hr coverage on 8th June 2012 
(with a 1.5\,hr gap in between) and another 3\,hr the following
night. Table~\ref{tbl:grond_obslog} provides a list of observations
taken including heliocentric distance r, observatory-centric distance
$\Delta$ and the phase angle $\alpha$.
                              
The GROND data were reduced and analysed
with the standard tools and methods described in Kr\"uhler et al.\
(\cite{kruehler08}). The $g$$^\prime$, $r$$^\prime$, $i$$^\prime$
photometry was obtained using aperture photometry with aperture
sizes corresponding to 1.5$^{\prime \prime}$ (approximately 2$\times$ the
width of the PSF) to include the complete flux of the slightly
elongated source images (Ryugu was moving approximately 0.8$^{\prime \prime}$
during the $\approx$35\,s integration times).
The $z$, $J$, $H$, and $K$-band measurements were not used for the analysis
due to the substantially lower S/N.

The photometry and photometric calibration of 162173 Ryugu
was achieved as follows. First, the moving target was extracted from
the list of detected sources for each band and filter by its
observatory-centric position. For each sequence number in each
observation ID we extracted the object's magnitude and its statistical
photometric error for each of the four target dither positions.
We additionally extracted several reference stars which were close to
the path of Ryugu and similar in brightness (or brighter). As not every star
was seen throughout the night (due to the rapid movement of Ryugu
of around 100$^{\prime \prime}$/h which required frequent re-pointing
of the telescope), we selected several stars so that we had
three to four stars at each given time. We monitored the magnitudes of these
stars, calibrated against USNO ($g^\prime$, $r^\prime$,
$i^\prime$), in parallel to the object's magnitudes.
This resulted in 1-$\sigma$ accuracies of approximately 0.02\,mag in the given bands.
As the absolute photometric calibration of USNO is not very
accurate, only the relative photometry is used in the remainder
of the analysis. The results of the GROND lightcurve measurements
(one value per TDP\footnote{Telescope Dither Position},
i.e. for each GROND pointing) are given in Tables~\ref{tbl:grond_day1_r}
and ~\ref{tbl:grond_day2_gri} in the appendix.
The magnitude errors include the photon noise (approximately 0.01-0.03\,mag depending
on the time of measurements during the nights) and the error from
the corrections related to the multiple star measurements (typically
well below 0.01\,mag). There is also a flat-field error due to the
varying position of Ryugu on the GROND camera (0.01 to 0.02\,mag)
which is difficult to quantify for an individual observation, but
explains the scatter for some of the data points.

\section{Searching for the spin-axis orientation}
\label{sec:search}

The standard lightcurve inversion technique (Kaasalainen \&
Torppa \cite{kaasalainen01}; Kaasalainen et al.\ \cite{kaasalainen01a})
derives the asteroid's shape, siderial rotation period
and spin-axis direction from the observed lightcurves.
Earlier work by M\"uller et al.\ (\cite{mueller11a}), based on a smaller
sample of lightcurve observations of \object{Ryugu,}  found
many solutions with different shapes, spin-orientations and rotation periods,
all being compatible in the $\chi^2$-sense with the available
visual data. This analysis was followed by using the derived solutions
in the context of the available thermal data. The radiometric
technique could eliminate wide ranges of solutions but did not
lead to a unique spin-axis orientation.
We repeated this procedure, now using many more lightcurve observations
and additional thermal data points. However, 
the almost spherical shape and
the insufficient quality of some data sets made this exercise very challenging
and no unique solution was found. We therefore tried a different path.

\subsection{Spherical shape model}
\label{sec:spherical_shape}

Knowing that the observed lightcurve amplitudes are small (the amplitude
is approximately 0.1\,mag) we simply assumed a spherical shape model with the spin-axis orientation, size,
albedo and thermal inertia as free parameters. Also, instead of using the
visual lightcurves first, we started by using the thermal measurements.
This approach worked very well in the case of (101955)~Bennu where the radiometrically
established spin-axis orientation (M\"uller et al.\ \cite{mueller12}) agreed within error bars
with the radar derived value (Nolan et al.\ \cite{nolan13}). Furthermore, in the case of
the very elongated object (25143)~Itokawa, a careful analysis of thermal data
in combination with a spherical shape model led to realistic predictions for
the orientation of the object's spin vector within approximately 10$^{\circ}$ of
the in-situ measured values (M\"uller et al.\ \cite{mueller14}).

  \begin{table}[h!tb]
    \begin{center}
    \caption{Summary of general TPM input parameters and applied ranges.
             \label{tbl:tpm_params}}
    \begin{tabular}{lcl}
      \hline
      \hline
      \noalign{\smallskip}
      Param.\  &  Value/Range & Remarks \\
      \noalign{\smallskip}
      \hline
      \noalign{\smallskip}
    $\Gamma$            & 0...2500                  & [J\,m$^{-2}$\,s$^{-0.5}$\,K$^{-1}$], thermal inertia \\
    $\rho$              & 0.0...0.9                 & r.m.s. of the surface slopes \\
    $f$                 & 0.6                       & surface fraction covered by craters \\
    $\epsilon$          & 0.9                       & emissivity \\
    $H_{\rm{V}}$-mag.\  & 19.25\,$\pm$\,0.03        & [mag], Ishiguro et al.\ (\cite{ishiguro14}) \\
    G-slope             & 0.13\,$\pm$\,0.02         & Ishiguro et al.\ (\cite{ishiguro14}) \\
    shape               & spherical shape           & see Sect.~\ref{sec:spherical_shape} \\
                        & convex shapes             & see Sect.~\ref{sec:convex_all} \\
    spin-axis           & 4\,$\pi$                  & see Sect.~\ref{sec:search} \\
    P$_{\mathrm{sid}}$  & 7.6312                    & [h], see Sect.~\ref{sec:spherical_shape} \\
                        & free parameter            & see Sect.~\ref{sec:convex_all} \\
     \noalign{\smallskip}
     \hline
     \noalign{\smallskip}
    \end{tabular}
    \end{center}
  \end{table}

The analysis was performed by means of a TPM code (Lagerros \cite{lagerros96}; \cite{lagerros97}; \cite{lagerros98}).
The target is described by a given size, shape, spin-state
and albedo and placed at the true, epoch-specific observing and
illumination geometry. The TPM considers a 1-d heat conduction
into the surface with the thermal inertia $\Gamma$ being the critical
parameter. Surface roughness is also included, described by
`f' the fraction of the surface covered by spherical crater
segments and `$\rho$', the r.m.s.\ of the surface slopes, connected
to the crater width-to-depth ratio (Lagerros \cite{lagerros98}).
For each surface facet, the energy balance between solar insolation,
reflected light and thermal emission is treated individually.
The reflected light contribution is calculated by multiplying
the solar irradiance with the bidirectional reflectance and the
transition phase between reflected light and thermal emission is
estimated using Lambert's scattering law (Lagerros \cite{lagerros96}
and references therein).
We use the H-G system (Bowell et al.\ \cite{bowell89}) to describe the
amount of reflected light. We used H$_V$ = 19.25 $\pm$ 0.03\,mag
and G = 0.13 $\pm$ 0.02 which was derived from calibrated
V-band observations covering a very wide phase angle range
(Ishiguro et al.\ \cite{ishiguro14}). Table~\ref{tbl:tpm_params}
summarises the general TPM settings for our radiometric analysis.

\subsubsection{Using only mid-/far-IR thermal data}
\label{sec:sph_set1}

In the first step we used the above-mentioned thermal measurements
by Herschel-PACS (Tbl.~\ref{tbl:herschel_obs}), Subaru-COMICS (Tbl.~\ref{tbl:subaru_obs}),
AKARI, and Spitzer-IRS (see M\"uller et al.\ \cite{mueller11a}), but we
excluded the Spitzer-IRAC measurements (Tbl.~\ref{tbl:spitzer_obs}).
The IRAC measurements are taken at shorter wavelengths (3.6 and 4.5\,$\mu$m)
where reflected light contributions might play a role.
In addition, the individual short-wavelength (absolute) thermal fluxes are
dominated by the hottest areas on the surface (related to local shape features
near the sub-solar point), and global shape and spin properties are less relevant.

We used a spherical shape model with approximately 100 different
spin axis orientations distributed over the entire 4\,$\pi$ solid angle.
In a first round, we used a grid of pole directions with 30$^{\circ}$
steps, and later on we refined the steps around some of the promising pole
directions (when the $\chi^2$-values were within 3$\sigma$ of an
acceptable solution). This iterative approach worked well, knowing that
the surface temperature distribution on a spherical shape model changes
only gradually when changing the pole direction by one step.
Size, albedo and thermal inertia are the dominating object properties when
trying to reproduce the observed fluxes; the roughess has only a minor influence.
We started by assuming a moderate surface roughness (r.m.s.-slopes of 0.2)
to lower the number of free parameters in the first iteration.

\begin{figure}[h!tb]
 \rotatebox{90}{\resizebox{!}{\hsize}{\includegraphics{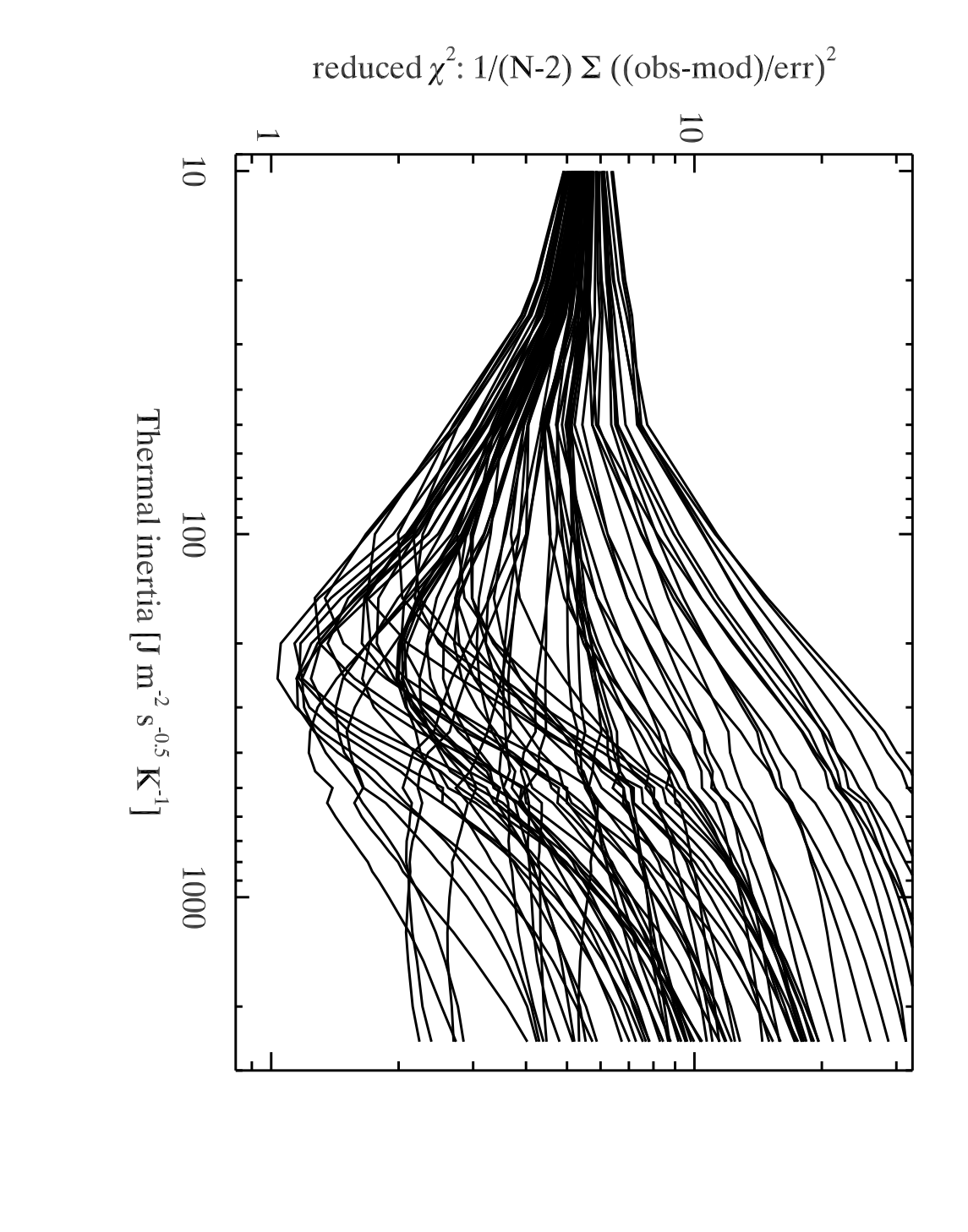}}}
 \rotatebox{90}{\resizebox{!}{\hsize}{\includegraphics{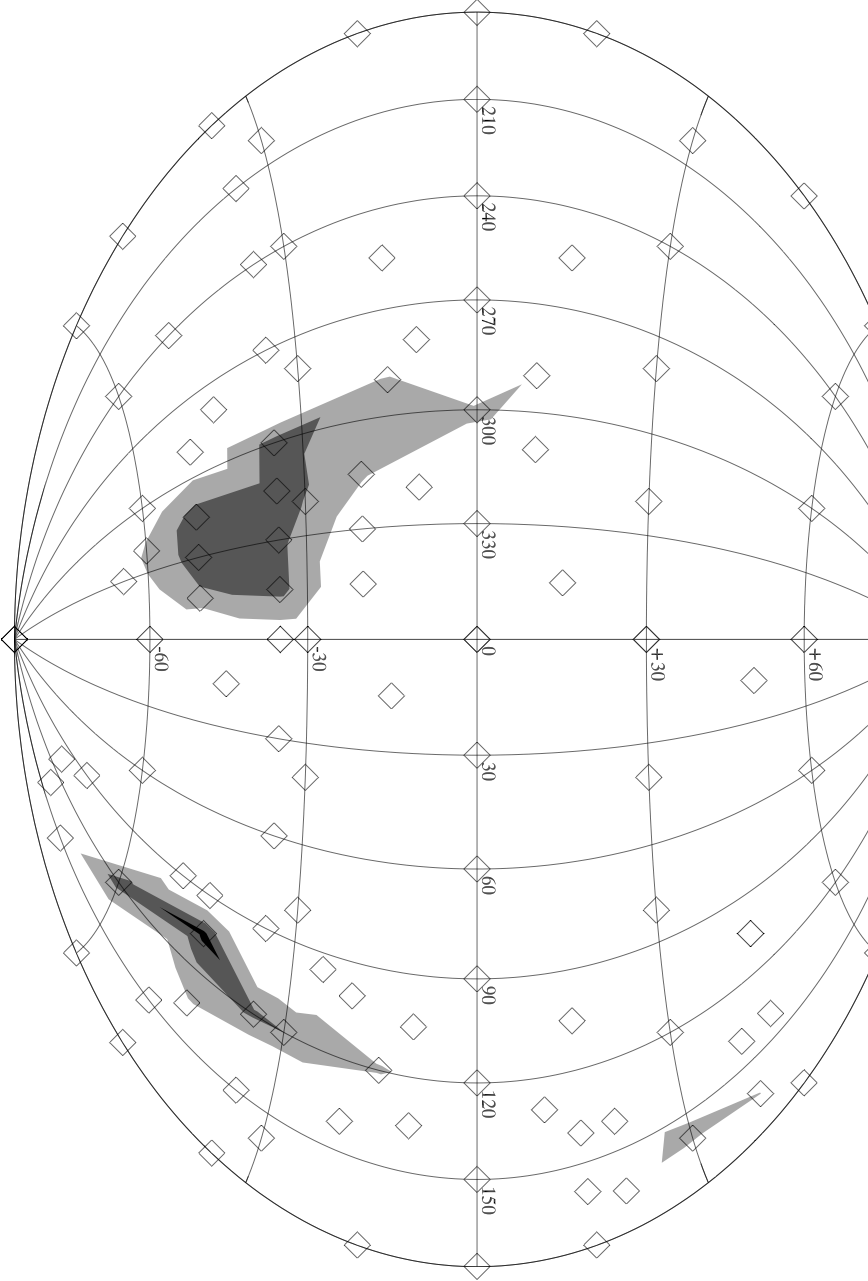}}}
  \caption{$\chi^2$ test for the complete set of spin-vector orientations and
           the full range of thermal inertias. Values close to 1.0
           indicate acceptable solutions. Top: Reduced $\chi^2$ values
           as a function of thermal inertia. Bottom: 1$\sigma$- (dark grey),
           2$\sigma$- (intermediate grey), and 3$\sigma$- (light grey) $\chi^2$
           values as a function of spin-axis ecliptic longitude and latitude
           in aitoff projection. The ($\lambda$; $\beta$) ecliptic coordinates
           of all tested orientations are shown as diamonds.
     \label{fig:chi2_1}}
\end{figure}

For each value of the spin axis, we solved for the effective size and
geometric albedo for a wide range of thermal inertias ranging from
0 (regolith with extremely low heat conductivity) to 2500\,J\,m$^{-2}$\,s$^{-0.5}$\,K$^{-1}$
(bare rock surface with very high conductivity). The results are shown in
Figure~\ref{fig:chi2_1} (top) with one curve per spin-axis orientation, each with
a specific size/albedo solution as a function of thermal inertia. Reduced
$\chi^2$ values close to 1.0 indicate a good fit to the thermal data set.
In the bottom part of the figure we show the standard $\chi^2$ values in
the ecliptic longitude and latitude space:
$\chi^2$ = $\Sigma ((FD_{TPM} - FD_{Obs})/err_{Obs})^2$, with the confidence levels
indicated by grey colours: 1-$\sigma$ level agreement (dark grey) where $\chi^2$ $<$ $\chi_{min}^2$ + 1$^2$,
2-$\sigma$ level (intermediate grey) where $\chi^2$ $<$ $\chi_{min}^2$ + 2$^2$, and
3-$\sigma$ level (light grey) where $\chi^2$ $<$ $\chi_{min}^2$ + 3$^2$.
Many orientations of the spin axis can be excluded with very high probability;
they simply do not allow us to explain all thermal measurements simultaneously,
independent of the selected thermal inertia (and size/albedo). However there
are still two remaining zones
for the possible spin vector: (1) longitudes 90-130$^{\circ}$ and a wide
range of latitudes from -15$^{\circ}$ to -70$^{\circ}$; (2) longitudes
290-350$^{\circ}$ combined with latitudes +10$^{\circ}$ to -60$^{\circ}$.
At the 3-$\sigma$ level there is even a third small zone at an approximate longitude
160$^{\circ}$ and latitude +30$^{\circ}$. The true zones are probably
slightly larger considering that we assumed a spherical shape and a fixed
level of surface roughness.

\subsubsection{Using Spitzer-IRAC averaged lightcurve measurements}
\label{sec:sph_set2}

Before investigating the full IRAC data set, we looked into the
most crucial aspect for our spherical-shape analysis;
namely the target's flux increase between 10/11th Feb.\  and 2nd May, 2013, the
epochs when full lightcurves were taken by Spitzer-IRAC.
The flux increase is related to two effects: (i) The changed observing
geometry: the Spitzer-centric distance decreased from approximately 0.23\,AU in
Feb.\ 2013 to approximately 0.11\,AU in May 2013 (while the heliocentric distances
and phase angles were very similar); (ii) The object moved on the sky by approximately
88$^{\circ}$ between February and May 2013, and depending on the orientation
of the spin axis, the aspect angle (the angle under which the observer
sees the rotation axis) was very different. The flux change between the
two Spitzer lightcurve epochs is therefore very much related to the
object's spin-axis orientation.
Table~\ref{tbl:spitzer_obs} contains details of these measurements.

\begin{figure}[h!tb]
 \rotatebox{90}{\resizebox{!}{\hsize}{\includegraphics{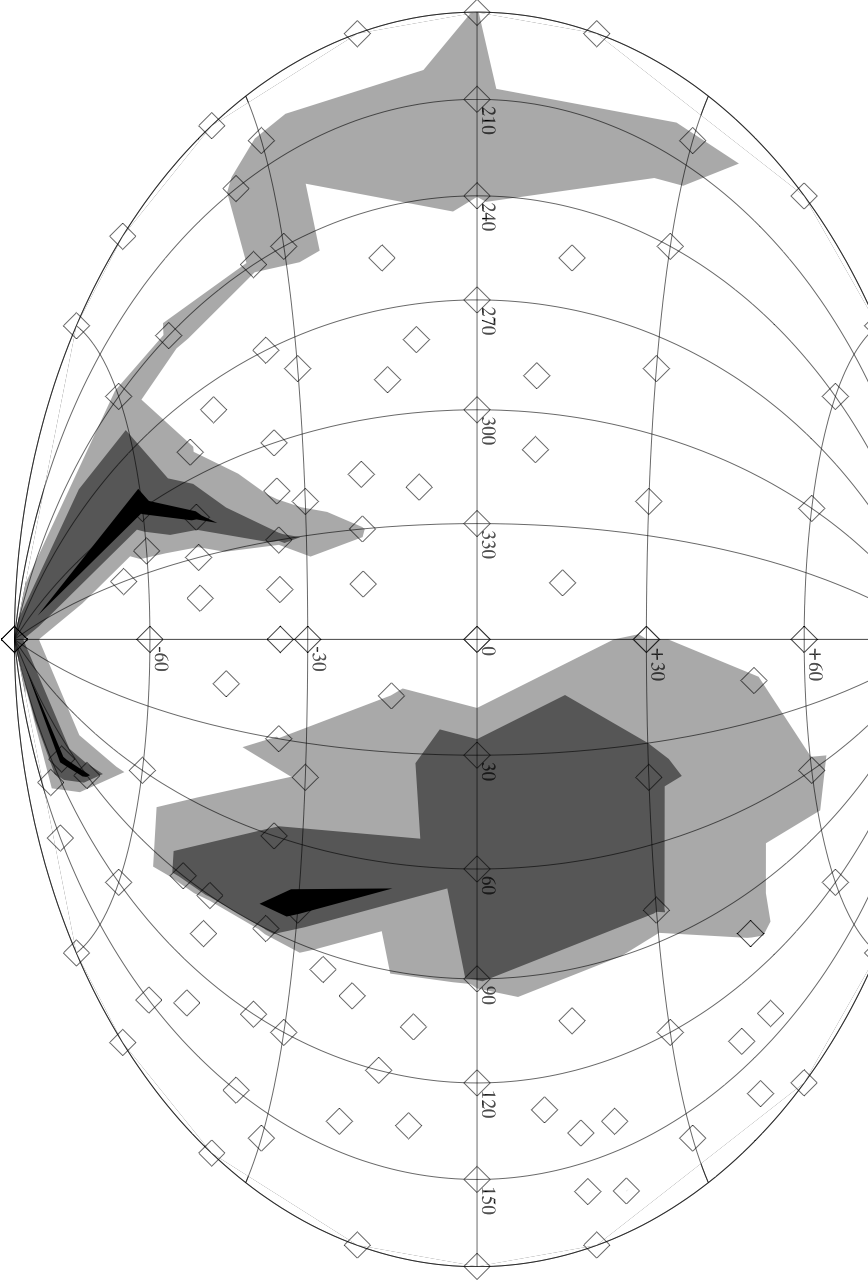}}}
  \caption{Comparison between model flux ratios and the observed Spitzer-IRAC
   flux ratio of 3.6 $\pm$ 0.2 in the ($\lambda$; $\beta$)-space.
   The grey zones indicate spin-axis orientations
   which can reproduce the observed flux ratio on a 1-$\sigma$ (dark-grey),
   2-$\sigma$ (intermediate-grey), and 3-$\sigma$ (light-grey) level.
     \label{fig:chi2_2}}
\end{figure}

We use the lightcurve-averaged observed fluxes to avoid shape effects in our
search for the spin vector (see entries $<$lc1$>$ and $<$lc2$>$ in Tbl.~\ref{tbl:spitzer_obs}).
We also did not convert the observed in-band fluxes
into monochromatic flux densities at the reference wavelength due to unknowns
in the object's SED at these short wavelengths where reflected light starts to
play a role. The SED shape at wavelengths below 5\,$\mu$m is, in general,
very sensitive to the highest surface temperatures and this depends on unknown
surface structures, roughness and possibly also on material properties.

The flux changes between the two visits are independent of all corrections.
The flux ratio between the second and the first epoch is 3.58 ($\pm$0.1) in
the 3.6\,$\mu$m band and 3.61 ($\pm$0.05) in the 4.5\,$\mu$m band. For our
calculations we used a conservative average flux ratio of 3.6 $\pm$ 0.2,
independent of wavelengths.

Starting out again with the spherical shape model, we determined the size, albedo and thermal inertia
solution connected to the minima in the reduced-$\chi^2$ curves,
for
each of the 100 spin orientations (top part of Fig.~\ref{fig:chi2_1}) to guarantee in each case the optimal
solution with respect to our full thermal data set.
Now we can predict the model fluxes and flux
ratios for the two Spitzer epochs  for each spin orientation (see Table~\ref{tbl:spitzer_obs}). We
compare the model flux ratios with the observed flux ratio of 3.6 $\pm$ 0.2
by calculating the $\chi^2$-values $((FD^{TPM}_{ep2}/FD^{TPM}_{ep1})-3.6)/0.2)^2$
and the corresponding 1-, 2-, and 3-$\sigma$ thresholds (see formula above).
Figure~\ref{fig:chi2_2} shows the result of this $\chi^2$-test on the basis
of the observed IRAC flux ratio. Here again, large ranges of spin-axis
orientations can be excluded because of a mismatch with the observed
flux ratio. There remain, however, many possible pro- and retrograde solutions.

\subsubsection{Summary spherical shape approach}
\label{sec:summary}

\begin{figure}[h!tb]
 \rotatebox{90}{\resizebox{!}{\hsize}{\includegraphics{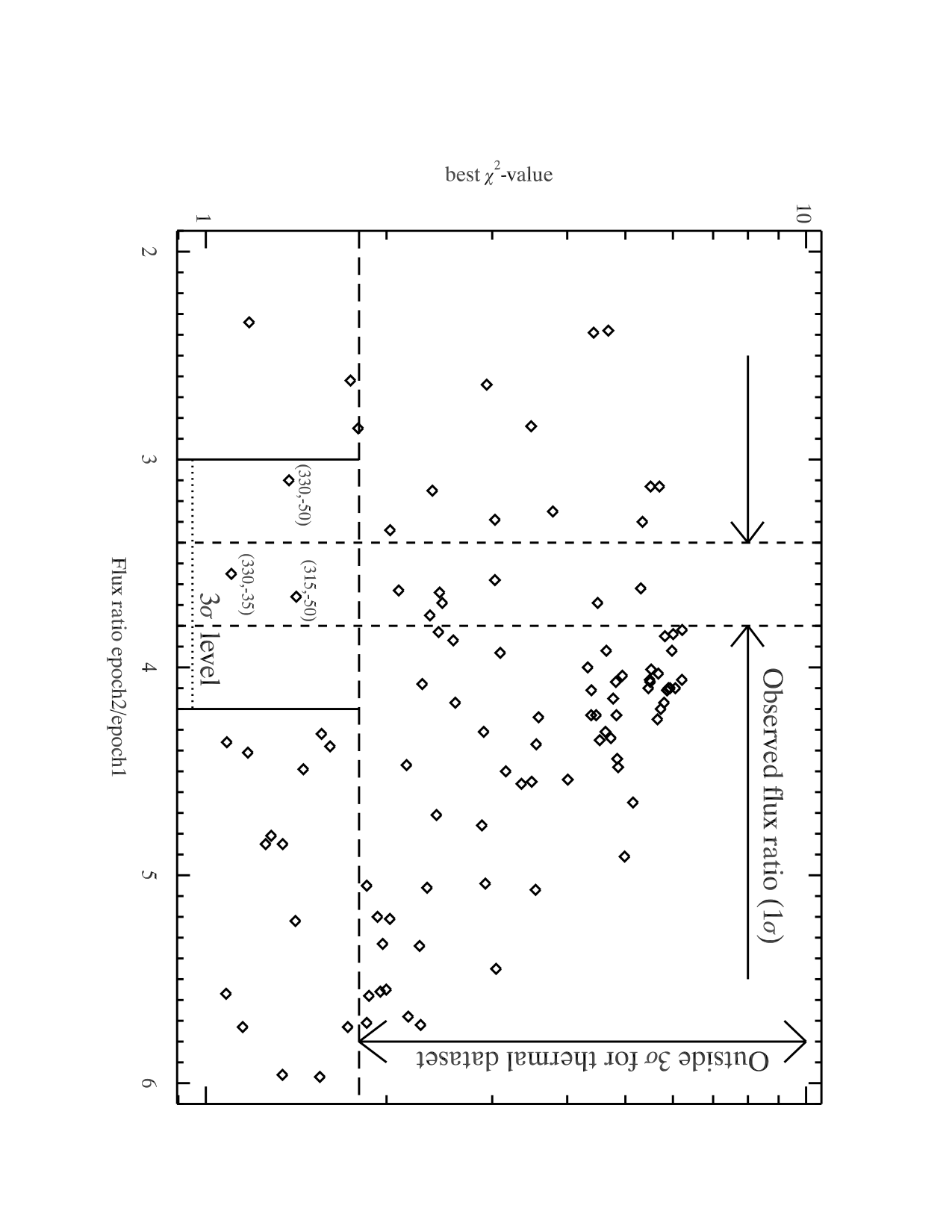}}}
  \caption{Calculated flux ratios for the two IRAC epochs on 2nd May and
   10/11th Feb, 2013, as a function of the minimum $\chi^2$ values
   from Figure~\ref{fig:chi2_1}. The observed IRAC flux ratio of 3.6 $\pm$ 0.2
   (and the 3-$\sigma$ level)
   is indicated by vertical lines and the acceptance level from the analysis
   of the thermal data set is given by the dashed horizontal line.
     \label{fig:flxratio1}}
\end{figure}

Figure~\ref{fig:flxratio1} shows the summary of both constraints with each
spin-vector orientation represented by a diamond. The y-axis shows
the values of the reduced $\chi^2$ minima from Fig.~\ref{fig:chi2_1} (top),
the x-axis shows the corresponding model prediction for the IRAC
flux ratio. The points below the horizontal long-dashed line show all solutions
which are compatible (3$\sigma$) with the thermal data set, the vertical
dashed lines indicate the observed flux ratio and the solid lines indicate
the 3-$\sigma$ range). There are only three solutions
compatible with both constraints: ($\lambda$; $\beta$)$_{ecl}$ =
(315$^{\circ}$, -50$^{\circ}$), (330$^{\circ}$, -35$^{\circ}$), and (330$^{\circ}$, -50$^{\circ}$),
with the first two being within both
1-$\sigma$ levels. Due to coarse sampling of the 4$\pi$ space and the
various assumptions for the shape and surface roughness,
we estimated that the true spin-vector solution must lie somewhere
in the range 310$^{\circ}$ to 335$^{\circ}$ in ecliptic longitude and
-65$^{\circ}$ to -30$^{\circ}$ in ecliptic latitude. Our analysis from
Sections \ref{sec:sph_set1} and \ref{sec:sph_set2} for this range of pole
solutions points to a low-roughness surface (r.m.s.\ of surface slopes below 0.2) and
thermal inertias of approximately 200-300\,J\,m$^{-2}$\,s$^{-0.5}$\,K$^{-1}$, but 
still with significant uncertainties in size (approximately 815-900\,m) and
albedo (between 0.04 and 0.06).

\subsection{Using thermal and visual-wavelength photometric data together}
\label{sec:convex_all}

To confirm the results obtained by the method described in the previous section, we tried another independent
approach to determine the spin axis orientation of Ryugu, using both photometric data and
all thermal data together in one inversion procedure.
The thermal data are the ones previously published and presented in Section~\ref{sec:obs},
the visual lightcurves are presented in M.-J.\ Kim et al.\ (in preparation) together with
a set of ground-based observations taken with GROND mounted on the MPI 2.2m telescope
at the ESO La Silla observatory (Chile) which is described in Section~\ref{sec:grond} and
in the appendix~\ref{app:grond}.
 The method is a modification of the standard lightcurve
inversion of Kaasalainen et al.\ (\cite{kaasalainen01a}) including also a thermophysical model. 
Its output is a convex shape model together with all relevant geometrical and 
physical parameters (spin axis direction, rotation period, size, thermal inertia of the surface, albedo and surface roughness) 
that best fit the lightcurves and thermal fluxes. It uses the 
Levenberg-Marquardt algorithm to minimize the $\chi^2$ difference between the data and the model by 
optimising the model parameters. The total $\chi^2$ of the fit is composed of two parts 
\begin{center}
\begin{equation}
 \label{eq:chisq}
 \chi^2 = \chi^2_\mathrm{LC} + w\,\chi^2_\mathrm{IR}\,,
\end{equation}
\end{center}
where $\chi^2_\mathrm{LC}$ corresponds to the 
difference between the model and photometric data and $\chi^2_\mathrm{IR}$ corresponds to the difference between the model and thermal data. 
The relative weight $w$ of one data source with respect to the
other is set such that the best-fit model gives an acceptable fit to both data sets. Formally, the optimum weight can be found with the 
approach proposed by Kaasalainen (\cite{kaasalainen11}).
Note that, overall, we use approximately 3100 data points from lightcurve measurements 
(Kim et al., in preparation, and appendix~\ref{app:grond}) in addition to
the large collection of thermal measurements (see Sect.~\ref{sec:obs}), while the pole, rotation period, size, thermal inertia and shape are used as
free parameters. For our calculations we used the rebinned Spitzer-IRS spectrum,
18 Subaru-COMICS measurements, 2 AKARI \& 1 Herschel-PACS data points, as well as the 2 Spitzer-IRAC thermal lightcurves
and the 10-epoch IRAC point-and-shoot sequence, with IRAC data always taken in two channels.

The thermo-physical part of the code is based on solving 1D heat conduction to get the temperature of each surface facet, for which the flux is computed. 
It is an independent implementation of the TPM code described in Sect.~\ref{sec:spherical_shape} based on the thermal model of 
Lagerros (\cite{lagerros96}; \cite{lagerros97}; \cite{lagerros98}). 
The hemispherical albedo needed for the boundary condition at the surface is computed in each step from the parameters of the Hapke's photometric model 
that is also used in the photometric part. 
This new method has been developed by \v{D}urech et al.\ (\cite{durech12}) and tested on asteroid (21)~Lutetia,
as well as on several other asteroids.
We describe the method and its results in detail in a separate paper (\v{D}urech et al., in prep.).

\begin{figure}[h!tb]
 \includegraphics[width=\columnwidth]{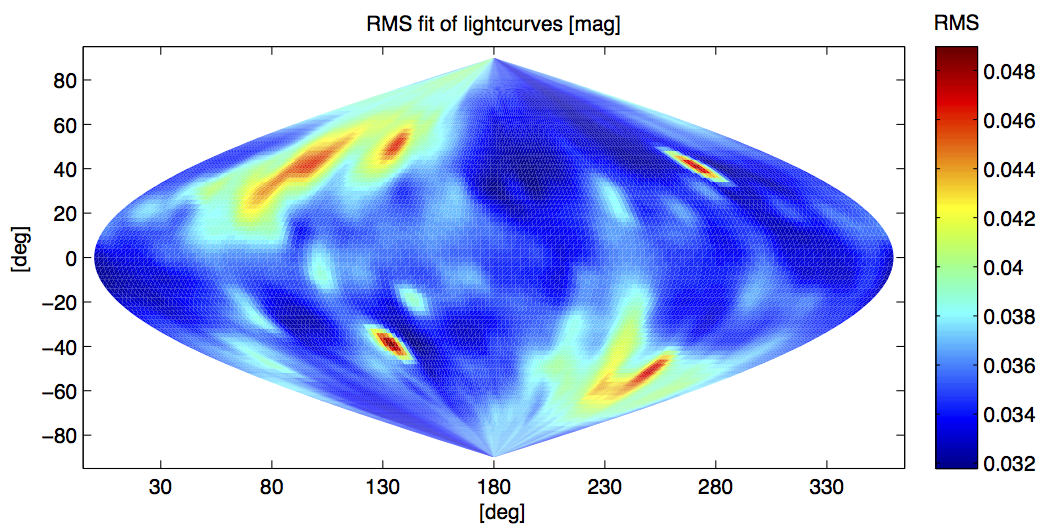}
 \includegraphics[width=\columnwidth]{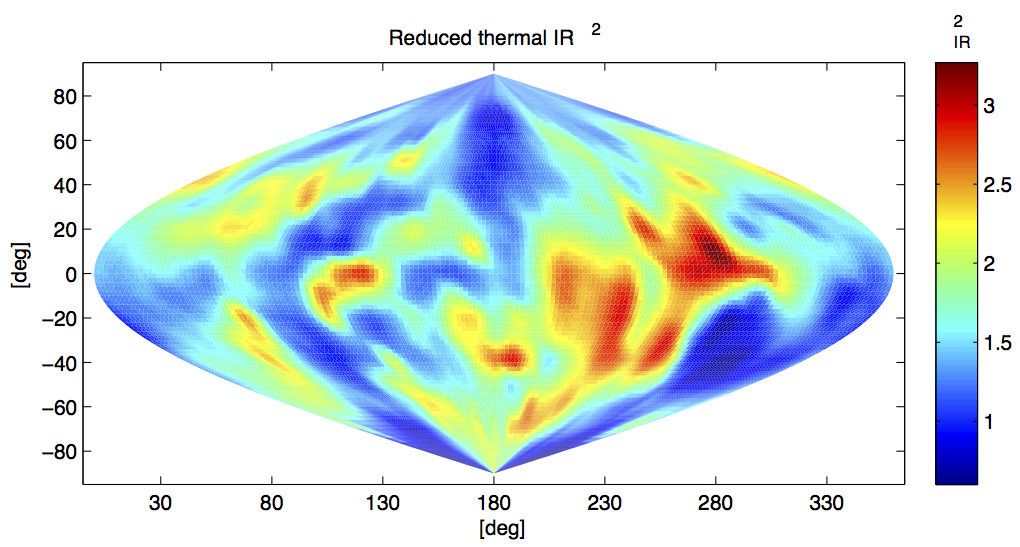}
 \includegraphics[width=\columnwidth]{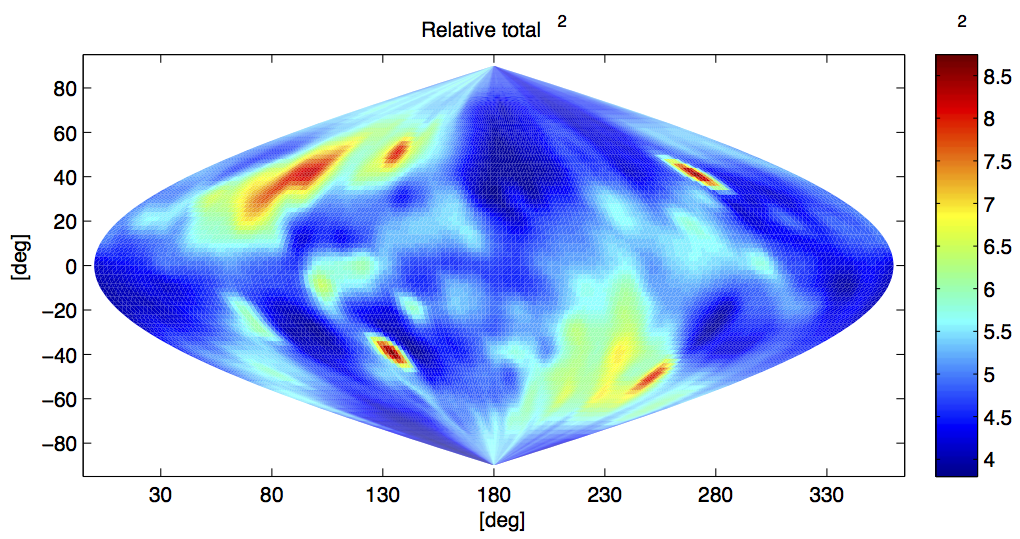}
  \caption{The colour map of the goodness of fit for the lightcurves, thermal data and combined data for all possible orientations of the spin axis 
   given in ecliptic coordinates $(\lambda, \beta)$. The top panel shows the rms residuals between the model and observed lightcurves, 
   the middle plot shows the reduced $\chi^2$ for the thermal data and the bottom plot shows the total $\chi^2$ (Eq.~\ref{eq:chisq}).
     \label{fig:chisq_total}}
\end{figure}

\begin{figure}[h!tb]
 \includegraphics[width=\columnwidth]{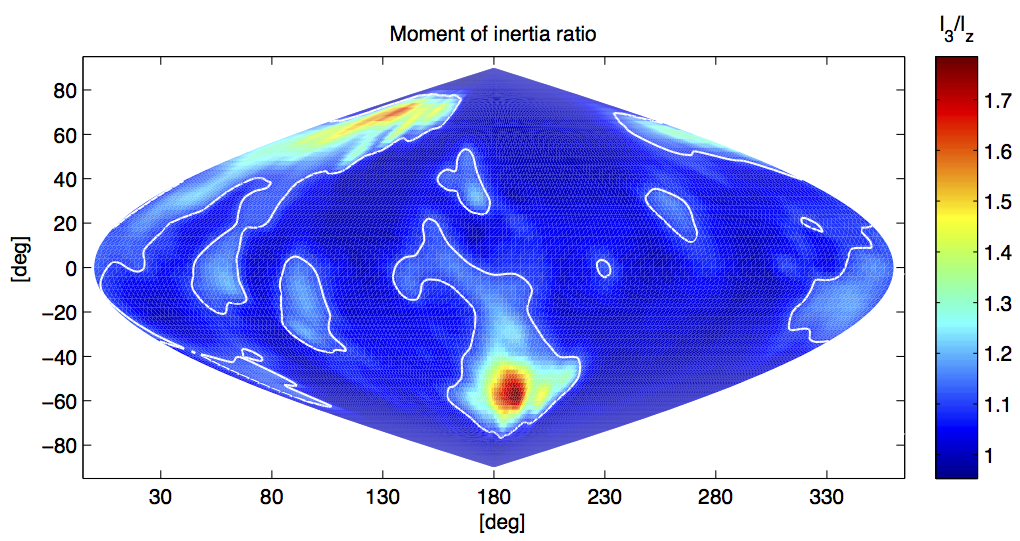}
  \caption{The colour map of the ratio $I_3 / I_z$ for models corresponding to all possible orientations of the spin axis 
   given in ecliptic coordinates $(\lambda, \beta)$. The white contour corresponds to the level 1.1.
     \label{fig:inertia}}
\end{figure}

The results for Ryugu are shown in Fig~\ref{fig:chisq_total}, where the fit to the lightcurves, thermal data 
and the combined total $\chi^2$ is plotted for a grid of pole directions
with a $5^\circ$ step and the rotation period $P = 7.6300\,$h. Because the model is rather flexible and the quality of the lightcurves is poor,
there is no clear minimum in $\chi^2$ that would define a unique spin direction. Moreover, even the sidereal
rotation period cannot be determined unambiguosly from the current data set.
There are more possible values ($P = 7.6300$, 7.6311, 7.6326\,h, for example) that give
essentially the same fit to the data and very similar $\chi^2$ maps for the pole direction.
However, some solutions with low values of $\chi^2$ 
are not correct in the physical sense; the model fits the data well, but the corresponding shape 
is too elongated along the rotation axis, which does not agree with the assumption of a relaxed rotation around the principal axis with the maximum inertia.
Because the inversion algorithm works with the Gaussian image of the body (see Kaasalainen \&
Torppa \cite{kaasalainen01}), 
the inertia tensor cannot be constrained during the shape optimisation. The alignment of rotation and principal inertia axes has to be checked 
afterwards and unphysical models must be rejected. For each model, we computed the moment of inertia $I_z$ corresponding to the rotation 
around the actual $z$ axis and moment of inertia $I_3$ corresponding to the rotation along the shortest axis (minimum energy). We plot the ratio $I_3 / I_z$  
for all poles in Fig.~\ref{fig:inertia}. The artificial boundary of 1.1 (based on our experience with models from lightcurves) divides those solutions that are formally acceptable ($I_3 / I_z < 1.1$) from those that are not. 

\begin{figure}[h!tb]
 \includegraphics[width=\columnwidth]{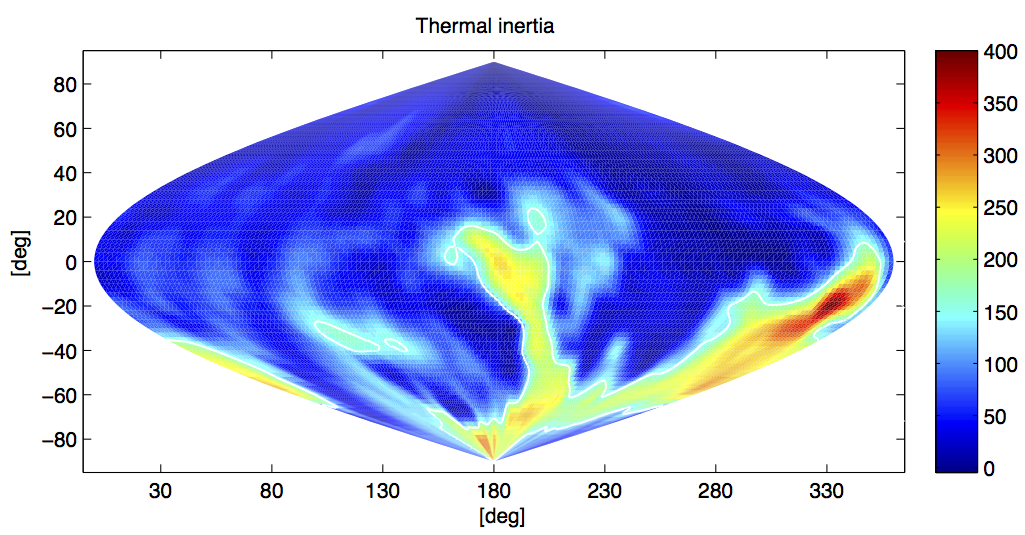}
  \caption{The colour map of the thermal inertia $\Gamma$ for models corresponding to all possible orientations of the spin axis 
   given in ecliptic coordinates $(\lambda, \beta)$. The white contour corresponds to $\Gamma = 150$\,J\,m$^{-2}$\,s$^{-0.5}$\,K$^{-1}$.
     \label{fig:thermal_inertia}}
\end{figure}

\begin{figure}[h!tb]
 \includegraphics[width=\columnwidth]{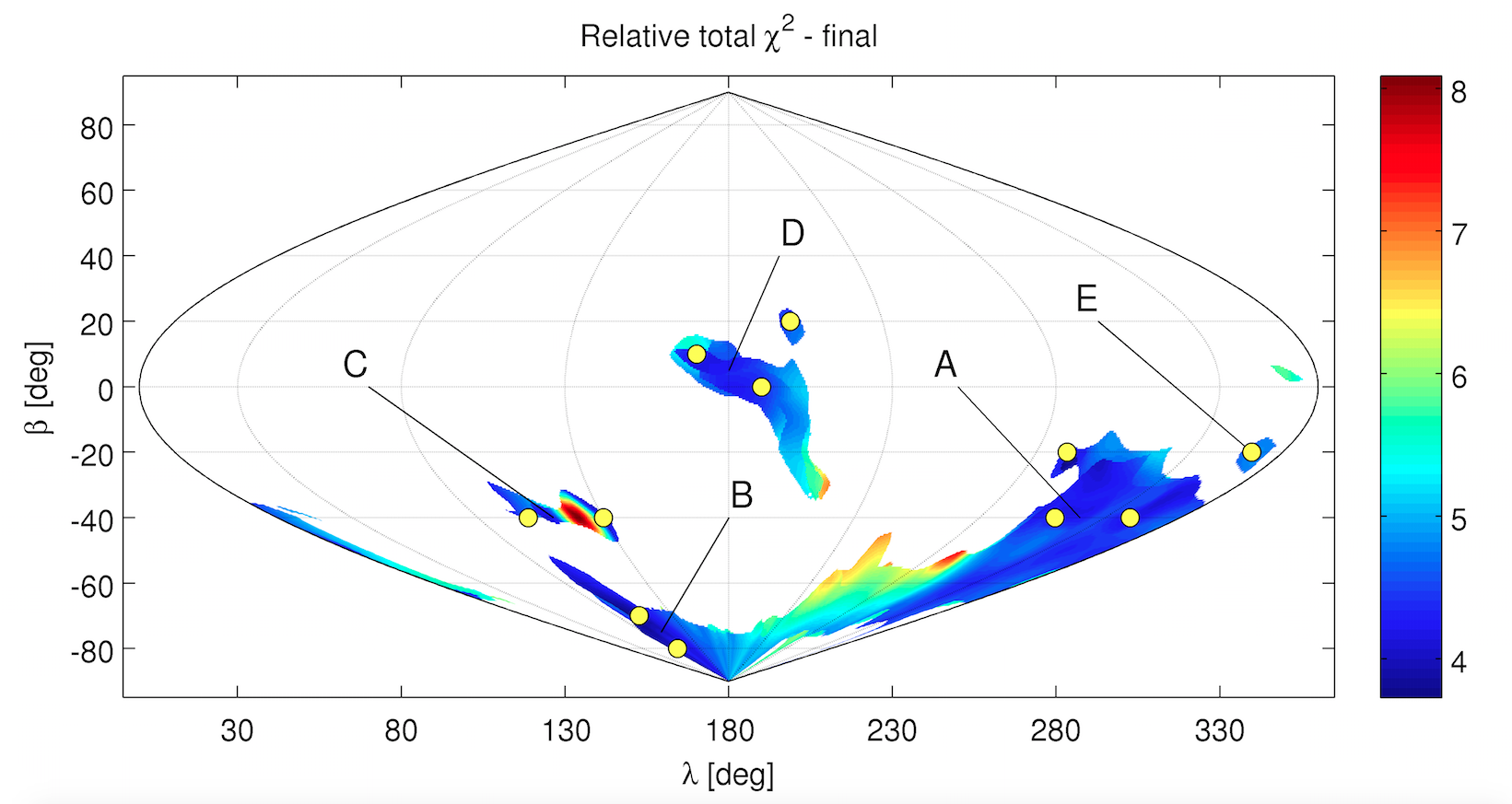}
  \caption{The intersection between the total $\chi^2$ colour maps (Fig.~\ref{fig:chisq_total}) for three possible rotation periods and the conditions $I_3 / I_z < 1.1$ and $\Gamma > 150$\,J\,m$^{-2}$\,s$^{-0.5}$\,K$^{-1}$.
     \label{fig:final_chisq}}
\end{figure}

Another constraint we used was related to thermal inertia $\Gamma$ of our models that is shown in Fig.~\ref{fig:thermal_inertia}. 
According to Campins et al.\ \cite{campins09}, thermal inertia of Ryugu is higher than 
150\,J\,m$^{-2}$\,s$^{-0.5}$\,K$^{-1}$, which is also confirmed by our analysis in Section~\ref{sec:ti}. 
Therefore, from all solutions in Fig.~\ref{fig:chisq_total} (and also from similar maps for two other acceptable periods of 7.6311 and 7.6326\,h), we selected only those that had $I_3 / I_z < 1.1$ and $\Gamma >150$\,J\,m$^{-2}$\,s$^{-0.5}$\,K$^{-1}$. 
These are shown in Fig.~\ref{fig:final_chisq}, where the blue areas show plausible pole solutions with low total $\chi^2$. Still the pole direction is not well defined.

\begin{figure}[h!tb]
 \includegraphics[width=\columnwidth]{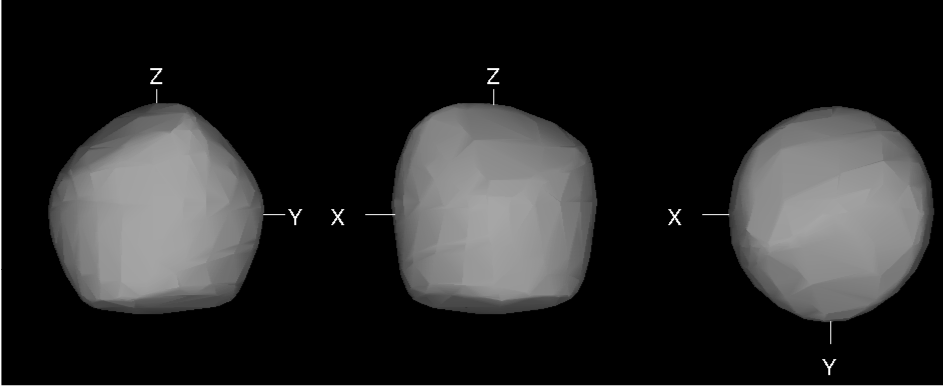}
  \caption{The formally best-fit shape model of Ryugu for pole direction $(340^\circ, -40^\circ)$.
     \label{fig:shape}}
\end{figure}

\begin{figure*}[t]
 \includegraphics[width=\textwidth]{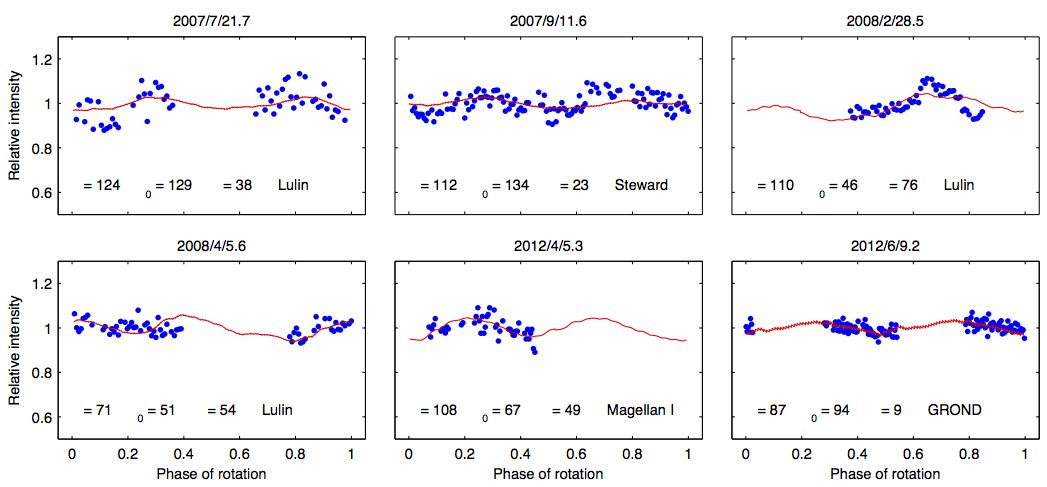}
  \caption{Comparison between the model (red curves) and the data (points) for a subset of visual lightcurves.
   The viewing and illumination geometry for the corresponding pole $(340^\circ, -40^\circ)$ is given
   by the latitude of sub-Earth point $\theta$, latitude of the subsolar point $\theta_0$, and the
   solar phase angle $\alpha$.
     \label{fig:lightcurves}}
\end{figure*}

Because this convex model is more flexible than the spherical model from Sections~\ref{sec:sph_set1} and
\ref{sec:sph_set2}, the range of possible pole directions is larger. 
Solutions with $\lambda$ between 310 and 340$^\circ$ and $\beta \sim -40^\circ$ (denoted A in Fig.~\ref{fig:final_chisq}) are preferred because they not only provide the lowest $\chi^2$ but are also stable against the limit of inertia ratio, 
the minimum value of $\Gamma$, particular value of the weight $w$ in Eq.~(\ref{eq:chisq}) and the resolution of the shape model. 
Moreover, they show no systematic trends in the distribution of residuals for the fit of thermal data and are also consistent with the results of the previous section. 
The formally best-fit model for period $P = 7.6311\,$h has the pole direction $(340^\circ, -40^\circ)$, a volume-equivalent diameter of 853\,m, a surface-equivalent diameter of 862\,m, 
a thermal inertia of 220\,J\,m$^{-2}$\,s$^{-0.5}$\,K$^{-1}$ and a very smooth surface (assuming no surface roughness in the model setup). The corresponding shape model is
shown in Fig.~\ref{fig:shape} and the synthetic lightcurves produced by this model are compared with the observed lightcurves in Fig.~\ref{fig:lightcurves}.

The other formally possible pole directions correspond to other blue `islands' (denoted A, B, C, D, E) in Fig.~\ref{fig:final_chisq} with the
approximate pole coordinates B: $(100^\circ, -70^\circ)$, C: $(100^\circ, -40^\circ)$, D: $(180^\circ, 0^\circ)$, and E: $(350^\circ, -20^\circ)$ 
(see Table~\ref{tbl:jd_solutions}). 

  \begin{table}[h!tb]
    \begin{center}
    \caption{Summary of possible pole solutions in ecliptic coordinates (see Fig.~\ref{fig:final_chisq}) from our
             analysis of the combined visual and thermal data set. Zone 'A' is our preferred solution
             because this zone is connected to the lowest $\chi^2$ and stable against the limit of inertia ratio,
             the minimum value of $\Gamma$, and different resolutions of the shape model (see also Section~\ref{sec:convex_all}).
             \label{tbl:jd_solutions}}
    \begin{tabular}{rrrlc}
      \hline
      \hline
      \noalign{\smallskip}
         & \multicolumn{2}{c}{Pole solution}  &  P$_{sid}$ & \\
      ID & $\lambda$ [$^{\circ}$] & $\beta$ [$^{\circ}$]  &  [h] & Zone \\
      \noalign{\smallskip}
      \hline
      \noalign{\smallskip}                   
       1 & 290 & -20 & 7.63108 & A \\ 
       2 & 340 & -40 & 7.63109 & A \\ 
       3 & 310 & -40 & 7.63001 & A \\ 
            \noalign{\smallskip}
       4 & 100 & -70 & 7.63254 & B \\ 
       5 &  90 & -80 & 7.62997 & B \\ 
            \noalign{\smallskip}
       6 & 130 & -40 & 7.63256 & C \\ 
       7 & 100 & -40 & 7.63005 & C \\ 
            \noalign{\smallskip}
       8 & 170 &  10 & 7.63123 & D \\ 
       9 & 190 &   0 & 7.63001 & D \\ 
      10 & 200 &  20 & 7.63001 & D \\ 
            \noalign{\smallskip}
      11 & 350 & -20 & 7.63256 & E \\ 
     \noalign{\smallskip}
     \hline
     \noalign{\smallskip}
    \end{tabular}
    \end{center}
  \end{table}

\section{Discussion and final thermophysical model analysis}
\label{sec:tpm}

\subsection{Analysis of previously published solutions}

\begin{figure*}[h!tb]
\rotatebox{90}{\resizebox{!}{9cm}{\includegraphics{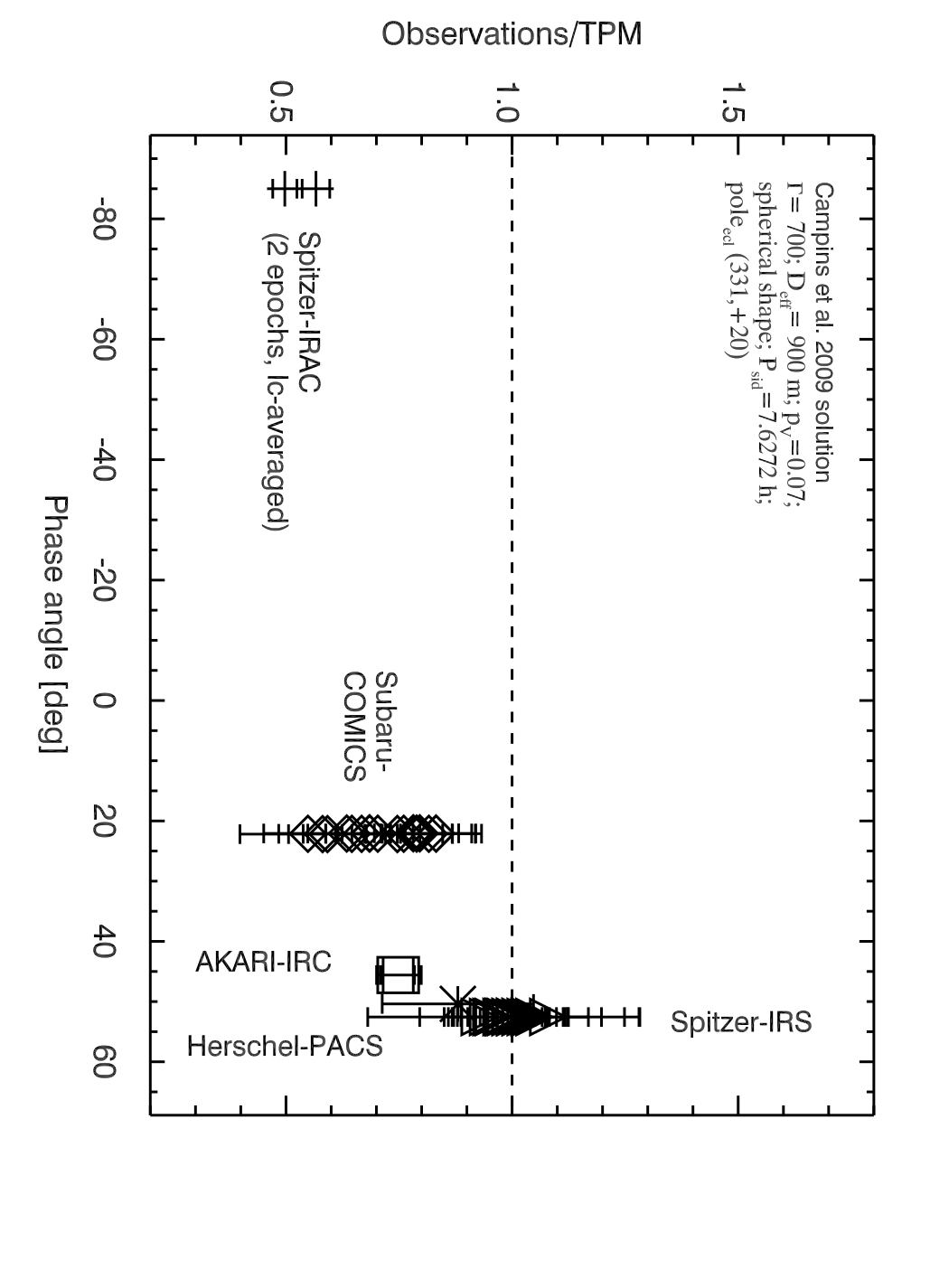}}}
\rotatebox{90}{\resizebox{!}{9cm}{\includegraphics{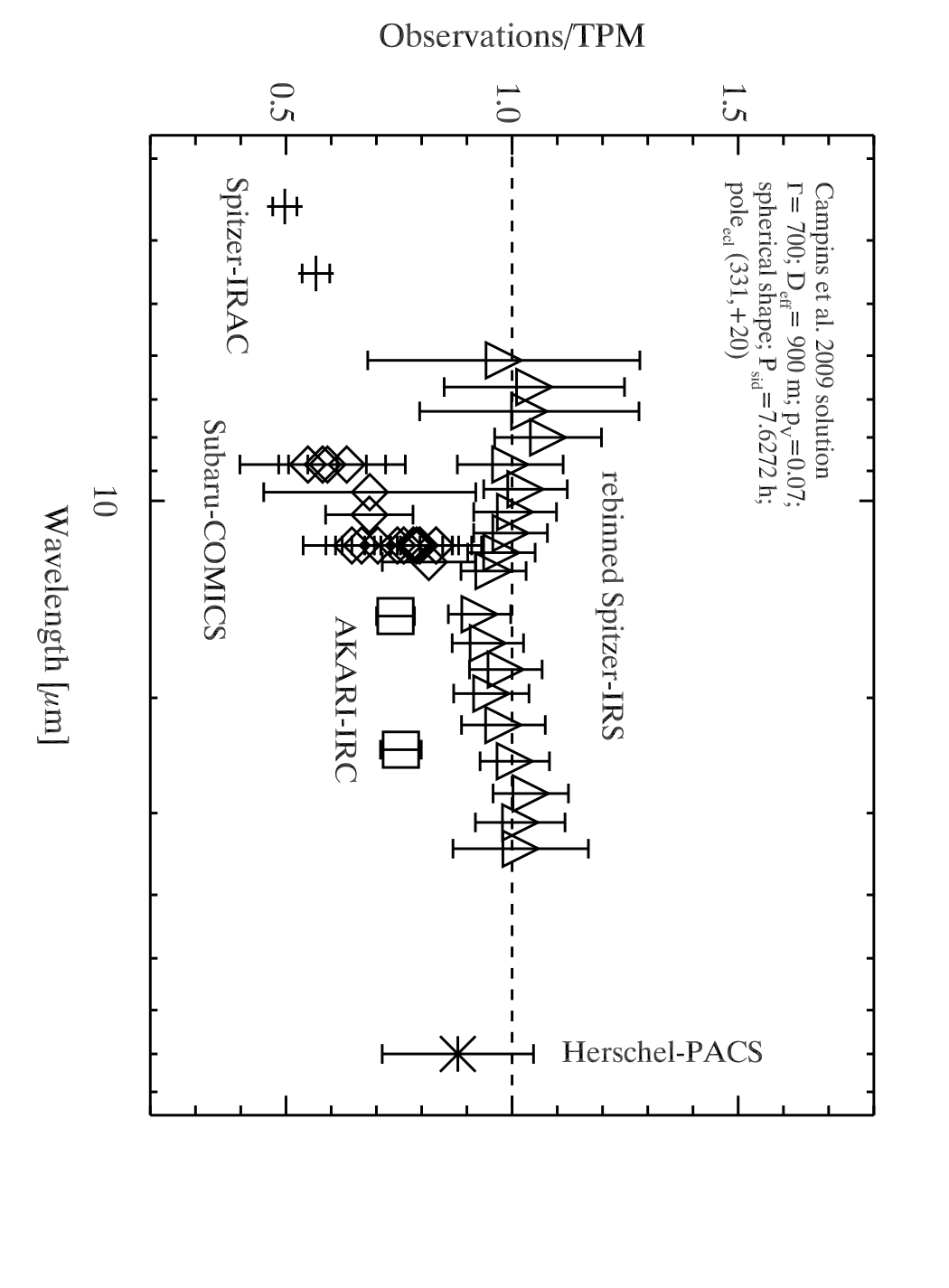}}}
\rotatebox{90}{\resizebox{!}{9cm}{\includegraphics{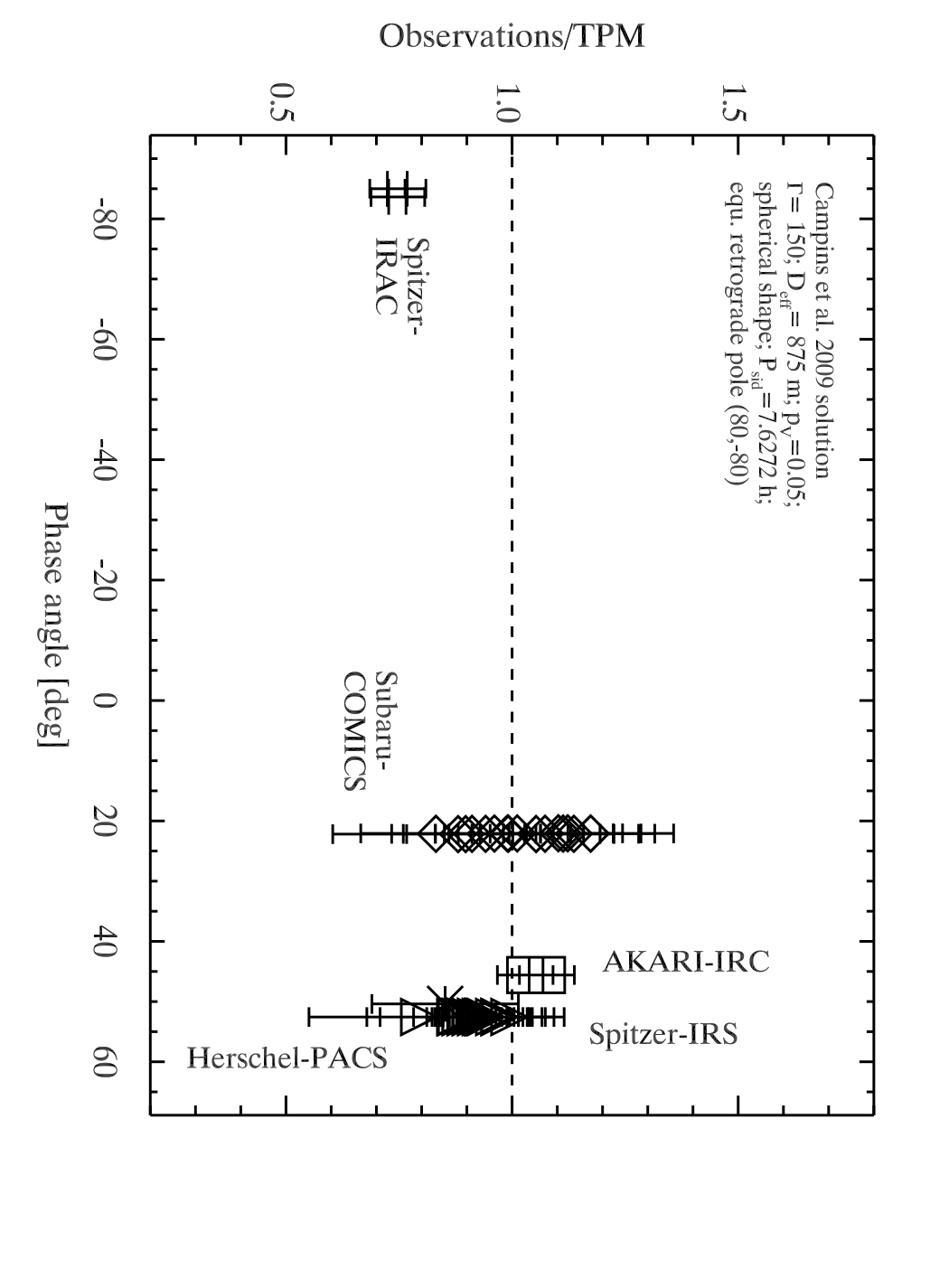}}}
\rotatebox{90}{\resizebox{!}{9cm}{\includegraphics{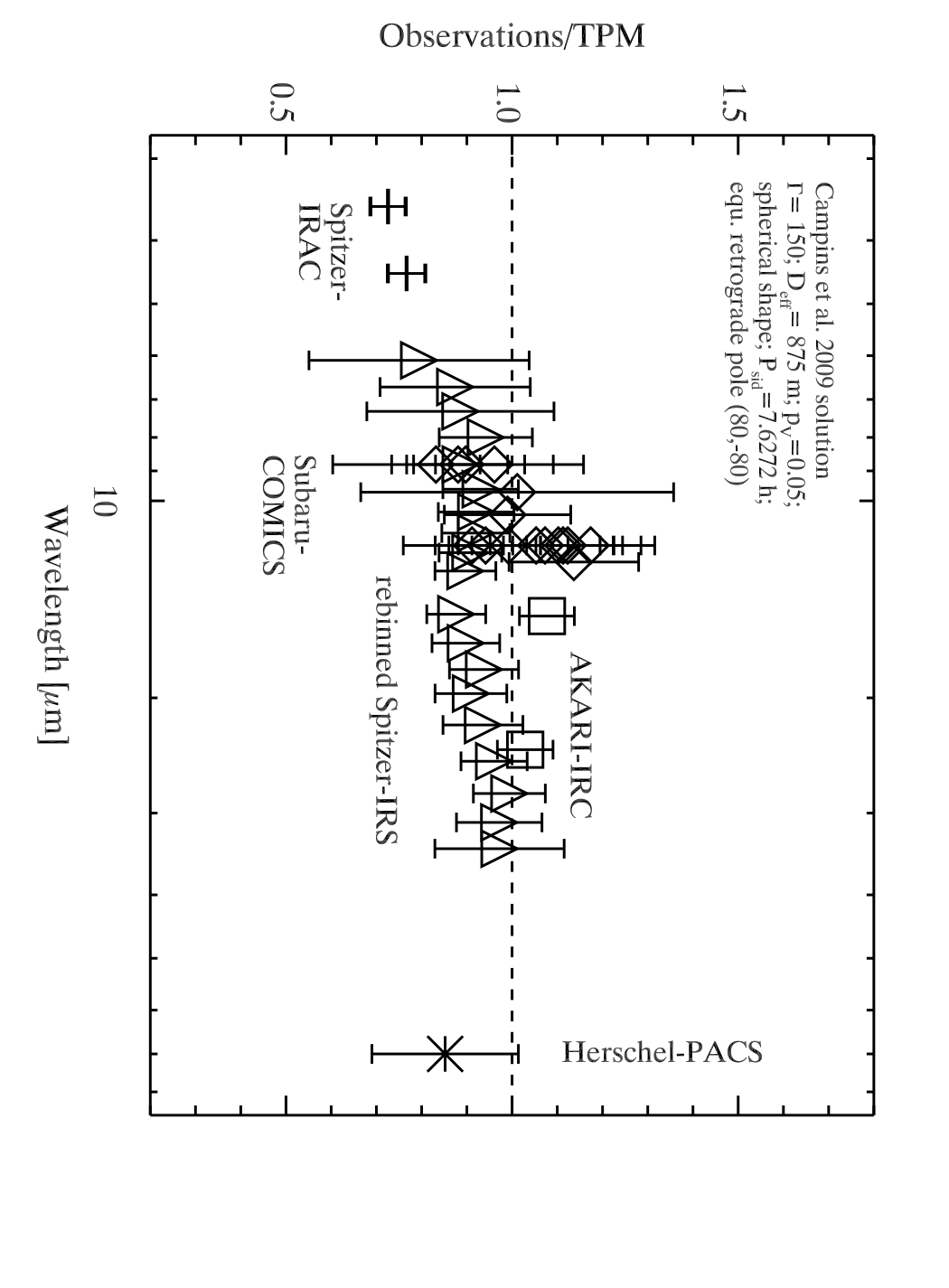}}}
\caption{Test of the Campins et al.\ (\cite{campins09}) solutions against our large
           thermal data set. Top left: Observation-to-TPM flux ratios as a function of phase angle;
           top right: as a function of wavelength; bottom left/right: same as top figure, but now 
           calculated for the lower thermal inertia limit of 150\,J\,m$^{-2}$\,s$^{-0.5}$\,K$^{-1}$
           and the extreme case of an equatorial retrograde geometry (at the time of IRS observations and
           as seen from Spitzer).
     \label{fig:obsmod_campins}}
\end{figure*}

Campins et al.\ (\cite{campins09}) analysed a single-epoch IRS measurement and found
a thermal inertia of 700 $\pm$ 200\,J\,m$^{-2}$\,s$^{-0.5}$\,K$^{-1}$ under the assumption
of the published spin-axis orientation by Abe et al.\ (\cite{abe08}). They also investigated the
reliability of the IRS data and found a rigorous lower limit of 150\,J\,m$^{-2}$\,s$^{-0.5}$\,K$^{-1}$
for an extreme case of an equatorial retrograde geometry. We tested the Campins et al.\ (\cite{campins09}) solution
($\Gamma$ = 700$\pm$ 200\,J\,m$^{-2}$\,s$^{-0.5}$\,K$^{-1}$, D$_{eff}$ = 0.90 $\pm$ 0.14\,km,
p$_V$ = 0.07 $\pm$ 0.01, spherical shape, spin axis with ($\lambda$, $\beta$)$_{ecl}$ = (331$^{\circ}$, +20$^{\circ}$),
P$_{sid}$ = 7.6272\,h) against our thermal data set. Their solution explains the Spitzer-IRS measurements
very well, with observation-to-model flux ratios close to one (see Fig.~\ref{fig:obsmod_campins} top left/right),
confirming the Campins et al.\ (\cite{campins09}) analysis.
However, this solution fails to reproduce measurements at other phase angles.
The high thermal inertia of 700\,J\,m$^{-2}$\,s$^{-0.5}$\,K$^{-1}$ (in combination with the spin-axis
orientation) leads to model predictions which are up to a factor of two higher than the measurements,
with the largest offsets being connected to the before-opposition Spitzer-IRAC
measurements. Using the limiting thermal inertia of 150\,J\,m$^{-2}$\,s$^{-0.5}$\,K$^{-1}$ in
combination with the extreme case of an equatorial retrograde geometry\footnote{The extreme case of an
equatorial retrograde geometry at the time of IRS observations and as seen from Spitzer corresponds to
a spin-axis orientation of ($\lambda$, $\beta$) = (80$^{\circ}$, -80$^{\circ}$).}
helps to explain part of the phase-angle-dependent offsets seen before (in combination
with a radiometrically optimised size and albedo), but overestimates the IRAC lightcurves by
approximately 25\% and also introduces trends with wavelengths and phase angle in the ratio plots
(Fig.~\ref{fig:obsmod_campins} bottom left/right). We tested the full range of thermal inertias from
0-2500\,J\,m$^{-2}$\,s$^{-0.5}$\,K$^{-1}$ for smooth and rough surfaces, but it was not possible
to simultaneously explain  all thermal measurements with either of these two spin-axis orientations.
The remaining offsets are too big to be explained by shape effects without violating the
constraints from the shallow visual lightcurves.
The analysis of the Campins et al.\ (\cite{campins09}) solutions
in the context of our much larger data set shows that
(i) the high thermal inertia of 700\,J\,m$^{-2}$\,s$^{-0.5}$\,K$^{-1}$ in combination with the spin axis by Abe et al.\ \cite{abe08}) is not correct;
(ii) the low thermal inertia of 150\,J\,m$^{-2}$\,s$^{-0.5}$\,K$^{-1}$ in combination with an extreme case of an equatorial retrograde geometry
is also problematic;
(iii) the thermal data include crucial information about thermal inertia and the
orientation of the spin axis;
(iv) a single-epoch measurement, even if covering a wide wavelength range, is not sufficient
to determine the object's thermal inertia;
(v) offsets and trends between measurements and model predictions are clearly connected to
incorrect assumptions in the model surface temperature distribution and not due to shape
effects which could possibly explain offsets of 10\%, but not the huge discrepancies
seen in Fig.\ \ref{fig:obsmod_campins}.

It is also worth taking a closer look at the three proposed solutions by M\"uller et al.\ (\cite{mueller11a}).
Here, the authors used more complex shapes, but their thermal data were limited
to a phase angle range between $+$20$^{\circ}$ and $+$55$^{\circ}$ and the focus of the TPM analysis was
on high-roughness surfaces. We tested all three shape-spin settings
against our much larger thermal data set (including the IRAC point-and-shoot measurements),
and now considering much smoother surfaces with r.m.s.\ surface slopes below 0.4.
The calculations show the following:
(i) all solutions seem to point to a much smoother surface than what was assumed in M\"uller et al.\ (\cite{mueller11a});
(ii) their best solution ($\lambda$, $\beta$) = (73.1$^{\circ}$, -62.3$^{\circ}$) in ecliptic coordinates leads to a
thermal inertia of 200\,J\,m$^{-2}$\,s$^{-0.5}$\,K$^{-1}$, and their solutions \#2 and
\#3 would require lower thermal inertias below the Campins et al.\ limit of 150\,J\,m$^{-2}$\,s$^{-0.5}$\,K$^{-1}$;
(iii) solution \#1 is close to our solutions in zone B (see Fig.~\ref{fig:final_chisq} and Table~\ref{tbl:jd_solutions})
and explains most of the observational data within the given observational error bars.
Only the PACS measurement and one of the IRAC lightcurve measurements are off by approximately 15-20\%.
This model also fails to explain part of the IRAC point-and-shoot sequence and produces observation-to-model
flux ratios ranging from 0.78 to 1.55, some ratios even being well outside the 3-$\sigma$ error bars. 
Interestingly, the point-and-shoot measurements in early 2013, when the phase angle was changing from
-70$^{\circ}$ to -90$^{\circ}$ , are nicely matched, while some of the measured fluxes in May 2013
are up to 55\% higher than the corresponding model predictions. In this time period, the phase angle was
in the range -90$^{\circ}$ to -55$^{\circ}$ and the object had its closest approach to Spitzer ($\approx$0.11\,AU).
Overall, solution \#1 from M\"uller et al.\ (\cite{mueller11a}) might still be acceptable from a
statistical point of view (depending on the weight
of the short-wavelength IRAC data in the radiometric analysis), but the mismatch to some IRAC measurements
is obvious and we exclude that solution.

Yu et al.\ (\cite{yu2014}) calculated a shape solution from MPC (Minor Planet Center) lightcurves,
but their shape solution is not publically available. Looking at the MPC data, we doubt
that a reliable shape can be reconstructed for the existing data. Also, the effective diameter
of 1.13 $\pm$ 0.03\,km (p$_V$=0.042 $\pm$ 0.003) is unrealistically large. Using the Yu et
al.\ (\cite{yu2014}) solution (in combination with a spherical shape) produces observation-to-model
flux ratios in the range 0.43 to 0.88, that is, the model predictions are far too high.

\subsection{New solutions and constraints on Ryugu's properties}

\subsubsection{Thermal inertia}
\label{sec:ti}

\begin{figure}[t]
 \rotatebox{90}{\resizebox{!}{\columnwidth}{\includegraphics{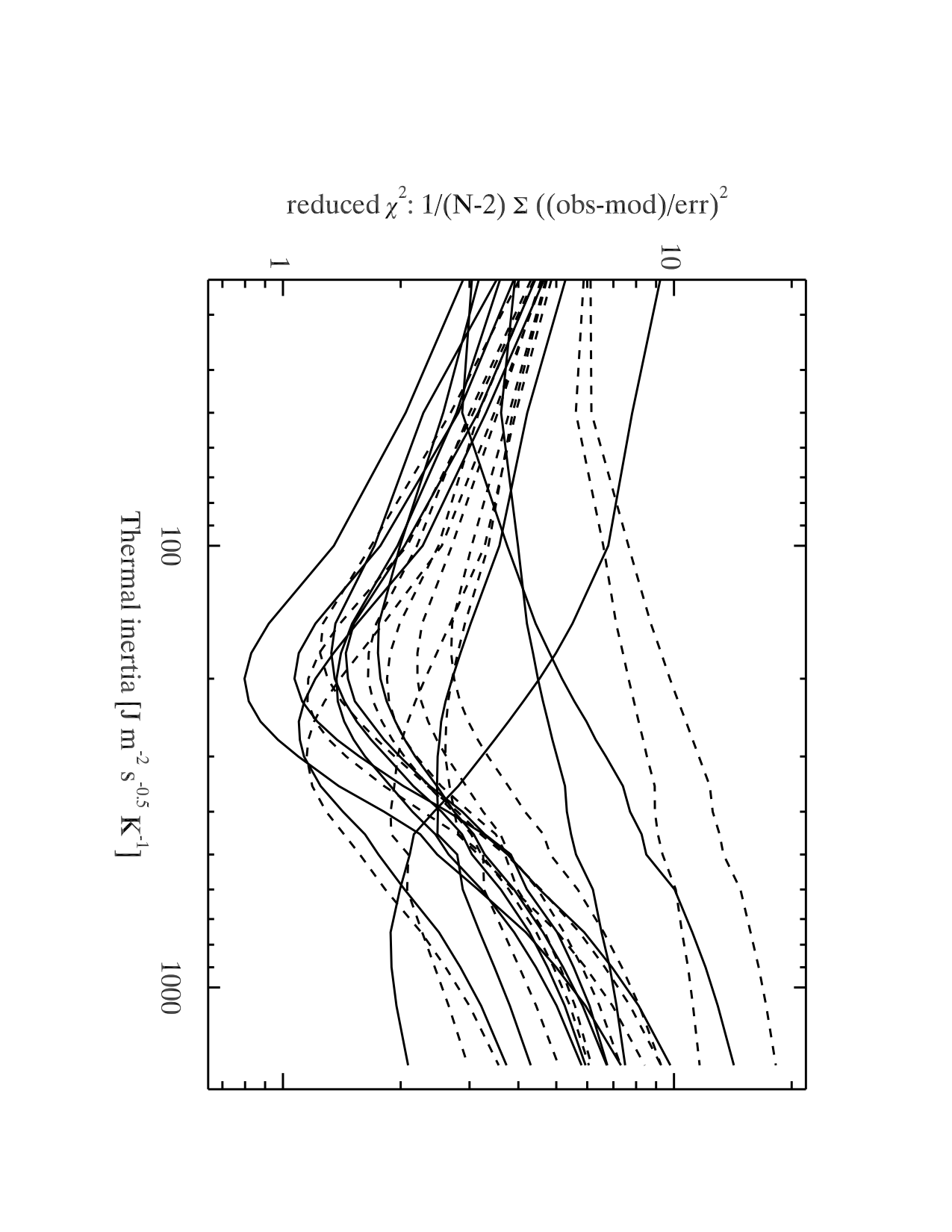}}}
  \caption{Reduced $\chi^2$ values as a function of thermal inertia for
   all 11 shape and spin-pole solutions. True shape solutions are shown as
   solid lines, spherical shapes as dashed lines. Acceptable solutions should
   have reduced $\chi^2$ values close to 1.
     \label{fig:chi2_final}}
\end{figure}

The $\chi^2$ analysis in the sections \ref{sec:spherical_shape} and \ref{sec:convex_all}
point towards a pole direction close to $(310^\circ, -40^\circ)$, but it cannot completely rule out
other solutions from a statistical point of view.
This can also be seen in Fig.~\ref{fig:chi2_final} where we show $\chi^2$ curves
as a function of thermal inertia for all remaining 11 shape and spin-pole solutions (Tbl.~\ref{tbl:jd_solutions}).
The corresponding spherical shape solutions are shown as dashed lines. The most
striking thing is that the lowest $\chi^2$ values are all connected to thermal
inertias of 200-300 \,J\,m$^{-2}$\,s$^{-0.5}$\,K$^{-1}$. There is another group
of solutions with higher $\chi^2$ values close to our thermal-inertia boundary
of 150\,J\,m$^{-2}$\,s$^{-0.5}$\,K$^{-1}$ (see analysis by Campins et al.\ (\cite{campins09})
of the Spitzer-IRS data), but no acceptable solutions anymore at Itokawa-like and higher values
for the thermal inertia.

Our combined thermal data set, including the IRAC measurements, constrains
the thermal inertia very well: we can confirm the
thermal-inertia boundary of 150\,J\,m$^{-2}$\,s$^{-0.5}$\,K$^{-1}$ (see analysis by
Campins et al.\ (\cite{campins09})) and starting at approximately 300\,J\,m$^{-2}$\,s$^{-0.5}$\,K$^{-1}$
the models cannot explain the IRAC measurement sequences anymore.

\subsubsection{Surface roughness}

The statistical analysis in Section~\ref{sec:convex_all} led to a formally best solution
for smooth surfaces. If we introduce roughness at a low level, we still find
the best solutions connected to zone A, but now the best spin pole seems to move towards
a higher longitude approaching the solution at $(340^\circ, -40^\circ)$. However, as soon as
the r.m.s.\ of the surface slopes goes above 0.1 there are
problems in fitting the IRAC point-and-shoot sequence, and also in fitting the two IRAC
thermal lightcurves. Overall, higher levels of roughness are connected to size-albedo
solutions with higher thermal inertias and vice versa. This degeneracy problem is present
in most radiometric solutions  (see also Rozitis \& Green \cite{rozitis11} for a discussion
on the degeneracy between roughness and thermal inertia), but here the low-roughness
solutions are clearly favoured.

\subsubsection{Size and albedo}

Radiometric size and albedo constraints depend heavily on the shape-spin solution. In
the central part of zone A (solution \#2 in Table~\ref{tbl:jd_solutions})
we find sizes (of an equal volume sphere) of 850 to 880\,m, and albedos of 0.044 to 0.050
(connected to H$_V$ = 19.25\,$\pm$\,0.03\,mag). Including the zone-A boundaries and
considering the full thermal inertia and roughness range, as well as the corresponding
shape solutions, we estimate a size range of approximately 810 to 900\,m (of an equal-volume sphere)
and geometric V-band albedos of approximately 0.042 to 0.055. However, all the solutions (\#1 and \#3 in
Table~\ref{tbl:jd_solutions}) have issues with fitting part of the thermal data
(see Section~\ref{sec:tpm}) and we restrict our size-albedo values to solution \#2
in the central zone A.

\subsubsection{Grain sizes}

We use a well-established method (Gundlach \& Blum \cite{gundlach13}) to determine the
grain size of the surface regolith of Ryugu.
First, the thermal inertia $\Gamma$ can be translated into a possible range of
thermal conductivities $\lambda$ with
\begin{equation}
\lambda = \frac{\Gamma^2}{\phi \rho c},
\label{E:lambdaAst}
\end{equation}
where $c$ is the specific heat capacity, $\rho$ the material density, and $\phi$ the
regolith volume-filling factor, which is typically unknown. This last parameter
is varied between 0.6 (close to the densest packing or "random close pack (RCP)"
of equal-sized particles) and 0.1 (extremely fluffy packing or "random ballistic
deposition (RBD)", plausible only for small regolith particles where the
van-der-Waals forces are larger than local gravity).
For the calculation, we used the CM2 meteoritic sample properties from Opeil et
al.\ (\cite{opeil10}), with a density $\rho$ = 1700 kg\,m$^{-3}$, and a specific
heat capacity of the regolith particles $c$ = 500\,J\,kg$^{-1}$\,K$^{-1}$.
This leads to heat conductivities $\lambda$ in the range 0.1\,W\,K$^{-1}$\,m$^{-1}$ (average
thermal inertia combined with the highest volume-filling factor) to 0.6\,W\,K$^{-1}$\,m$^{-1}$
assuming the lowest filling factor (also considering the full thermal inertia range 
would result in a $\lambda$ range of 0.04 to 1.06\,W\,K$^{-1}$\,m$^{-1}$).

We combine this information with the heat conductivity model by Gundlach \& Blum
(\cite{gundlach13}), again by using properties of CM2 meteorites, to estimate
possible grain sizes on the surface.
First, we calculated maximum surface temperatures in the range $\approx$320 to 375\,K,
considering the derived object properties and heliocentric distances
(r$_{helio}$: 1.00 - 1.41\,AU) of our observational (thermal) data set.
At the object's semi-major axis distance (a = 1.18\,AU) we find a reference 
maximum temperature of 350\,K. It is worth noting that changing the surface 
temperature by a few tens of degrees does not significantly affect the results.
In a second step, we determine the mean free path of the photons, the Hertzian
dilution factor for granular packing for the specified regolith volume-filling
factors. Our estimated grain sizes are in the range 1 to 10\,mm, in excellent
agreement with indpendent calculations by Gundlach (priv.\ communications)
which led to 0.7 to approximately 7\,mm grain sizes. We note that our value is different
than values given in  Gundlach \& Blum (\cite{gundlach13}), mainly due to
different assumptions for the thermal inertia.
At the derived grain-size level, the heat transport on the surface is still
dominated by radiation, with increasing heat conduction for lower thermal
inertias.

\subsubsection{Reference solution}

\begin{figure}[t]
 \rotatebox{90}{\resizebox{!}{\columnwidth}{\includegraphics{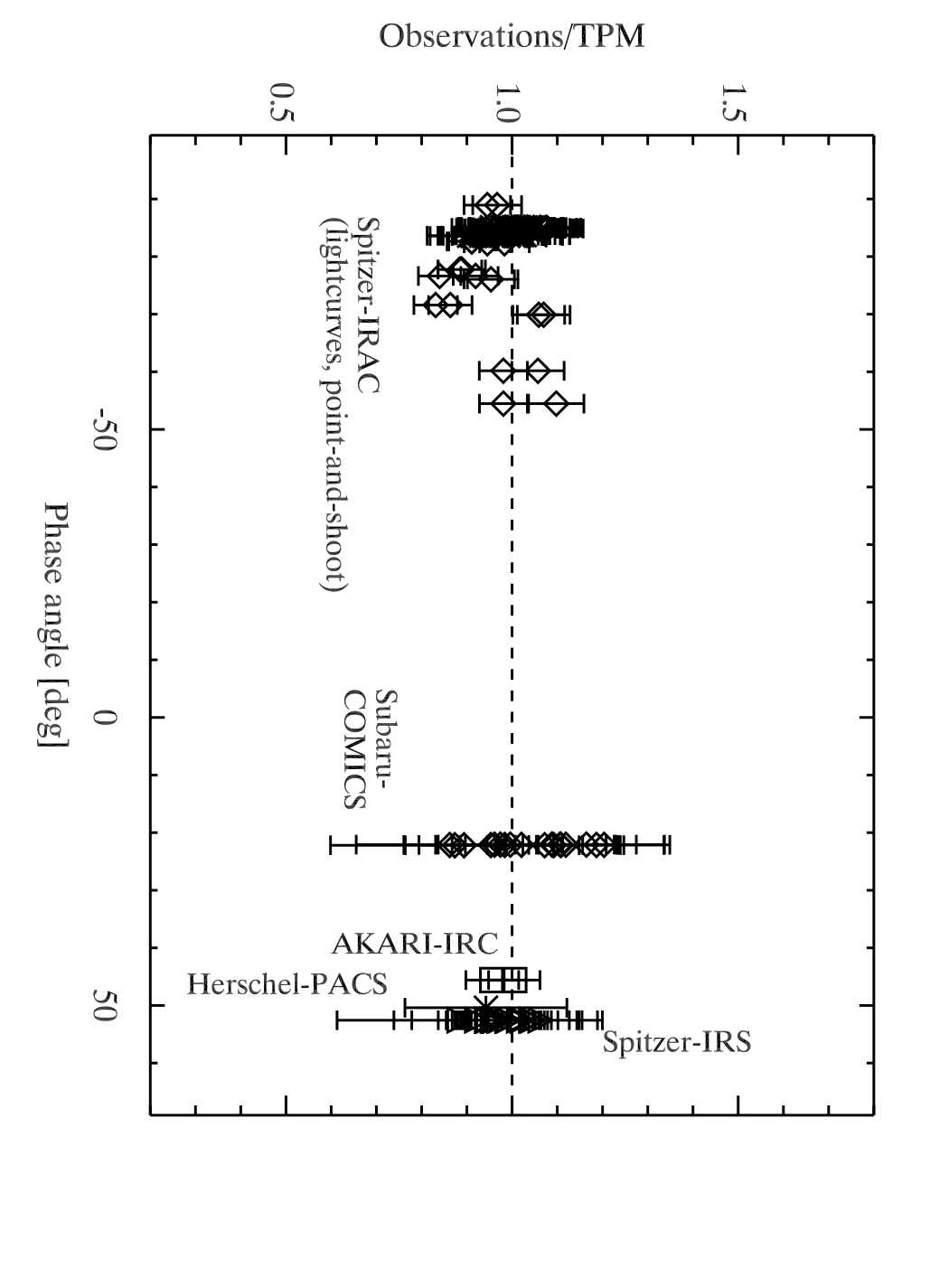}}}
 \rotatebox{90}{\resizebox{!}{\columnwidth}{\includegraphics{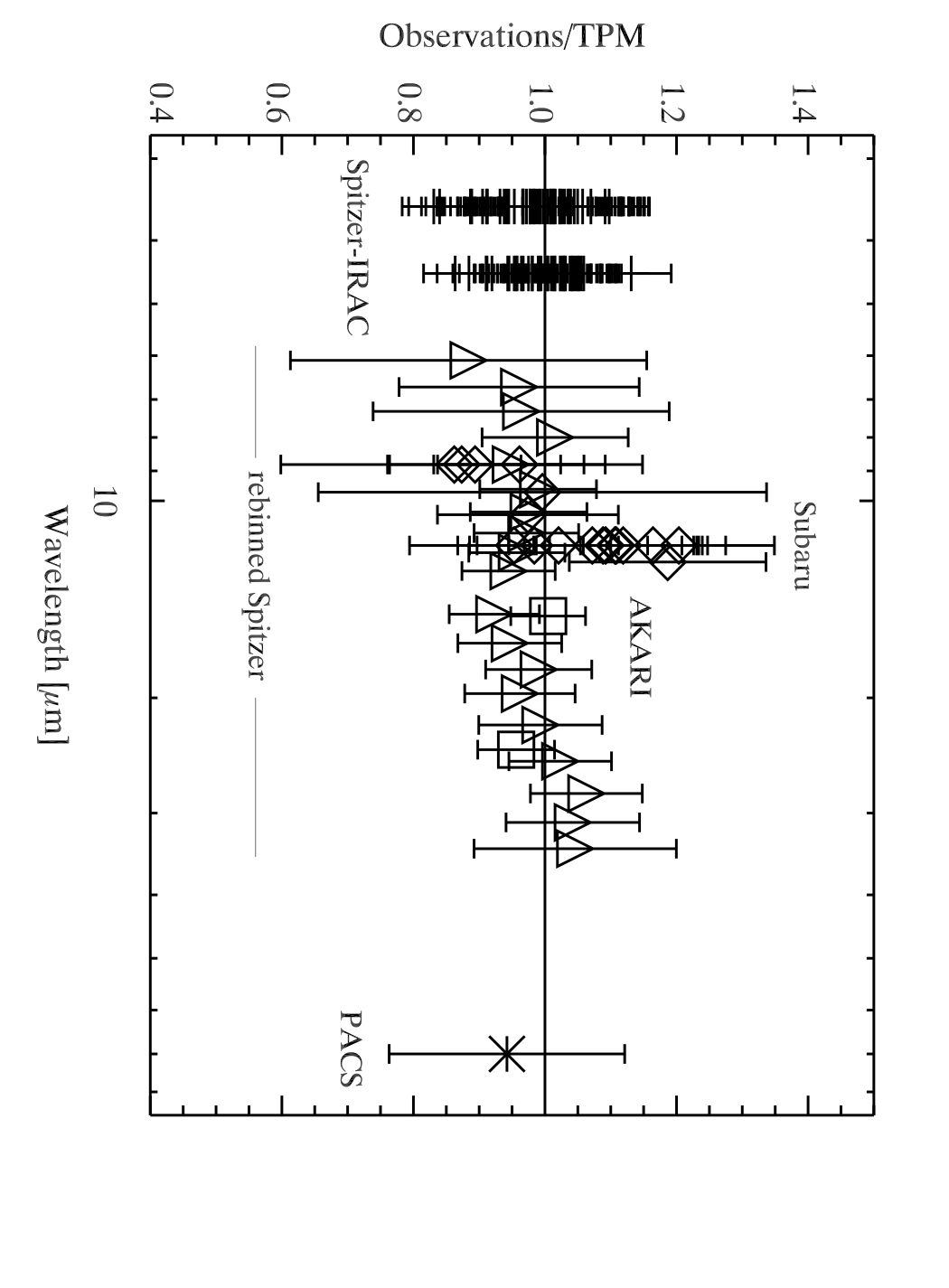}}}
  \caption{All thermal observations divided by the corresponding TPM prediction based on solution
           \#2 in Tbl.~\ref{tbl:jd_solutions} as a function of phase angle (top), and as a function of wavelengths (bottom).
           No trend in observation-to-model flux ratios is visible over the very wide phase angle range from $-89^{\circ}$ to $+53^{\circ}$,
           nor over the wide range of wavelengths from 3.55 to 70\,$\mu$m. The rebinned IRS data are
           shown as triangles together with the absolute flux error of each individual data point.
     \label{fig:obsmod1}}
\end{figure}

\begin{figure}[t]
 \rotatebox{90}{\resizebox{!}{\columnwidth}{\includegraphics{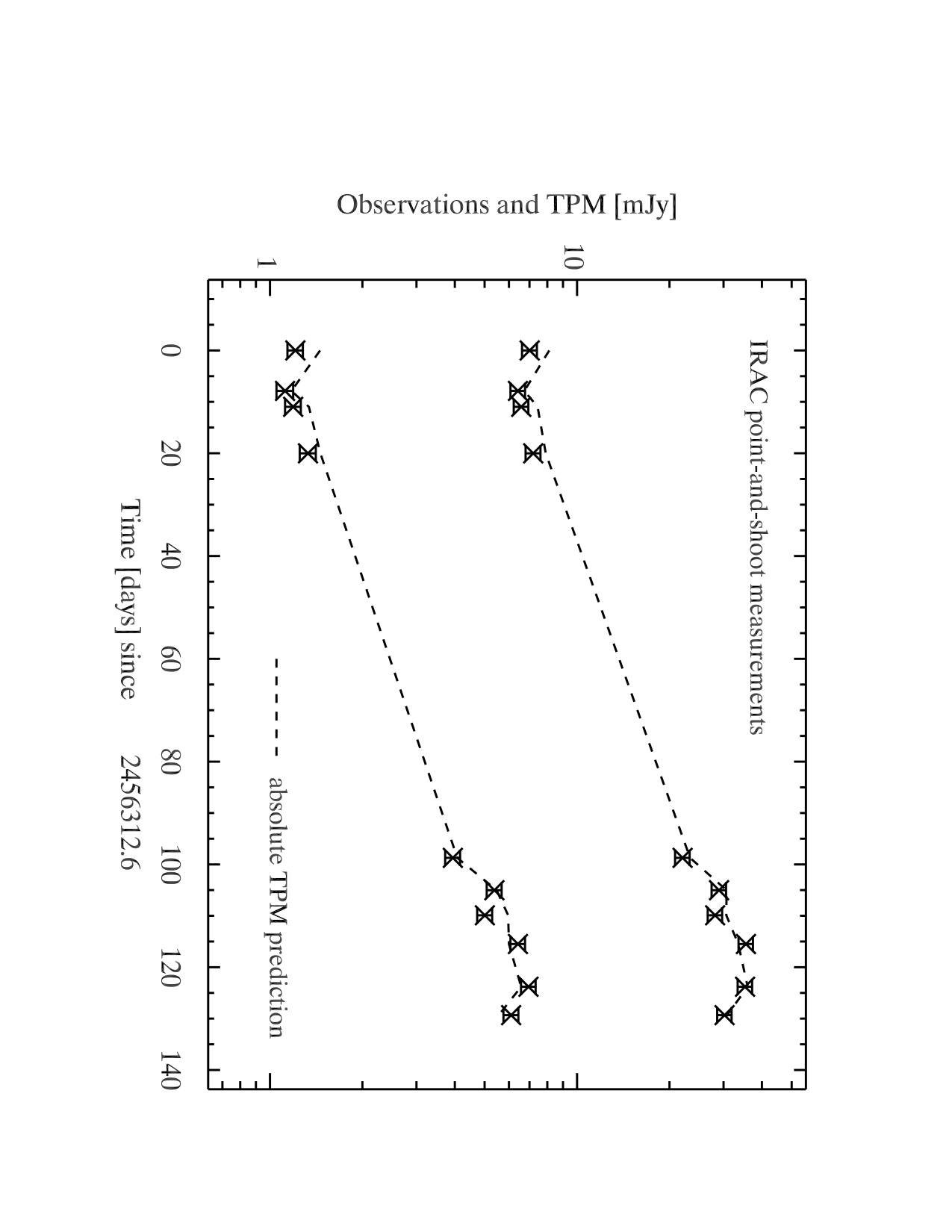}}}
  \caption{Absolutely calibrated Spitzer-IRAC point-and-shoot fluxes at 3.550 and 4.493\,$\mu$m
     taken between 20th Jan, and  30th May, 2013 (phase angles go from -71.6$^{\circ}$ to -88.9$^{\circ}$
     and back to -54.5$^{\circ}$ during that period. The absolute TPM predictions (\#2 n Tbl.~\ref{tbl:jd_solutions})
     for both channels are shown as dashed lines.
     \label{fig:irac_pas}}
\end{figure}

\begin{figure}[t]
  \rotatebox{90}{\resizebox{!}{\columnwidth}{\includegraphics{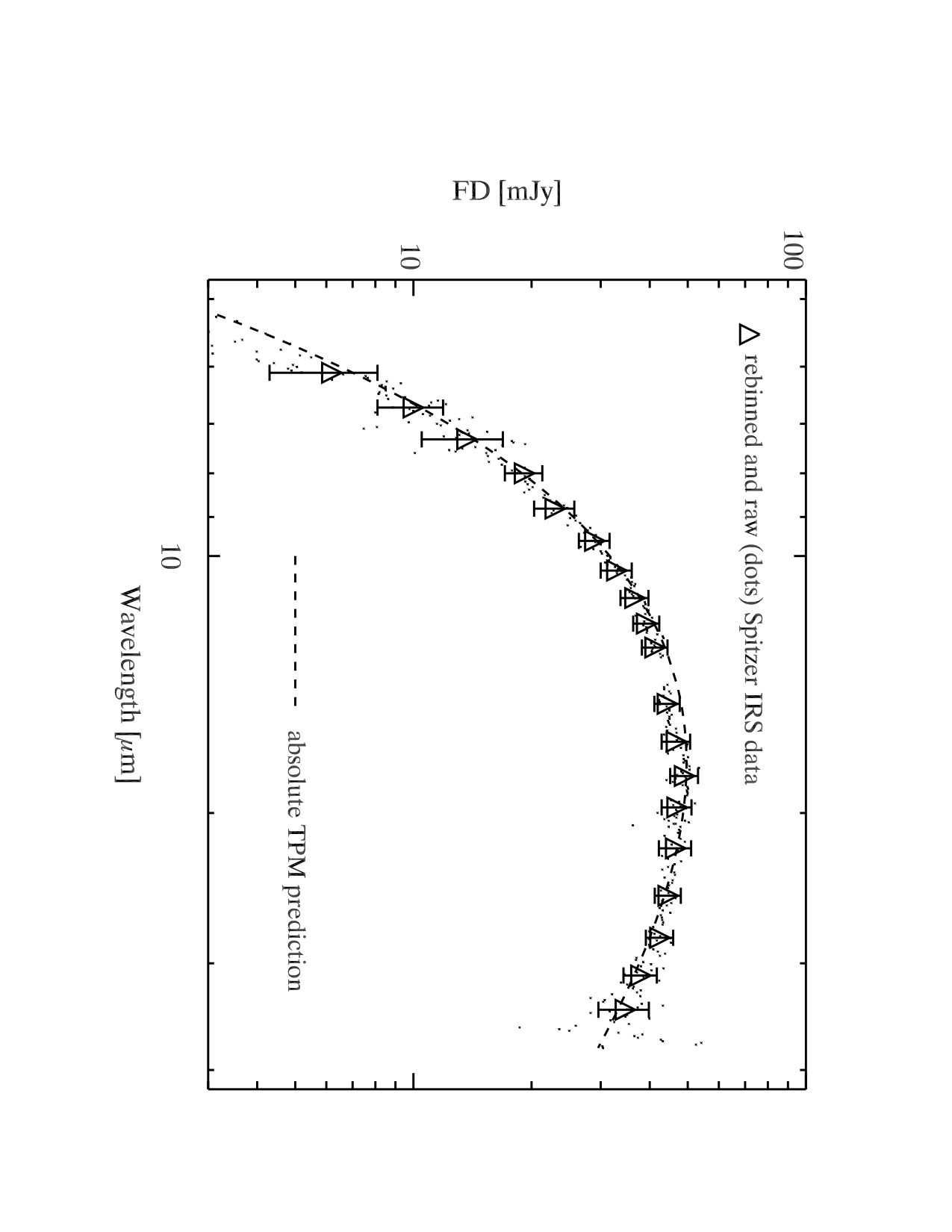}}}
  \caption{The absolutely calibrated Spitzer-IRS spectrum from  2nd May, 2008 (see Campins
           et al.\ \cite{campins09}), rebinned data are shown as triangles, the full data
           set as dots. The absolute TPM prediction (\#2 n Tbl.~\ref{tbl:jd_solutions})
           is shown as a dashed line.
     \label{fig:irs}}
\end{figure}

\begin{figure}[t]
 \rotatebox{90}{\resizebox{!}{\columnwidth}{\includegraphics{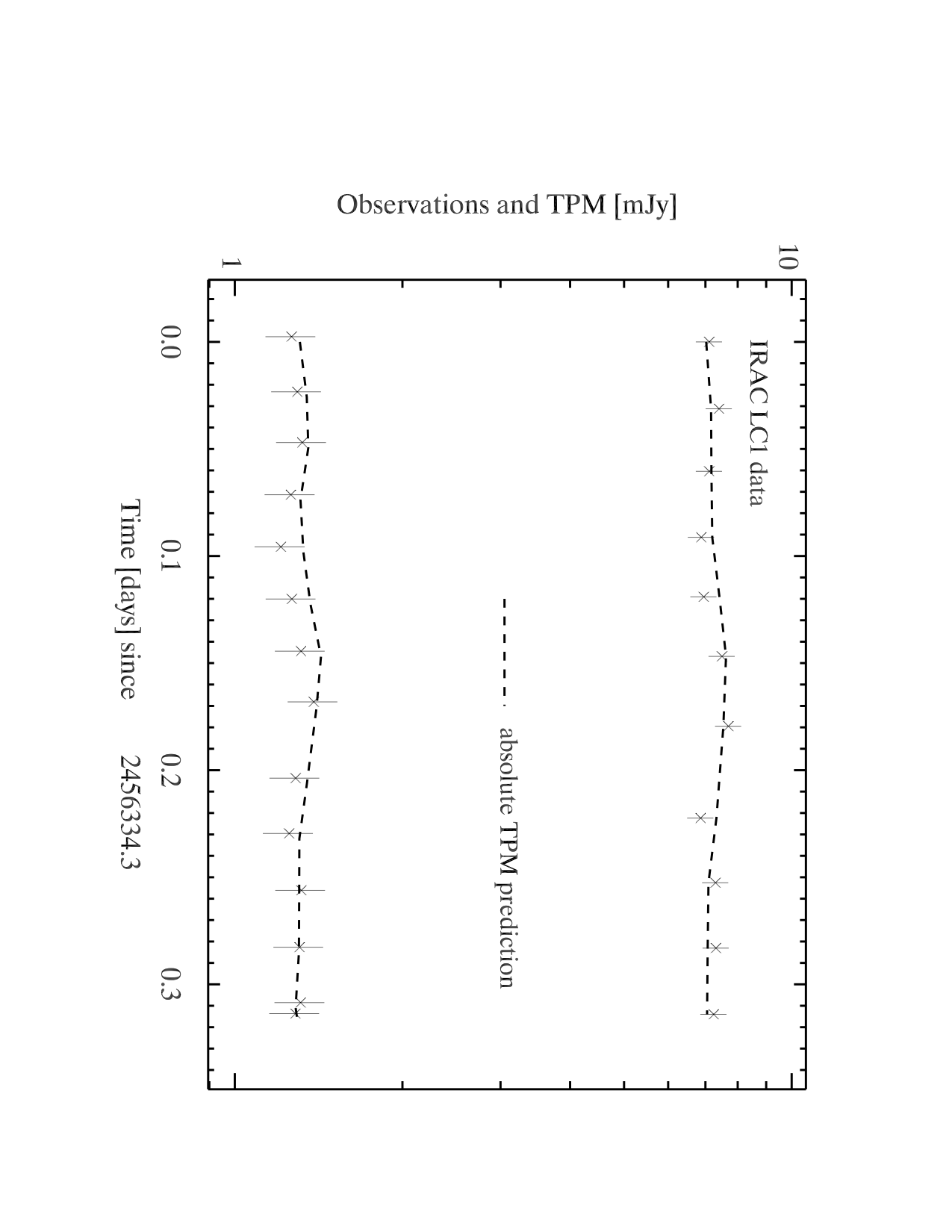}}}
 \rotatebox{90}{\resizebox{!}{\columnwidth}{\includegraphics{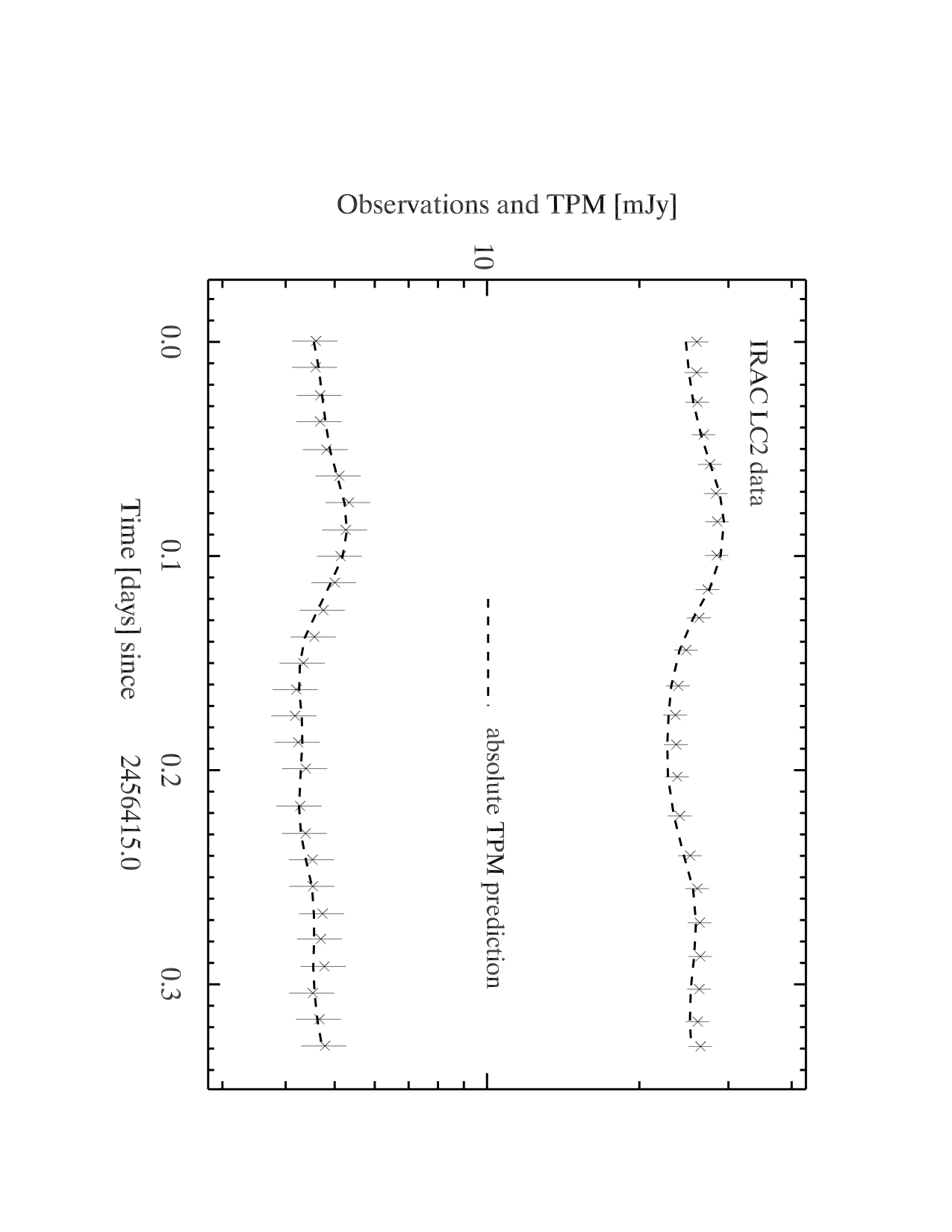}}}
  \caption{The absolutely calibrated Spitzer-IRAC lightcurves in both channels from
            10/11th Feb, 2013 (phase angle: -83.6$^{\circ}$, aspect angle: 137.8$^{\circ}$) (top),
            and from 2nd May, 2013 (phase angle: -85.0$^{\circ}$, aspect angle: 129.5$^{\circ}$) (bottom).
            The absolute TPM predictions (\#2 n Tbl.~\ref{tbl:jd_solutions}) are shown as dashed lines.
     \label{fig:irac_lc1}}
\end{figure}

Figures~\ref{fig:obsmod1}, \ref{fig:irac_pas}, \ref{fig:irac_lc1},
and \ref{fig:irs} inform us on the quality
of our TPM predictions and the thermal IR data, in combination with wavelengths,
phase angle and time of the individual measurements. For these figures we calculated
the TPM predictions for each data point using the true observing and illumination
geometry as seen from the specific observatory. The TPM fluxes are used in terms of
absolute times and absolute fluxes; {\bf no shifting or scaling was applied}. The model
has the following settings:
\begin{itemlist}
\item[$\bullet$] shape solution with ($\lambda$, $\beta$) =
                 (340$^{\circ}$, -40$^{\circ}$) in ecliptic coordinates;
                 P$_{sid}$ = 7.63109\,h (\#2 in Tbl.~\ref{tbl:jd_solutions})
                 from inversion technique using visual and thermal lightcurves
                 and infrared photometric data points
\item[$\bullet$] thermal inertia (top layer) of 200\,J\,m$^{-2}$\,s$^{-0.5}$\,K$^{-1}$
\item[$\bullet$] low surface roughness with r.m.s.\ of surface slopes of 0.05
\item[$\bullet$] size (of an equal-volume sphere): 856\,m
\item[$\bullet$] geometric V-band albedo: 0.049 (connected to H$_V$=19.25\,mag)
\item[$\bullet$] emissivity of 0.9
\end{itemlist}

This TPM solution matches the different thermal observational data sets very well
over wide ranges of phase angles, times, wavelengths and
rotational phases. For the IRAC data (very short thermal wavelengths at the Wien-part
of the SED) there are still small residuals,
but here individual shape facets and local temperature anomalies can easily
change the total fluxes and significantly influence the interpretation.
At longer wavelengths, that is, in the Rayleigh-Jeans part of the spectrum,
these small-scale shape and temperature features are less relevant. Here,
the shape of the SED is connected to the disk-averaged temperatue on the surface.
The observed long-wavelength fluxes are therefore crucial for the
determination of the object's size.

The observation-to-model plots in Figure~\ref{fig:obsmod1} are very
sensitive to changes in thermal inertia.  Smaller thermal inertias lead to a
peaked temperature distribution (close to the sub-solar point) and the corresponding
disk-integrated object flux is
dominated by the hottest surface temperature; at least at short wavelengths below 5\,$\mu$m.
Larger thermal inertias `transport' the surface heat to the `evening' parts of the
surface which are not directly illuminated by the Sun. This redistribution leads to
a slightly different shape of the spectral energy distribution since the warm regions
at the evening terminator contribute noticably to the total flux. As a result, the slope
in observation-to-model flux ratios changes (best visible in Fig.~\ref{fig:obsmod_campins}, right side).

The difference between a warm evening side and a cold morning side is largest
for large phase angles and at wavelengths close to or beyond the thermal emission peak
at approximately 20\,$\mu$m (see also figures and discussion in M\"uller \cite{mueller02}).
Our thermal data set has observations taken before and after opposition, covering
a wide range of phase angles, but the data are very diverse in wavelength and
quality. The Spitzer IRAC data are all taken before opposition (negative phase angles,
leading the Sun) and, in addition, these very short wavelengths are less sensitive to
the morning-evening effects. All crucial data sets (AKARI, Subaru, Herschel) for constraining
the thermal inertia via before-after-opposition asymmetries are taken either at a very
limited phase angle range (AKARI-IRC, Herschel-PACS) or have substantial
error bars (Subaru-COMICS, Herschel-PACS).

The most direct thermal inertia influence can be seen in the Spitzer-IRS spectrum
(see Fig~\ref{fig:obsmod_campins}, right side).
The ratios between observed
fluxes and the corresponding model prediction show a clear trend with wavelengths
at low (and also at high) values of thermal inertia.
The mismatch between observed
and modeled fluxes is evident, and would be even larger in cases of higher thermal
inertias. From a statistical point of view, both cases are still acceptable, but
there are two reasons why we have higher confidence in \#2 of Table~\ref{tbl:jd_solutions}
which is connected to a thermal inertia of around 200\,J\,m$^{-2}$\,s$^{-0.5}$\,K$^{-1}$:
(1) TPM predictions with lower or higher thermal inertias produce 
    wavelength-dependent ratios increasing or decreasing with wavelength, respectively;
    these trends can be attributed to unrealistic model temperature distributions
    on the surface;
(2) For a comparison of model SED slopes with the observed IRS slope,
    it is more appropriate to use only the IRS measurement errors
    without adding an absolute calibration error. In this case, the low/high
    thermal inertia trends are statistically significant with many flux ratios at
    short and long wavelengths being outside the 3-sigma limit.
However, for a realistic comparison with all other data sets we also used absolute flux
errors for IRS data in Figs.~\ref{fig:obsmod_campins}, \ref{fig:obsmod1},
and \ref{fig:irs}.

\section{Conclusions}
\label{sec:conclusion}

Despite our extensive experience in reconstructing rotational and physical
properties of many small bodies and the large observational data set for Ryugu,
this case remains challenging. The visual lightcurves have very low amplitudes
and the data quality is not sufficient to find unique shape and spin properties
in a standard way by lightcurve inversion techniques.

We have collected all available data sets, published and unpublished,
and obtained a large data set of new visual and thermal measurements from ground and space,
including Herschel, Spitzer, and AKARI measurements, with multi-epoch, multi-phase angle
and wavelength coverage. We also re-reduced previously published data (Subaru, AKARI)
with improved methods.

We combined all data and analysed them using different methods and thermophysical
model codes with the goal being to determine the object's size, albedo, shape,
surface, thermal and spin properties. In addition to standard (visual) lightcurve
inversion techniques, we applied TPM radiometric techniques assuming spherical
and more complex shapes, and also a new radiometric-inversion technique using
all data simultaneously.
Our results are thus summarised:
 This C-class asteroid has a retrograde rotation with the most likely
 axis orientation of ($\lambda$, $\beta$)$_{ecl}$ = (340$^{\circ}$, -40$^{\circ}$),
 a rotation period of P$_{sid}$ = 7.63109\,h and a very low surface roughness
 (r.m.s.\ of surface slopes $<$ 0.1).
 The object's spin-axis orientation has an obliquity of 136$^{\circ}$ with
 respect to Ryugu's orbital plane normal (full possible range: 114$^{\circ}$ to 136$^{\circ}$).
 We find a thermal inertia of the top-surface layer of 150 -
 300\,J\,m$^{-2}$\,s$^{-0.5}$\,K$^{-1}$, and, based on estimated heat conductivities in
 the range 0.1 to 0.6\,W\,K$^{-1}$\,m$^{-1}$, 
 we find grain sizes of $\approx$1-10\,mm on the top-layer surface.
 We derived a radiometric size (of an equal-volume sphere) of approximately 850 to 880\,m
 (connected to the above rotational and thermal properties). Considering also
 the less-likely solutions from zones B and C in Table~\ref{tbl:jd_solutions} would widen
 the size range to approximately 810 to 905\,m.
 The convex shape model has approximate axis ratios of a/b = 1.025 and b/c = 1.014.
 Some of the solutions in Table~\ref{tbl:jd_solutions} have more elongated shapes
 with a/b rising to approximately 1.06, and b/c axis ratios ranging from 1.01
 to 1.07; one of the solutions (\#5 in Table~\ref{tbl:jd_solutions}) shows a more
 extreme b/c ratio of 1.21.
 Using an absolute magnitude in V-band of H$_V$ = 19.25\,$\pm$\,0.03\,mag
 we find a geometric V-band albedo of p$_V$ = 0.044-0.050. Less probably, solutions
 from Table~\ref{tbl:jd_solutions} would result in an albedo maximum of 0.06.
 A radiometric analysis using a simple spherical shape points to a very
 similar spin-vector solution lying somewhere in the range
 310$^{\circ}$ to 335$^{\circ}$ in ecliptic longitude and
 -65$^{\circ}$ to -30$^{\circ}$ in ecliptic latitude, connected
 to a low-roughness surface (r.m.s.\ of surface slopes below 0.2),
 thermal inertias of approximately 200 to 300\,J\,m$^{-2}$\,s$^{-0.5}$\,K$^{-1}$,
 but still with significant uncertainties in size (approximately 815-900\,m)
 and albedo (between 0.04 and 0.06), but in excellent agreement with
 our final solution.
 Automatic procedures using radiometric and lightcurve inversion techniques
 simultaneously led to a range of possible spin solutions (from the
 statistical point of view), grouped into five different zones
 (see Fig.~\ref{fig:final_chisq}).
 A more careful (manual) testing of all solutions within the five zones
 was required to find the most likely axis orientation in the context of
 our large, but complex thermal data set.

Our analysis shows that thermal data can help to reconstruct an object’s rotational properties,
in addition to its physical and thermal characteristics. But the Ryugu case also shows that:
(i) high-quality multi-aspect visual lightcurves are crucial for reconstructing
    shape and spin properties, and even more for cases with low-amplitude
    lightcurves;
(ii) thermal data can help to reconstruct the spin properties, also in cases of
    low-amplitude lightcurve objects;
(iii) there are many ways of combining visual (originating from the illuminated
    surface only) and thermal data (related to the warm surface areas) in a single
    inversion technique, and the outcome depends strongly
    on the quality of individual data sets and the weights given to each data set;
(iv) thermal data are very important for constraining size, albedo, and
    thermal inertia, but one must consider the degeneracy between thermal
    inertia and surface roughness: often low-roughness combined with low-inertia
    solutions fit the data equally well as high-roughness combined with high-inertia
    settings;
(v) thermal data also carry information about the object’s spin-axis orientation,
    but the interpretation is not straight forward: first of all, shape features can be misleading,
    secondly, short-wavelength fluxes (such as the short-wavelenth Spitzer-IRAC data) are
    originating from the hottest terrains on the surface and global shape and
    spin properties have weaker influences;
(vi) long-wavelength thermal data (beyond the thermal emission peak) are
    crucial for constraining size, albedo and thermal inertia, and they
    are important for reconstructing spin properties (if error bars are not too large);
(vii) single-epoch thermal observations, even in cases with a full thermal emission
    spectrum as obtained by Spitzer-IRS, can lead to incorrect thermal inertias.

\begin{figure}[t]
 \rotatebox{0}{\resizebox{\hsize}{!}{\includegraphics{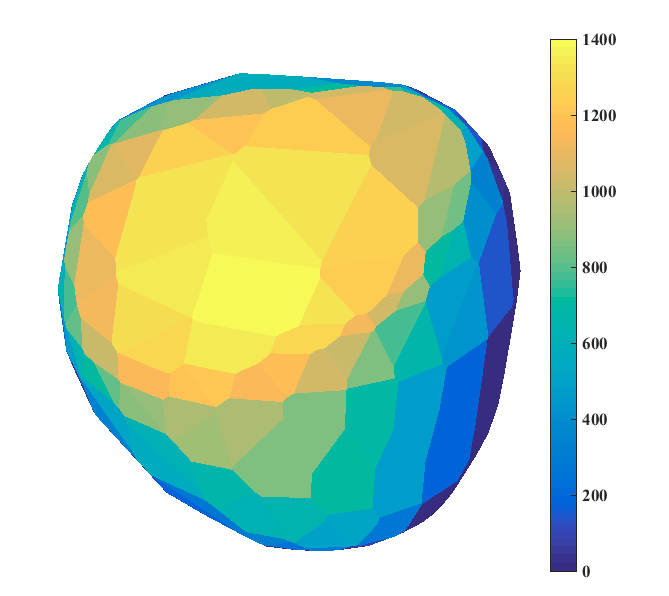}}}
 \rotatebox{0}{\resizebox{\hsize}{!}{\includegraphics{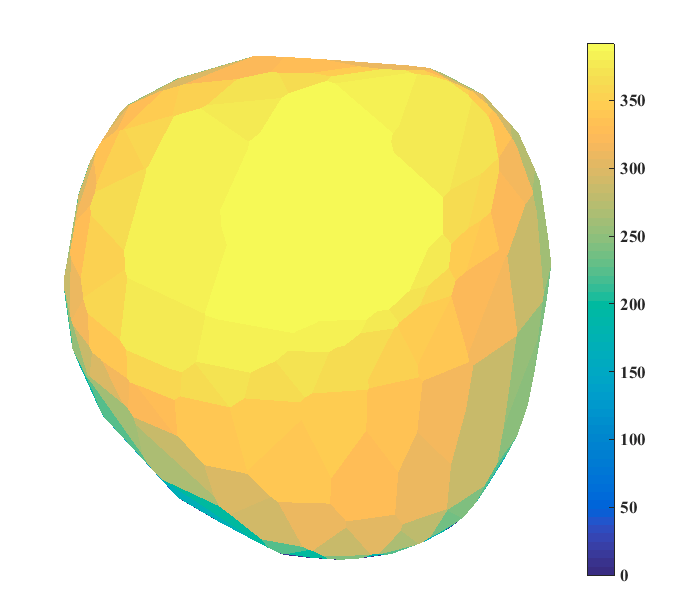}}}
  \caption{Ryugu as seen from Earth on 1st July, 2018, close to the arrival time
           of the Hayabusa-2 mission at the asteroid (r=0.988\,AU, $\Delta$=1.903\,AU,
           $\alpha$=18.6$^{\circ}$), calculated using our
           reference shape, spin (\#2 in Tbl.~\ref{tbl:jd_solutions}),
           size, thermal solution, with the z-axis pointing in the direction
           of the spin axis (along the insolation/temperature
           colour bars). Top: Solar insolation in W/m$^{2}$,
           with the sub-solar point located at the peak insolation.
           Bottom: The TPM-calculated temperatures (in Kelvin).
     \label{fig:haya2arrival}}
\end{figure}

Our favoured solution (\#2 in Tbl.~\ref{tbl:jd_solutions}) can now be used to make visual
and thermal predictions for future observations. This solution is also the
"design reference model" for the preparation and conduction of the Hayabusa2 mission.
Figure~\ref{fig:haya2arrival} shows the thermal picture of Ryugu close to the arrival
time of the Hayabusa-2 mission in July 2018.


\begin{acknowledgements}
  The work of CK has been supported by the PECS-98073 contract of the
  European Space Agency and the Hungarian Space Office, the K-104607 grant
  of the Hungarian Research Fund (OTKA), the Bolyai Research Fellowship
  of the Hungarian Academy of Sciences, and the NKFIH grant GINOP-2.3.2-15-2016-00003. 
  EV acknowledges the support of the German {\it DLR} project number 50~OR~1108.
  Part of the funding for GROND (both hardware as well as personnel) was generously
  granted by the Leibniz-Prize to Prof.\ G.\ Hasinger (DFG grant HA 1850/28-1).
  The work of J\v{D} was supported by the grant 15-04816S of the Czech Science Foundation.
  MJK and YJC were partially supported by the National Research Council of Fundamental
  Science \& Technology through a National Agenda Project "Development of
  Electro-optic Space Surveillance System" and by matching funds from the Korea
  Astronomy and Space Science Institute.
  The work of SH was supported by the Hypervelocity Impact Facility, ISAS, JAXA.
  The work of MI and HY was supported by research programs through the National Research
  Foundation of Korea (NRF) funded by the Korean government (MEST)
  (No.\ 2012R1A4A1028713 and 2015R1D1A1A01060025).
  PS acknowledges support through the Sofja Kovalevskaja Award from the Alexander von
  Humboldt Foundation of Germany.
  The work of MD was performed in the context of the NEOShield-2 project, which has received
  funding from the European Union's Horizon 2020 research and innovation programme under grant
  agreement No. 640351. MD was also supported by the project 11-BS56-008 (SHOCKS) of the
  French Agence National de la Recherche (ANR).\\
  We thank to M.\ Yoshikawa (ISAS) who offered his continuous advice and encouragement,
  and B.\ Gundlach for support in calculating grain sizes.
  We thank J.\ Elliott for the GROND test observations on May 27/28, 2012
  and S.\ Larson for providing additional lightcurve observations.\\
  The research leading to these results has received funding from the European
  Union's Horizon 2020 Research and Innovation Programme, under Grant Agreement no 687378.
\end{acknowledgements}

\clearpage

\begin{appendix}

\section{Spitzer-IRAC point-and-shoot observations of 162173 (1999~JU$_{3}$)}
\label{app:spitzer_ps}

 The Spitzer-IRAC point-and-shoot fluxes of Ryugu from Spitzer PID~90145 (PI: M.\ Mueller)
 are given in Table~\ref{tbl:spitzer_ps}. All fluxes given below are absolutely calibrated in-band IRAC fluxes,
 the solar reflection is not subtracted, and the thermal flux component not colour-corrected.
 The times given are AOR start time as measured onboard Spitzer (UTC), that is, not corrected
 for light travel time between asteroid and spacecraft.

  \begin{table*}[h!tb]
    \begin{center}
    \caption{Spitzer-IRAC point-and-shoot observation of 162173 Ryugu
             from January to May 2013.
             \label{tbl:spitzer_ps}}
    \begin{tabular}{ll|llrrr|llrrr}
      \hline
      \hline
      \noalign{\smallskip}
      &           & \multicolumn{3}{l}{Channel 1} & Flux & Flux$_{err}$ & \multicolumn{3}{l}{Channel 2} & Flux & Flux$_{err}$ \\
Label & DATE\_OBS & mag & mag$_{err}$ & S/N & [$\mu$Jy] & [$\mu$Jy] & mag & mag$_{err}$ & S/N & [$\mu$Jy] & [$\mu$Jy] \\
      \noalign{\smallskip}
      \hline
      \noalign{\smallskip}
1999\_JU3c    & 2013-01-20T02:05:04 & 13.32 & 0.03 & 38 & 1318.7 & 34.3 &            10.978 & 0.011 &  91 &  7300.4 &  80.3 \\
1999\_JU3d    & 2013-01-27T23:05:00 & 13.41 & 0.03 & 37 & 1218.3 & 32.9 &            11.074 & 0.011 &  91 &  6682.6 &  73.5 \\
1999\_JU3e    & 2013-01-31T01:13:30 & 13.34 & 0.03 & 38 & 1293.5 & 33.6 &            11.046 & 0.011 &  91 &  6857.2 &  75.4 \\
1999\_JU3f    & 2013-02-09T02:28:22 & 13.22 & 0.03 & 40 & 1442.0 & 36.0 &            10.948 & 0.011 &  91 &  7505.0 &  82.6 \\
      \noalign{\smallskip}
1999\_JU3-p2a & 2013-04-28T19:26:46 & 12.04 & 0.01 & 71 & 4287.0 & 60.0 & \hphantom{0}9.73  & 0.01  & 167 & 22979.9 & 137.9 \\
1999\_JU3-p2b & 2013-05-05T02:39:45 & 11.70 & 0.01 & 83 & 5847.3 & 70.2 & \hphantom{0}9.44  & 0.01  & 200 & 30209.8 & 151.0 \\
1999\_JU3-p2c & 2013-05-09T23:44:05 & 11.78 & 0.01 & 77 & 5446.9 & 70.8 & \hphantom{0}9.47  & 0.01  & 200 & 29332.4 & 146.7 \\
1999\_JU3-p2d & 2013-05-15T13:44:49 & 11.51 & 0.01 & 91 & 6978.3 & 76.8 & \hphantom{0}9.22  & 0.01  & 200 & 36961.4 & 184.8 \\
1999\_JU3-p2e & 2013-05-23T21:18:11 & 11.43 & 0.01 & 91 & 7553.6 & 83.1 & \hphantom{0}9.22  & 0.01  & 200 & 36723.8 & 183.6 \\
1999\_JU3-p2f & 2013-05-29T09:47:38 & 11.57 & 0.01 & 91 & 6627.5 & 72.9 & \hphantom{0}9.39  & 0.01  & 200 & 31488.2 & 157.4 \\
      \noalign{\smallskip}
     \hline
    \end{tabular}
    \end{center}
  \end{table*}


\section{Spitzer-IRAC lightcurve observations of 162173 Ryugu}
\label{app:spitzer_lc}
 
 The Spitzer-IRAC lightcurve fluxes of Ryugu from Spitzer PID~90145 (PI: M.\ Mueller)
 are given in Tables~\ref{tbl:spitzer_lc1_ch1}, \ref{tbl:spitzer_lc1_ch2}, \ref{tbl:spitzer_lc2_ch1},
 and \ref{tbl:spitzer_lc2_ch2}. All fluxes given below are absolutely calibrated in-band IRAC fluxes,
 the solar reflection is not subtracted and the thermal flux component is not colour corrected.
 The times given are AOR start time as measured onboard Spitzer (UTC), that is, not corrected
 for light travel time between asteroid and spacecraft.

  \begin{table*}[h!tb]
    \begin{center}
    \caption{Spitzer-IRAC lightcurve observation of 162173 Ryugu
             from 10/11th February, 2013, channel 1. The estimated flux
             uncertainty is 18\,$\mu$Jy.
             \label{tbl:spitzer_lc1_ch1}}
    \begin{tabular}{llrc|cllr}
      \hline
      \hline
      \noalign{\smallskip}
     DATE\_OBS & MJD\_OBS & Flux [$\mu$Jy] & &  &         DATE\_OBS & MJD\_OBS & Flux [$\mu$Jy] \\ 
      \noalign{\smallskip}
      \hline            
      \noalign{\smallskip}
      2013-02-10T20:11:48.655 & 56333.84154 & 1394.5  &   &  &   2013-02-10T23:57:08.084 & 56333.99801 & 1495.0 \\ 
      2013-02-10T20:18:44.643 & 56333.84635 & 1374.6  &   &  &   2013-02-11T00:04:06.072 & 56334.00285 & 1503.1 \\ 
      2013-02-10T20:25:40.639 & 56333.85116 & 1392.2  &   &  &   2013-02-11T00:11:02.466 & 56334.00767 & 1498.2 \\ 
      2013-02-10T20:32:39.041 & 56333.85601 & 1411.2  &   &  &   2013-02-11T00:17:59.266 & 56334.01249 & 1539.9 \\ 
      2013-02-10T20:41:34.231 & 56333.86220 & 1399.5  &   &  &   2013-02-11T00:24:18.067 & 56334.01688 & 1506.3 \\ 
      2013-02-10T20:48:51.430 & 56333.86726 & 1385.0  &   &  &   2013-02-11T00:31:14.461 & 56334.02170 & 1487.6 \\ 
      2013-02-10T20:55:47.019 & 56333.87207 & 1407.6  &   &  &   2013-02-11T00:38:11.651 & 56334.02652 & 1492.7 \\ 
      2013-02-10T21:02:45.816 & 56333.87692 & 1430.3  &   &  &   2013-02-11T00:45:08.452 & 56334.03135 & 1526.2 \\ 
      2013-02-10T21:09:42.214 & 56333.88174 & 1418.7  &   &  &   2013-02-11T00:52:04.029 & 56334.03616 & 1449.2 \\ 
      2013-02-10T21:16:39.002 & 56333.88656 & 1438.5  &   &  &   2013-02-11T01:15:35.614 & 56334.05250 & 1398.2 \\ 
      2013-02-10T21:22:55.396 & 56333.89092 & 1478.4  &   &  &   2013-02-11T01:22:57.215 & 56334.05761 & 1337.4 \\ 
      2013-02-10T21:29:52.997 & 56333.89575 & 1437.0  &   &  &   2013-02-11T01:30:17.605 & 56334.06270 & 1371.7 \\ 
      2013-02-10T21:36:50.180 & 56333.90058 & 1433.1  &   &  &   2013-02-11T01:37:39.991 & 56334.06782 & 1306.8 \\ 
      2013-02-10T21:43:46.988 & 56333.90541 & 1396.0  &   &  &   2013-02-11T01:45:01.596 & 56334.07294 & 1309.3 \\ 
      2013-02-10T21:50:42.570 & 56333.91021 & 1359.5  &   &  &   2013-02-11T01:52:46.181 & 56334.07831 & 1360.8 \\ 
      2013-02-10T21:57:59.769 & 56333.91528 & 1345.3  &   &  &   2013-02-11T02:00:08.981 & 56334.08344 & 1335.0 \\ 
      2013-02-10T22:04:59.370 & 56333.92013 & 1370.6  &   &  &   2013-02-11T02:07:29.778 & 56334.08854 & 1364.3 \\ 
      2013-02-10T22:12:54.162 & 56333.92563 & 1355.6  &   &  &   2013-02-11T02:15:56.176 & 56334.09440 & 1390.8 \\ 
      2013-02-10T22:19:50.962 & 56333.93045 & 1359.5  &   &  &   2013-02-11T02:24:20.573 & 56334.10024 & 1409.3 \\ 
      2013-02-10T22:26:46.548 & 56333.93526 & 1348.6  &   &  &   2013-02-11T02:31:01.171 & 56334.10487 & 1431.5 \\ 
      2013-02-10T22:33:06.946 & 56333.93966 & 1349.2  &   &  &   2013-02-11T02:39:27.158 & 56334.11073 & 1441.4 \\ 
      2013-02-10T22:40:02.934 & 56333.94448 & 1315.2  &   &  &   2013-02-11T02:46:49.146 & 56334.11585 & 1468.5 \\ 
      2013-02-10T22:46:59.730 & 56333.94930 & 1348.4  &   &  &   2013-02-11T02:54:11.950 & 56334.12097 & 1453.8 \\ 
      2013-02-10T22:53:54.925 & 56333.95411 & 1336.6  &   &  &   2013-02-11T03:01:31.938 & 56334.12606 & 1420.7 \\ 
      2013-02-10T23:01:52.116 & 56333.95963 & 1372.0  &   &  &   2013-02-11T03:09:16.527 & 56334.13144 & 1420.5 \\ 
      2013-02-10T23:08:11.318 & 56333.96402 & 1365.3  &   &  &   2013-02-11T03:17:43.323 & 56334.13731 & 1396.7 \\ 
      2013-02-10T23:15:08.111 & 56333.96884 & 1375.9  &   &  &   2013-02-11T03:25:04.120 & 56334.14241 & 1436.7 \\        
      2013-02-10T23:22:04.907 & 56333.97367 & 1404.1  &   &  &   2013-02-11T03:32:26.920 & 56334.14753 & 1407.6 \\        
      2013-02-10T23:28:59.301 & 56333.97846 & 1391.8  &   &  &   2013-02-11T03:39:48.509 & 56334.15264 & 1415.8 \\        
      2013-02-10T23:35:59.297 & 56333.98333 & 1402.8  &   &  &   2013-02-11T03:46:29.899 & 56334.15729 & 1427.5 \\        
      2013-02-10T23:42:55.284 & 56333.98814 & 1465.4  &   &  &   2013-02-11T03:53:53.899 & 56334.16243 & 1397.2 \\        
      2013-02-10T23:50:12.885 & 56333.99320 & 1429.0  &   &  &   2013-02-11T04:01:14.683 & 56334.16753 & 1408.5 \\ 
      \noalign{\smallskip}
     \hline
    \end{tabular}
    \end{center}
  \end{table*}

  \begin{table*}[h!tb]
    \begin{center}
    \caption{Spitzer-IRAC lightcurve observation of 162173 Ryugu
             from 10/11th February, 2013, channel 2. The estimated flux
             uncertainty is 41\,$\mu$Jy.
             \label{tbl:spitzer_lc1_ch2}}
    \begin{tabular}{llrc|cllr}
      \hline
      \hline
      \noalign{\smallskip}
     DATE\_OBS & MJD\_OBS & Flux [$\mu$Jy] & &  &         DATE\_OBS & MJD\_OBS & Flux [$\mu$Jy] \\ 
      \noalign{\smallskip}
      \hline            
      \noalign{\smallskip}
      2013-02-10T20:14:16.651 & 56333.84325 &  7498.1 & & & 2013-02-11T00:03:36.873 & 56334.00251 &  7938.7 \\
      2013-02-10T20:22:13.041 & 56333.84876 &  7408.8 & & & 2013-02-11T00:11:33.271 & 56334.00802 &  8012.1 \\
      2013-02-10T20:30:09.439 & 56333.85428 &  7491.1 & & & 2013-02-11T00:21:48.067 & 56334.01514 &  7982.7 \\
      2013-02-10T20:39:04.630 & 56333.86047 &  7436.1 & & & 2013-02-11T00:30:44.855 & 56334.02135 &  7990.0 \\
      2013-02-10T20:48:21.825 & 56333.86692 &  7491.1 & & & 2013-02-11T00:40:39.655 & 56334.02824 &  8019.5 \\
      2013-02-10T20:56:17.426 & 56333.87242 &  7567.4 & & & 2013-02-11T00:50:34.842 & 56334.03513 &  7975.3 \\
      2013-02-10T21:07:13.015 & 56333.88001 &  7715.2 & & & 2013-02-11T01:16:07.016 & 56334.05286 &  7456.7 \\
      2013-02-10T21:15:08.600 & 56333.88552 &  7651.5 & & & 2013-02-11T01:24:31.007 & 56334.05869 &  7273.6 \\
      2013-02-10T21:24:25.798 & 56333.89197 &  7779.4 & & & 2013-02-11T01:33:58.206 & 56334.06526 &  7180.4 \\
      2013-02-10T21:34:21.790 & 56333.89886 &  7665.6 & & & 2013-02-11T01:42:24.990 & 56334.07112 &  7147.4 \\
      2013-02-10T21:41:17.387 & 56333.90367 &  7525.7 & & & 2013-02-11T01:50:09.990 & 56334.07650 &  7334.1 \\
      2013-02-10T21:49:14.976 & 56333.90920 &  7408.8 & & & 2013-02-11T01:57:30.782 & 56334.08161 &  7206.9 \\
      2013-02-10T21:58:29.765 & 56333.91562 &  7340.9 & & & 2013-02-11T02:09:05.176 & 56334.08964 &  7320.6 \\
      2013-02-10T22:06:28.159 & 56333.92116 &  7200.3 & & & 2013-02-11T02:16:27.164 & 56334.09476 &  7402.0 \\
      2013-02-10T22:16:21.760 & 56333.92803 &  7193.6 & & & 2013-02-11T02:25:55.558 & 56334.10134 &  7602.4 \\
      2013-02-10T22:23:19.747 & 56333.93287 &  7088.4 & & & 2013-02-11T02:33:40.553 & 56334.10672 &  7679.8 \\
      2013-02-10T22:33:35.743 & 56333.94000 &  7167.2 & & & 2013-02-11T02:44:12.552 & 56334.11403 &  7616.4 \\
      2013-02-10T22:41:32.547 & 56333.94552 &  7094.9 & & & 2013-02-11T02:51:34.149 & 56334.11915 &  7750.8 \\
      2013-02-10T22:49:27.331 & 56333.95101 &  7134.2 & & & 2013-02-11T03:01:00.938 & 56334.12571 &  7736.6 \\
      2013-02-10T22:57:24.132 & 56333.95653 &  7167.2 & & & 2013-02-11T03:09:50.734 & 56334.13184 &  7616.4 \\
      2013-02-10T23:05:21.318 & 56333.96205 &  7134.2 & & & 2013-02-11T03:19:17.526 & 56334.13840 &  7553.5 \\
      2013-02-10T23:13:39.720 & 56333.96782 &  7240.2 & & & 2013-02-11T03:27:41.514 & 56334.14423 &  7456.7 \\
      2013-02-10T23:22:34.110 & 56333.97401 &  7402.0 & & & 2013-02-11T03:35:05.119 & 56334.14936 &  7463.6 \\
      2013-02-10T23:29:29.692 & 56333.97882 &  7539.6 & & & 2013-02-11T03:44:56.903 & 56334.15621 &  7532.7 \\
      2013-02-10T23:37:28.093 & 56333.98435 &  7665.6 & & & 2013-02-11T03:54:24.903 & 56334.16279 &  7546.5 \\
      2013-02-10T23:45:44.081 & 56333.99009 &  7679.8 & & & 2013-02-11T04:02:48.488 & 56334.16862 &  7487.7 \\
      2013-02-10T23:53:41.291 & 56333.99562 &  7808.1 & & &                         &             &         \\
      \noalign{\smallskip}    
     \hline                   
    \end{tabular}             
    \end{center}
  \end{table*}

  \begin{table*}[h!tb]
{\tiny
    \begin{center}
    \caption{Spitzer-IRAC lightcurve observation of 162173 Ryugu
             from 2nd May, 2013, channel 1. The estimated flux
             uncertainty is 49\,$\mu$Jy.
             \label{tbl:spitzer_lc2_ch1}}
    \begin{tabular}{llrc|cllr}
      \hline
      \hline
      \noalign{\smallskip}
     DATE\_OBS & MJD\_OBS & Flux [$\mu$Jy] & &  &         DATE\_OBS & MJD\_OBS & Flux [$\mu$Jy] \\ 
      \noalign{\smallskip}
      \hline            
      \noalign{\smallskip}
      2013-05-02T11:48:43.708 & 56414.49217 & 5148.0  & & &   2013-05-02T15:50:18.343 & 56414.65993 & 4534.4 \\
      2013-05-02T11:52:43.704 & 56414.49495 & 4992.9  & & &   2013-05-02T15:54:17.542 & 56414.66270 & 4407.4 \\
      2013-05-02T11:55:56.512 & 56414.49718 & 5049.7  & & &   2013-05-02T15:57:31.131 & 56414.66494 & 4549.1 \\
      2013-05-02T11:59:56.508 & 56414.49996 & 5009.1  & & &   2013-05-02T16:01:32.330 & 56414.66774 & 4511.6 \\
      2013-05-02T12:03:10.504 & 56414.50220 & 5015.3  & & &   2013-05-02T16:04:44.729 & 56414.66996 & 4539.4 \\
      2013-05-02T12:07:12.504 & 56414.50501 & 5134.1  & & &   2013-05-02T16:08:45.525 & 56414.67275 & 4428.4 \\
      2013-05-02T12:10:23.699 & 56414.50722 & 4985.1  & & &   2013-05-02T16:11:19.119 & 56414.67453 & 4688.8 \\
      2013-05-02T12:14:24.495 & 56414.51001 & 5104.7  & & &   2013-05-02T16:15:19.126 & 56414.67730 & 4566.4 \\
      2013-05-02T12:18:32.089 & 56414.51287 & 5069.8  & & &   2013-05-02T16:18:32.317 & 56414.67954 & 4565.8 \\
      2013-05-02T12:22:39.690 & 56414.51574 & 4989.2  & & &   2013-05-02T16:22:31.516 & 56414.68231 & 4604.6 \\
      2013-05-02T12:25:13.682 & 56414.51752 & 5203.0  & & &   2013-05-02T16:25:45.106 & 56414.68455 & 4681.8 \\
      2013-05-02T12:29:14.084 & 56414.52030 & 5088.8  & & &   2013-05-02T16:29:45.512 & 56414.68733 & 4706.6 \\
      2013-05-02T12:32:26.084 & 56414.52252 & 4967.6  & & &   2013-05-02T16:32:57.109 & 56414.68955 & 4520.7 \\
      2013-05-02T12:36:27.674 & 56414.52532 & 5027.8  & & &   2013-05-02T16:36:59.113 & 56414.69235 & 4604.1 \\
      2013-05-02T12:39:41.275 & 56414.52756 & 5116.3  & & &   2013-05-02T16:40:11.503 & 56414.69458 & 4767.7 \\
      2013-05-02T12:43:40.075 & 56414.53032 & 5090.1  & & &   2013-05-02T16:44:12.296 & 56414.69736 & 4529.5 \\
      2013-05-02T12:46:52.466 & 56414.53255 & 5086.9  & & &   2013-05-02T16:47:24.300 & 56414.69959 & 4640.6 \\
      2013-05-02T12:50:53.669 & 56414.53534 & 5256.6  & & &   2013-05-02T16:50:51.889 & 56414.70199 & 4439.5 \\
      2013-05-02T12:54:07.661 & 56414.53759 & 5188.3  & & &   2013-05-02T16:54:52.698 & 56414.70478 & 4722.7 \\
      2013-05-02T12:58:06.070 & 56414.54035 & 5083.9  & & &   2013-05-02T17:05:21.681 & 56414.71206 & 4644.1 \\
      2013-05-02T13:01:20.058 & 56414.54259 & 5396.6  & & &   2013-05-02T17:09:20.880 & 56414.71483 & 4652.0 \\
      2013-05-02T13:05:42.453 & 56414.54563 & 5236.3  & & &   2013-05-02T17:12:34.079 & 56414.71706 & 4628.0 \\
      2013-05-02T13:08:56.445 & 56414.54788 & 5375.9  & & &   2013-05-02T17:16:33.274 & 56414.71983 & 4679.8 \\
      2013-05-02T13:12:56.452 & 56414.55065 & 5504.7  & & &   2013-05-02T17:19:47.668 & 56414.72208 & 4755.1 \\
      2013-05-02T13:16:08.854 & 56414.55288 & 5258.0  & & &   2013-05-02T17:23:48.074 & 56414.72486 & 4759.9 \\
      2013-05-02T13:20:09.655 & 56414.55567 & 5404.2  & & &   2013-05-02T17:26:59.664 & 56414.72708 & 4797.6 \\
      2013-05-02T13:23:22.838 & 56414.55790 & 5549.7  & & &   2013-05-02T17:31:00.066 & 56414.72986 & 4748.1 \\
      2013-05-02T13:27:22.451 & 56414.56068 & 5472.2  & & &   2013-05-02T17:34:14.066 & 56414.73211 & 4822.0 \\
      2013-05-02T13:30:35.642 & 56414.56291 & 5343.2  & & &   2013-05-02T17:38:15.257 & 56414.73490 & 4913.4 \\
      2013-05-02T13:34:34.833 & 56414.56568 & 5507.5  & & &   2013-05-02T17:41:27.655 & 56414.73713 & 4914.3 \\
      2013-05-02T13:38:44.439 & 56414.56857 & 5600.2  & & &   2013-05-02T17:44:54.448 & 56414.73952 & 4885.0 \\
      2013-05-02T13:41:18.032 & 56414.57035 & 5801.8  & & &   2013-05-02T17:48:55.256 & 56414.74231 & 5008.9 \\
      2013-05-02T13:45:17.637 & 56414.57312 & 5548.8  & & &   2013-05-02T17:52:08.451 & 56414.74454 & 4954.5 \\
      2013-05-02T13:48:30.821 & 56414.57536 & 5631.8  & & &   2013-05-02T17:56:09.248 & 56414.74733 & 5093.9 \\
      2013-05-02T13:52:30.031 & 56414.57813 & 5506.6  & & &   2013-05-02T17:59:21.236 & 56414.74955 & 4922.1 \\
      2013-05-02T13:55:44.015 & 56414.58037 & 5735.9  & & &   2013-05-02T18:03:21.646 & 56414.75233 & 4931.0 \\
      2013-05-02T13:59:44.812 & 56414.58316 & 5716.3  & & &   2013-05-02T18:06:34.837 & 56414.75457 & 5091.7 \\
      2013-05-02T14:02:56.015 & 56414.58537 & 5568.5  & & &   2013-05-02T18:10:36.044 & 56414.75736 & 4952.8 \\
      2013-05-02T14:06:57.616 & 56414.58817 & 5709.6  & & &   2013-05-02T18:13:47.231 & 56414.75957 & 5125.6 \\
      2013-05-02T14:10:10.815 & 56414.59040 & 5749.7  & & &   2013-05-02T18:17:48.828 & 56414.76237 & 5144.2 \\
      2013-05-02T14:14:11.202 & 56414.59319 & 5591.1  & & &   2013-05-02T18:20:22.035 & 56414.76414 & 5212.9 \\
      2013-05-02T14:17:23.209 & 56414.59541 & 5585.7  & & &   2013-05-02T18:24:21.632 & 56414.76692 & 5180.0 \\
      2013-05-02T14:20:43.607 & 56414.59773 & 5453.1  & & &   2013-05-02T18:27:34.425 & 56414.76915 & 5135.8 \\
      2013-05-02T14:23:57.595 & 56414.59997 & 5631.5  & & &   2013-05-02T18:31:33.624 & 56414.77192 & 5226.4 \\
      2013-05-02T14:27:57.201 & 56414.60275 & 5452.7  & & &   2013-05-02T18:34:48.819 & 56414.77418 & 5102.8 \\
      2013-05-02T14:31:09.204 & 56414.60497 & 5388.4  & & &   2013-05-02T18:38:49.217 & 56414.77696 & 5174.1 \\
      2013-05-02T14:35:10.395 & 56414.60776 & 5438.7  & & &   2013-05-02T18:42:00.416 & 56414.77917 & 5284.0 \\
      2013-05-02T14:39:17.192 & 56414.61062 & 5309.8  & & &   2013-05-02T18:46:01.209 & 56414.78196 & 5114.8 \\
      2013-05-02T14:42:31.188 & 56414.61286 & 5339.8  & & &   2013-05-02T18:49:14.408 & 56414.78419 & 5189.6 \\
      2013-05-02T14:46:31.980 & 56414.61565 & 5107.4  & & &   2013-05-02T18:53:15.212 & 56414.78698 & 5187.1 \\
      2013-05-02T14:49:43.175 & 56414.61786 & 5222.6  & & &   2013-05-02T18:56:27.200 & 56414.78920 & 5204.9 \\
      2013-05-02T14:53:44.382 & 56414.62065 & 5161.8  & & &   2013-05-02T18:59:54.806 & 56414.79161 & 5053.2 \\
      2013-05-02T14:57:11.577 & 56414.62305 & 5175.6  & & &   2013-05-02T19:03:56.403 & 56414.79440 & 5167.3 \\
      2013-05-02T15:00:25.171 & 56414.62529 & 5047.9  & & &   2013-05-02T19:07:09.192 & 56414.79663 & 5100.9 \\
      2013-05-02T15:04:26.374 & 56414.62808 & 4862.1  & & &   2013-05-02T19:11:10.000 & 56414.79942 & 4921.1 \\
      2013-05-02T15:07:38.366 & 56414.63031 & 5031.6  & & &   2013-05-02T19:14:22.387 & 56414.80165 & 5025.2 \\
      2013-05-02T15:11:39.561 & 56414.63310 & 4956.4  & & &   2013-05-02T19:18:21.191 & 56414.80441 & 5198.8 \\
      2013-05-02T15:14:52.760 & 56414.63533 & 4793.7  & & &   2013-05-02T19:21:35.175 & 56414.80666 & 5002.6 \\
      2013-05-02T15:18:52.755 & 56414.63811 & 4865.5  & & &   2013-05-02T19:25:35.983 & 56414.80944 & 5065.2 \\
      2013-05-02T15:22:05.161 & 56414.64034 & 4835.9  & & &   2013-05-02T19:28:47.175 & 56414.81166 & 5073.1 \\
      2013-05-02T15:26:05.157 & 56414.64312 & 4637.7  & & &   2013-05-02T19:32:49.577 & 56414.81446 & 5106.5 \\
      2013-05-02T15:29:18.755 & 56414.64536 & 4713.8  & & &   2013-05-02T19:35:22.780 & 56414.81624 & 5101.3 \\
      2013-05-02T15:32:38.754 & 56414.64767 & 4713.3  & & &   2013-05-02T19:39:22.771 & 56414.81901 & 5126.8 \\
      2013-05-02T15:35:51.946 & 56414.64991 & 4574.8  & & &   2013-05-02T19:42:35.572 & 56414.82124 & 5051.4 \\
      2013-05-02T15:39:52.746 & 56414.65269 & 4486.3  & & &   2013-05-02T19:46:41.962 & 56414.82410 & 5201.0 \\
      2013-05-02T15:43:05.543 & 56414.65493 & 4670.2  & & &   2013-05-02T19:50:43.556 & 56414.82689 & 5070.2 \\
      2013-05-02T15:47:06.343 & 56414.65771 & 4565.6  & & &                           &             &        \\
      \noalign{\smallskip}
     \hline
    \end{tabular}
    \end{center}
}
  \end{table*}
                                                      
  \begin{table*}[h!tb]
{\tiny
    \begin{center}
    \caption{Spitzer-IRAC lightcurve observation of 162173 Ryugu
             from 2nd May, 2013, channel 2. The estimated flux
             uncertainty is 115\,$\mu$Jy.
             \label{tbl:spitzer_lc2_ch2}}
    \begin{tabular}{llrc|cllr}
      \hline
      \hline
      \noalign{\smallskip}
     DATE\_OBS & MJD\_OBS & Flux [$\mu$Jy] & &  &         DATE\_OBS & MJD\_OBS & Flux [$\mu$Jy] \\ 
      \noalign{\smallskip}
      \hline            
      \noalign{\smallskip}
      2013-05-02T11:48:13.708 & 56414.49183 & 26751.5 & & &   2013-05-02T15:44:37.945 & 56414.65599 & 24874.1 \\
      2013-05-02T11:53:15.708 & 56414.49532 & 27073.7 & & &   2013-05-02T15:48:39.144 & 56414.65879 & 24668.8 \\
      2013-05-02T11:56:28.509 & 56414.49755 & 26875.0 & & &   2013-05-02T15:52:45.933 & 56414.66164 & 24714.3 \\
      2013-05-02T12:02:32.102 & 56414.50176 & 27248.8 & & &   2013-05-02T15:55:59.530 & 56414.66388 & 24578.1 \\
      2013-05-02T12:05:44.496 & 56414.50399 & 27198.7 & & &   2013-05-02T15:59:58.330 & 56414.66665 & 24691.5 \\
      2013-05-02T12:09:45.691 & 56414.50678 & 26999.0 & & &   2013-05-02T16:04:13.119 & 56414.66960 & 24555.5 \\
      2013-05-02T12:13:53.289 & 56414.50964 & 27048.8 & & &   2013-05-02T16:08:22.721 & 56414.67249 & 24646.1 \\
      2013-05-02T12:18:09.288 & 56414.51261 & 26949.3 & & &   2013-05-02T16:12:51.513 & 56414.67560 & 24851.2 \\
      2013-05-02T12:22:16.089 & 56414.51546 & 27123.7 & & &   2013-05-02T16:16:51.919 & 56414.67838 & 24897.1 \\
      2013-05-02T12:26:38.483 & 56414.51850 & 27023.9 & & &   2013-05-02T16:20:05.517 & 56414.68062 & 24714.3 \\
      2013-05-02T12:29:51.682 & 56414.52074 & 27173.7 & & &   2013-05-02T16:24:05.520 & 56414.68340 & 24646.1 \\
      2013-05-02T12:33:52.088 & 56414.52352 & 27148.6 & & &   2013-05-02T16:28:19.110 & 56414.68633 & 24851.2 \\
      2013-05-02T12:37:05.271 & 56414.52576 & 27198.7 & & &   2013-05-02T16:32:26.715 & 56414.68920 & 25058.1 \\
      2013-05-02T12:43:08.872 & 56414.52996 & 26899.7 & & &   2013-05-02T16:38:32.707 & 56414.69343 & 24966.0 \\
      2013-05-02T12:46:22.071 & 56414.53220 & 27349.4 & & &   2013-05-02T16:41:43.902 & 56414.69565 & 25012.0 \\
      2013-05-02T12:51:24.067 & 56414.53570 & 27475.7 & & &   2013-05-02T16:45:45.097 & 56414.69844 & 24759.8 \\
      2013-05-02T12:55:39.657 & 56414.53865 & 27935.0 & & &   2013-05-02T16:49:19.089 & 56414.70092 & 24874.1 \\
      2013-05-02T12:59:39.656 & 56414.54143 & 28245.4 & & &   2013-05-02T16:53:20.694 & 56414.70371 & 24851.2 \\
      2013-05-02T13:03:16.054 & 56414.54394 & 28454.3 & & &   2013-05-02T17:04:51.283 & 56414.71170 & 24782.7 \\
      2013-05-02T13:08:26.050 & 56414.54752 & 28454.3 & & &   2013-05-02T17:08:50.079 & 56414.71447 & 25150.6 \\
      2013-05-02T13:12:25.644 & 56414.55030 & 28612.0 & & &   2013-05-02T17:12:02.876 & 56414.71670 & 25058.1 \\
      2013-05-02T13:15:38.050 & 56414.55252 & 28744.1 & & &   2013-05-02T17:17:04.473 & 56414.72019 & 25547.5 \\
      2013-05-02T13:19:38.444 & 56414.55531 & 28850.2 & & &   2013-05-02T17:22:14.078 & 56414.72377 & 25712.7 \\
      2013-05-02T13:23:52.842 & 56414.55825 & 28876.7 & & &   2013-05-02T17:27:32.074 & 56414.72745 & 25760.1 \\
      2013-05-02T13:27:54.439 & 56414.56105 & 29224.6 & & &   2013-05-02T17:33:43.663 & 56414.73176 & 26094.5 \\
      2013-05-02T13:31:07.236 & 56414.56328 & 28770.5 & & &   2013-05-02T17:38:45.663 & 56414.73525 & 26263.2 \\
      2013-05-02T13:35:13.634 & 56414.56613 & 29522.2 & & &   2013-05-02T17:43:22.456 & 56414.73845 & 26287.4 \\
      2013-05-02T13:39:22.024 & 56414.56900 & 29224.6 & & &   2013-05-02T17:47:23.248 & 56414.74124 & 26408.8 \\
      2013-05-02T13:43:46.028 & 56414.57206 & 29767.9 & & &   2013-05-02T17:50:36.045 & 56414.74347 & 26506.2 \\
      2013-05-02T13:46:57.629 & 56414.57428 & 29795.3 & & &   2013-05-02T17:56:39.642 & 56414.74768 & 26924.5 \\
      2013-05-02T13:50:58.418 & 56414.57707 & 29877.8 & & &   2013-05-02T18:00:54.439 & 56414.75063 & 27123.7 \\
      2013-05-02T13:54:11.226 & 56414.57930 & 29713.1 & & &   2013-05-02T18:06:59.235 & 56414.75485 & 27123.7 \\
      2013-05-02T13:59:20.816 & 56414.58288 & 29932.8 & & &   2013-05-02T18:10:12.036 & 56414.75708 & 27173.7 \\
      2013-05-02T14:03:29.214 & 56414.58575 & 29603.8 & & &   2013-05-02T18:15:13.637 & 56414.76057 & 27551.7 \\
      2013-05-02T14:09:40.007 & 56414.59005 & 29740.5 & & &   2013-05-02T18:19:44.027 & 56414.76370 & 27602.5 \\
      2013-05-02T14:13:40.014 & 56414.59282 & 29822.8 & & &   2013-05-02T18:23:50.828 & 56414.76656 & 27425.1 \\
      2013-05-02T14:16:51.998 & 56414.59505 & 29631.1 & & &   2013-05-02T18:29:07.620 & 56414.77023 & 27577.1 \\
      2013-05-02T14:22:16.803 & 56414.59881 & 29413.6 & & &   2013-05-02T18:34:10.417 & 56414.77373 & 27755.4 \\
      2013-05-02T14:25:29.591 & 56414.60104 & 29224.6 & & &   2013-05-02T18:38:18.018 & 56414.77660 & 27501.0 \\
      2013-05-02T14:29:29.197 & 56414.60381 & 29143.9 & & &   2013-05-02T18:42:33.217 & 56414.77955 & 27678.8 \\
      2013-05-02T14:34:46.001 & 56414.60748 & 28850.2 & & &   2013-05-02T18:46:32.010 & 56414.78231 & 27501.0 \\
      2013-05-02T14:39:48.391 & 56414.61098 & 28428.1 & & &   2013-05-02T18:50:46.400 & 56414.78526 & 27450.4 \\
      2013-05-02T14:43:55.985 & 56414.61384 & 28480.5 & & &   2013-05-02T18:54:55.599 & 56414.78814 & 27223.8 \\
      2013-05-02T14:47:09.984 & 56414.61609 & 27960.7 & & &   2013-05-02T19:00:19.997 & 56414.79190 & 27577.1 \\
      2013-05-02T14:51:17.179 & 56414.61895 & 27935.0 & & &   2013-05-02T19:03:32.399 & 56414.79413 & 27148.6 \\
      2013-05-02T14:55:40.382 & 56414.62200 & 27678.8 & & &   2013-05-02T19:08:33.996 & 56414.79762 & 27374.6 \\
      2013-05-02T14:58:53.573 & 56414.62423 & 27349.4 & & &   2013-05-02T19:13:51.187 & 56414.80129 & 27023.9 \\
      2013-05-02T15:03:54.772 & 56414.62772 & 27023.9 & & &   2013-05-02T19:17:50.789 & 56414.80406 & 27274.0 \\
      2013-05-02T15:08:02.772 & 56414.63059 & 27073.7 & & &   2013-05-02T19:21:04.378 & 56414.80630 & 26974.2 \\
      2013-05-02T15:11:16.760 & 56414.63283 & 26628.6 & & &   2013-05-02T19:27:08.374 & 56414.81051 & 26974.2 \\
      2013-05-02T15:16:16.767 & 56414.63631 & 26579.6 & & &   2013-05-02T19:30:21.178 & 56414.81275 & 27173.7 \\
      2013-05-02T15:20:33.162 & 56414.63927 & 25783.9 & & &   2013-05-02T19:34:52.373 & 56414.81588 & 26875.0 \\
      2013-05-02T15:26:35.954 & 56414.64347 & 25831.4 & & &   2013-05-02T19:38:51.978 & 56414.81866 & 27475.7 \\
      2013-05-02T15:30:43.950 & 56414.64634 & 25453.5 & & &   2013-05-02T19:42:04.369 & 56414.82088 & 27248.8 \\
      2013-05-02T15:35:21.153 & 56414.64955 & 25220.2 & & &   2013-05-02T19:47:05.962 & 56414.82437 & 27526.3 \\
      2013-05-02T15:41:26.347 & 56414.65378 & 25127.4 & & &   2013-05-02T19:50:20.353 & 56414.82662 & 27602.5 \\
      \noalign{\smallskip}    
     \hline                   
    \end{tabular}             
    \end{center}
}             
  \end{table*}

\section{Ground-based observations of 162173 Ryugu with GROND}
\label{app:grond}             
                              
The GROND data are explained in Section~\ref{sec:grond}. The reduced and calibrated
data are presented in Tables~\ref{tbl:grond_day1_r} and \ref{tbl:grond_day2_gri}.

\begin{table*}[h!tb]
{\tiny
    \begin{center}
\caption[]{GROND r$^{\prime}$ magnitudes and errors of 162173 Ryugu.
           The zero time in the table corresponds to MJD 56087.04942 (2012-Jun-09
           01:11:10 UT in the observer's reference frame) which is the mid time of
           the first GROND pointing (OB2\_1, TDP1).
           The observation identifiers are OB2\_1 ... OB2\_7, OB3\_1 ... OB3\_5,
           OB4\_1 ... OB4\_6, OB5\_1, OB6\_1 ... OB6\_7, OB7\_1 ... OB7\_8, with
           TDP1, TDP2, TDP3, and TDP4 in each observation. There are approximately 2\,h 
           lost (at approximately half time) due to bad weather. \label{tbl:grond_day1_r}}
\begin{tabular}{rcc||rcc||rcc}
\hline\hline\noalign{\smallskip}
   TDT           &  r$^{\prime}$ & r$^{\prime}$ err & TDT           &  r$^{\prime}$ & r$^{\prime}$ err & TDT               &  r$^{\prime}$ & r$^{\prime}$ err \\
   mid time [s]  &  [mag]        & [mag]            & mid time [s]  &  [mag]        & [mag]             & mid time [s]  &  [mag]        & [mag]            \\
\noalign{\smallskip}\hline\noalign{\smallskip}
    0.00  &    18.202 & 0.016  &   4587.84  &    18.238 & 0.023  &    16269.12  &    18.230 & 0.013 \\
  103.68  &    18.178 & 0.015  &   4691.52  &    18.235 & 0.017  &    16372.80  &    18.228 & 0.018 \\
  207.36  &    18.236 & 0.019  &   4795.20  &    18.230 & 0.020  &    16476.48  &    18.235 & 0.017 \\
  311.04  &    18.186 & 0.017  &   4907.52  &    18.196 & 0.018  &    16588.80  &    18.214 & 0.016 \\
  423.36  &    18.195 & 0.019  &   5019.84  &    18.236 & 0.020  &    16701.12  &    18.239 & 0.018 \\
  527.04  &    18.169 & 0.029  &   5123.52  &    18.203 & 0.018  &    16804.80  &    18.236 & 0.021 \\
  639.36  &    18.145 & 0.039  &   5227.20  &    18.235 & 0.023  &    16908.48  &    18.186 & 0.018 \\
  743.04  &    18.205 & 0.033  &   5348.16  &    18.239 & 0.025  &    17383.68  &    18.203 & 0.021 \\
  864.00  &    18.237 & 0.026  &   5451.84  &    18.223 & 0.021  &    17487.36  &    18.214 & 0.016 \\
  967.68  &    18.206 & 0.033  &   5555.52  &    18.232 & 0.034  &    17591.04  &    18.179 & 0.017 \\
 1071.36  &    18.179 & 0.021  &   5659.20  &    18.228 & 0.018  &    17703.36  &    18.210 & 0.015 \\
 1175.04  &    18.225 & 0.015  &   5780.16  &    18.270 & 0.035  &    17815.68  &    18.230 & 0.016 \\
 1296.00  &    18.197 & 0.019  &   5883.84  &    18.213 & 0.051  &    17919.36  &    18.250 & 0.016 \\
 1399.68  &    18.195 & 0.022  &   5987.52  &    18.237 & 0.038  &    18023.04  &    18.223 & 0.018 \\
 1503.36  &    18.209 & 0.017  &   6099.84  &    18.215 & 0.035  &    18126.72  &    18.231 & 0.016 \\
 1607.04  &    18.181 & 0.018  &   6212.16  &    18.239 & 0.057  &    18247.68  &    18.254 & 0.014 \\
 1728.00  &    18.200 & 0.027  &   6315.84  &    18.174 & 0.041  &    18351.36  &    18.244 & 0.017 \\
 1831.68  &    18.173 & 0.042  &   6419.52  &    18.200 & 0.029  &    18455.04  &    18.251 & 0.017 \\
 1935.36  &    18.218 & 0.055  &   6531.84  &    18.241 & 0.022  &    18558.72  &    18.262 & 0.019 \\
 2039.04  &    18.178 & 0.060  &   ---      &    ---    & ---    &    18679.68  &    18.256 & 0.016 \\
 2160.00  &    18.261 & 0.055  &  13564.80  &    18.195 & 0.019  &    18774.72  &    18.257 & 0.023 \\
 2263.68  &    18.153 & 0.033  &  13668.48  &    18.195 & 0.019  &    18878.40  &    18.290 & 0.020 \\
 2367.36  &    18.209 & 0.021  &  13772.16  &    18.208 & 0.018  &    18990.72  &    18.221 & 0.018 \\
 2471.04  &    18.198 & 0.019  &  13875.84  &    18.195 & 0.015  &    19103.04  &    18.226 & 0.016 \\
 2592.00  &    18.229 & 0.016  &  14454.72  &    18.229 & 0.025  &    19206.72  &    18.221 & 0.016 \\
 2704.32  &    18.192 & 0.015  &  14558.40  &    18.208 & 0.029  &    19310.40  &    18.239 & 0.017 \\
 2808.00  &    18.209 & 0.017  &  14662.08  &    18.190 & 0.023  &    19414.08  &    18.240 & 0.017 \\
 2911.68  &    18.208 & 0.016  &  14765.76  &    18.204 & 0.014  &    19535.04  &    18.235 & 0.016 \\
 3188.16  &    18.228 & 0.022  &  14878.08  &    18.231 & 0.017  &    19638.72  &    18.254 & 0.020 \\
 3291.84  &    18.195 & 0.020  &  14981.76  &    18.189 & 0.016  &    19742.40  &    18.245 & 0.017 \\
 3395.52  &    18.213 & 0.016  &  15085.44  &    18.234 & 0.017  &    19854.72  &    18.241 & 0.017 \\
 3499.20  &    18.184 & 0.021  &  15197.76  &    18.224 & 0.016  &    19967.04  &    18.250 & 0.017 \\
 3620.16  &    18.248 & 0.017  &  15310.08  &    18.200 & 0.017  &    20070.72  &    18.225 & 0.017 \\
 3723.84  &    18.215 & 0.022  &  15413.76  &    18.172 & 0.022  &    20174.40  &    18.201 & 0.018 \\
 3827.52  &    18.242 & 0.016  &  15517.44  &    18.239 & 0.018  &    20286.72  &    18.264 & 0.017 \\
 3939.84  &    18.220 & 0.023  &  15629.76  &    18.204 & 0.017  &    20399.04  &    18.264 & 0.019 \\
 4052.16  &    18.231 & 0.022  &  15742.08  &    18.224 & 0.018  &    20502.72  &    18.230 & 0.020 \\
 4155.84  &    18.185 & 0.021  &  15845.76  &    18.199 & 0.016  &    20606.40  &    18.204 & 0.017 \\
 4259.52  &    18.218 & 0.023  &  15949.44  &    18.246 & 0.017  &    20710.08  &    18.265 & 0.018 \\
 4363.20  &    18.215 & 0.022  &  16053.12  &    18.231 & 0.017  &              &           & \\
 4475.52  &    18.208 & 0.020  &  16165.44  &    18.207 & 0.018  &              &           & \\
\noalign{\smallskip}\hline         
\end{tabular}
    \end{center}
}
\end{table*}

  \begin{table*}[h!tb]
{\tiny
    \begin{center}
    \caption{GROND g$^{\prime}$, r$^{\prime}$, i$^{\prime}$ magnitudes and errors
             of 162173 Ryugu.
             The zero time in the table corresponds to MJD 56088.18661 (2012-Jun-10
             04:28:43 UT in the observer's reference frame) which is the mid time of
             the first GROND pointing (OB8\_1, TDP1).
             The observation identifiers are OB8\_1 ... OB8\_4, OB9\_1 ... OB9\_8,
             OB10\_1 ... OB10\_8, OB11\_1 ... OB11\_4, with
             TDP1, TDP2, TDP3, and TDP4 in each observation. Note, that one i$^{\prime}$
             measurement is missing due to technical problems.
             \label{tbl:grond_day2_gri}}
    \begin{tabular}{rcccccc||rcccccc}
      \hline
      \hline
      \noalign{\smallskip}
   TDT           &  g$^{\prime}$ & g$^{\prime}$ err &  r$^{\prime}$ & r$^{\prime}$ err &  i$^{\prime}$ & i$^{\prime}$ err &   TDT      &  g$^{\prime}$ & g$^{\prime}$ err &  r$^{\prime}$ & r$^{\prime}$ err &  i$^{\prime}$ & i$^{\prime}$ err \\
   mid time [s]  &  [mag]        & [mag]            &  [mag]        & [mag]            &  [mag]        & [mag]     &   mid time [s]  &  [mag]        & [mag]            &  [mag]        & [mag]     &  [mag]        & [mag]            \\
      \noalign{\smallskip}
      \hline
      \noalign{\smallskip}
    0.00 &   18.687 & 0.014 & 18.232 & 0.009 & 18.179 & 0.015 &  5468.17 &   18.719 & 0.018 & 18.289 & 0.015 & 18.234 & 0.019 \\
  103.68 &   18.685 & 0.013 & 18.211 & 0.011 & 18.170 & 0.015 &  5572.89 &   18.703 & 0.017 & 18.256 & 0.015 & 18.204 & 0.017 \\
  205.72 &   18.699 & 0.014 & 18.234 & 0.012 & 18.192 & 0.015 &  5677.43 &   18.715 & 0.018 & 18.279 & 0.015 & 18.250 & 0.020 \\
  311.64 &   18.679 & 0.014 & 18.238 & 0.011 & 18.185 & 0.015 &  5782.84 &   18.701 & 0.019 & 18.251 & 0.015 & 18.182 & 0.020 \\
  430.53 &   18.698 & 0.012 & 18.251 & 0.011 & 18.146 & 0.015 &  5893.00 &   18.684 & 0.016 & 18.239 & 0.015 & 18.154 & 0.020 \\
  535.16 &   18.691 & 0.015 & 18.255 & 0.012 & 18.190 & 0.015 &  5996.76 &   18.694 & 0.018 & 18.253 & 0.015 & 18.214 & 0.022 \\
  638.58 &   18.692 & 0.014 & 18.226 & 0.011 & 18.211 & 0.015 &  6099.75 &   18.716 & 0.022 & 18.272 & 0.015 & 18.198 & 0.025 \\
  744.77 &   18.697 & 0.014 & 18.230 & 0.011 & 18.171 & 0.015 &  6204.82 &   18.660 & 0.017 & 18.240 & 0.015 & 18.243 & 0.027 \\
  862.01 &   18.726 & 0.014 & 18.235 & 0.011 & 18.179 & 0.015 &  6314.98 &   18.707 & 0.019 & 18.212 & 0.018 & 18.213 & 0.026 \\
  966.56 &   18.715 & 0.015 & 18.243 & 0.011 & 18.205 & 0.012 &  6420.73 &   18.689 & 0.024 & 18.206 & 0.017 & 18.178 & 0.020 \\
 1070.58 &   18.703 & 0.014 & 18.235 & 0.011 & 18.204 & 0.015 &  6524.67 &   18.729 & 0.024 & 18.223 & 0.019 & 18.157 & 0.025 \\
 1178.76 &   18.720 & 0.014 & 18.269 & 0.011 & 18.227 & 0.014 &  6630.68 &   18.692 & 0.022 & 18.233 & 0.019 & 18.193 & 0.025 \\
 1296.17 &   18.685 & 0.015 & 18.245 & 0.011 & 18.189 & 0.016 &  6746.03 &   18.682 & 0.022 & 18.228 & 0.019 & 18.229 & 0.018 \\
 1400.54 &   18.693 & 0.014 & 18.264 & 0.012 & 18.194 & 0.015 &  6847.63 &   18.670 & 0.024 & 18.225 & 0.019 & 18.171 & 0.025 \\
 1506.82 &   18.719 & 0.015 & 18.239 & 0.012 & 18.214 & 0.016 &  6952.78 &   18.672 & 0.025 & 18.227 & 0.019 & 18.152 & 0.026 \\
 1612.66 &   18.730 & 0.015 & 18.257 & 0.012 & 18.227 & 0.013 &  7058.62 &   18.712 & 0.027 & 18.229 & 0.022 & 18.171 & 0.028 \\
 1923.70 &   18.717 & 0.016 & 18.287 & 0.013 & 18.218 & 0.014 &  7176.04 &   18.702 & 0.024 & 18.291 & 0.022 & 18.132 & 0.028 \\
 2029.19 &   18.744 & 0.016 & 18.262 & 0.012 & 18.228 & 0.015 &  7277.64 &   18.670 & 0.028 & 18.230 & 0.025 & 18.162 & 0.031 \\
 2134.86 &   18.730 & 0.016 & 18.280 & 0.012 & 18.238 & 0.018 &  7382.71 &   18.670 & 0.030 & 18.206 & 0.025 & 18.115 & 0.030 \\
 2240.61 &   18.752 & 0.018 & 18.285 & 0.012 & 18.192 & 0.018 &  7488.81 &   18.677 & 0.032 & 18.231 & 0.025 & 18.208 & 0.028 \\
 2354.92 &   18.724 & 0.016 & 18.270 & 0.013 & 18.219 & 0.018 &  7600.78 &   18.717 & 0.032 & 18.167 & 0.031 & 18.245 & 0.047 \\
 2460.41 &   18.704 & 0.013 & 18.287 & 0.013 & 18.217 & 0.014 &  7704.72 &   18.647 & 0.037 & 18.262 & 0.033 & 18.173 & 0.042 \\
 2564.70 &   18.726 & 0.018 & 18.278 & 0.013 & 18.197 & 0.018 &  7808.66 &   18.673 & 0.037 & 18.209 & 0.033 & 18.204 & 0.042 \\
 2670.62 &   18.729 & 0.017 & 18.275 & 0.013 & 18.167 & 0.018 &  7914.59 &   18.698 & 0.034 & 18.210 & 0.033 & 18.162 & 0.043 \\
 2789.16 &   18.724 & 0.017 & 18.263 & 0.013 & 18.237 & 0.019 &  8033.13 &   18.648 & 0.045 & 18.189 & 0.034 & 18.107 & 0.047 \\
 2891.55 &   18.715 & 0.018 & 18.310 & 0.014 & 18.237 & 0.019 &  8136.72 &   18.663 & 0.040 & 18.204 & 0.028 & 18.135 & 0.045 \\
 2996.61 &   18.730 & 0.017 & 18.281 & 0.014 & 18.245 & 0.018 &  8238.84 &   18.656 & 0.041 & 18.231 & 0.033 & 18.157 & 0.046 \\
 3102.62 &   18.739 & 0.015 & 18.279 & 0.014 & 18.210 & 0.019 &  8344.43 &   18.576 & 0.033 & 18.169 & 0.027 & 18.033 & 0.034 \\
 3217.02 &   18.715 & 0.015 & 18.271 & 0.014 & 18.191 & 0.019 &  8455.97 &   18.626 & 0.032 & 18.196 & 0.027 & 18.113 & 0.034 \\
 3318.71 &   18.723 & 0.016 & 18.277 & 0.014 & 18.241 & 0.022 &  8558.70 &   18.710 & 0.036 & 18.267 & 0.028 & 18.224 & 0.037 \\
 3420.84 &   18.720 & 0.018 & 18.297 & 0.014 & 18.205 & 0.021 &  8660.65 &   18.617 & 0.036 & 18.247 & 0.033 & 18.216 & 0.040 \\
 3525.72 &   18.720 & 0.018 & 18.296 & 0.014 & 18.229 & 0.020 &  8765.71 &   18.669 & 0.043 & 18.195 & 0.033 & 18.198 & 0.045 \\
 3636.14 &   18.729 & 0.018 & 18.252 & 0.014 & 18.212 & 0.019 &  9010.92 &   18.684 & 0.032 & 18.225 & 0.028 &  ---   & ---   \\
 3741.64 &   18.764 & 0.017 & 18.254 & 0.014 & 18.206 & 0.018 &  9114.60 &   18.632 & 0.044 & 18.240 & 0.031 & 18.064 & 0.045 \\
 3845.66 &   18.739 & 0.017 & 18.271 & 0.014 & 18.239 & 0.018 &  9218.28 &   18.635 & 0.035 & 18.171 & 0.025 & 18.198 & 0.040 \\
 3952.20 &   18.705 & 0.017 & 18.265 & 0.014 & 18.198 & 0.018 &  9321.96 &   18.664 & 0.034 & 18.237 & 0.025 & 18.242 & 0.033 \\
 4064.00 &   18.719 & 0.017 & 18.281 & 0.012 & 18.210 & 0.015 &  9434.28 &   18.683 & 0.031 & 18.200 & 0.025 & 18.145 & 0.033 \\
 4166.73 &   18.702 & 0.017 & 18.274 & 0.014 & 18.167 & 0.018 &  9537.96 &   18.689 & 0.028 & 18.235 & 0.025 & 18.158 & 0.026 \\
 4271.62 &   18.738 & 0.017 & 18.264 & 0.014 & 18.226 & 0.018 &  9641.64 &   18.632 & 0.027 & 18.239 & 0.020 & 18.212 & 0.028 \\
 4377.63 &   18.724 & 0.017 & 18.262 & 0.012 & 18.224 & 0.019 &  9753.96 &   18.713 & 0.027 & 18.169 & 0.020 & 18.159 & 0.022 \\
 4492.97 &   18.708 & 0.015 & 18.274 & 0.015 & 18.167 & 0.019 &  9866.28 &   18.639 & 0.022 & 18.223 & 0.020 & 18.178 & 0.027 \\
 4596.39 &   18.737 & 0.018 & 18.292 & 0.015 & 18.175 & 0.019 &  9969.96 &   18.646 & 0.024 & 18.205 & 0.020 & 18.198 & 0.026 \\
 4701.80 &   18.718 & 0.018 & 18.243 & 0.015 & 18.217 & 0.019 & 10073.64 &   18.694 & 0.029 & 18.228 & 0.020 & 18.165 & 0.028 \\
 4807.56 &   18.741 & 0.018 & 18.285 & 0.012 & 18.200 & 0.020 & 10177.32 &   18.679 & 0.030 & 18.212 & 0.020 & 18.173 & 0.029 \\
 4922.04 &   18.736 & 0.018 & 18.248 & 0.015 & 18.226 & 0.017 & 10298.28 &   18.688 & 0.022 & 18.241 & 0.013 & 18.144 & 0.020 \\
 5024.68 &   18.724 & 0.019 & 18.271 & 0.015 & 18.230 & 0.021 & 10401.96 &   18.717 & 0.025 & 18.247 & 0.017 & 18.188 & 0.022 \\
 5129.57 &   18.692 & 0.018 & 18.275 & 0.015 & 18.190 & 0.020 & 10505.64 &   18.666 & 0.021 & 18.242 & 0.013 & 18.195 & 0.022 \\
 5235.67 &   18.682 & 0.018 & 18.278 & 0.015 & 18.213 & 0.021 & 10609.32 &   18.644 & 0.021 & 18.220 & 0.013 & 18.188 & 0.021 \\
     \noalign{\smallskip}
     \hline
    \end{tabular}
    \end{center}
}
  \end{table*}

\end{appendix}
\end{document}